\documentclass{jpp}
\usepackage{epstopdf, epsfig}
\usepackage{hyperref}
\usepackage{amsmath,amsfonts,amssymb}
\usepackage{bm}
\usepackage{appendix}
\usepackage{orcidlink}

\newcommand*\pFq[2]{
        {}_{#1}F_{#2}\genfrac[]{0pt}{}
}

\shorttitle{Kinetic eigenfunctions}
\shortauthor{D.W. Crews, U. Shumlak}

\title{Phase space eigenfunctions with applications to continuum kinetic simulations} 

\author{Daniel W. Crews\orcidlink{0000-0003-0921-3515}\aff{1}\corresp{\email{daniel.crews@zap.energy}} and Uri Shumlak\orcidlink{0000-0002-2918-5446}\aff{1,2}}

\affiliation{\aff{1}Theory and Modeling Division, Zap Energy Inc., Everett, Washington 98203, USA
  \aff{2}Computational Plasma Dynamics Lab, Aerospace and Energetics Research Program,
  University of Washington, Seattle, Washington 98195--2400, USA}

\begin{document}

\maketitle

\begin{abstract}
  Continuum kinetic simulations are increasingly capable of resolving high-dimensional phase space
  with advances in computing.
  These capabilities can be more fully explored by using linear kinetic theory to initialize
  the self-consistent field and phase space perturbations of
  kinetic instabilities.
  The phase space perturbation of a kinetic eigenfunction in unmagnetized plasma has a simple analytic form,
  and in magnetized plasma may be well approximated by truncation of a cyclotron-harmonic expansion.
  We catalogue the most common use cases with a historical discussion of kinetic eigenfunctions
  and by conducting nonlinear Vlasov-Poisson and Vlasov-Maxwell simulations
  of single- and multi-mode two-stream, loss-cone, and Weibel instabilities
  in unmagnetized and magnetized plasmas with one- and two-dimensional geometries.
  Applications to quasilinear kinetic theory are discussed and applied to the bump-on-tail instability.
  In order to compute eigenvalues we present novel representations of the dielectric function for ring distributions
  in magnetized plasmas with power series, hypergeometric, and trigonometric integral forms.
  Eigenfunction phase space fluctuations are visualized for prototypical cases such as the Bernstein modes
  to build intuition.
  In addition, phase portraits are presented for the magnetic well associated
  with nonlinear saturation of the Weibel instability,
  distinguishing current-density-generating trapping structures from charge-density-generating ones.
  
  
  
\end{abstract}

\keywords{Plasma Instabilities, Plasma Simulation, Plasma Waves}

\section{Introduction}\label{sec:intro}
    Kinetic equations describe fundamental dynamics in collisionless plasma,
    so their linear analysis and nonlinear simulation is a
    perennial topic~(\cite{bertrand_waterbag, cheng_knorr, birdsall2004plasma, heath_dg, morrison_metriplectic}).
    Plasma kinetic theory and non-equilibrium thermodynamics are closely connected,
    and the basic problem in collisionless kinetic theory is to describe the stability of a
    distribution to collective perturbations~(\cite{penrose}).
    In the simplest spatially homogeneous plasmas instability modes arise from nonthermal distributions
    due to streaming, anisotropic pressure~(\cite{weibel_orig}), loss-cones~(\cite{rosenbluth_losscone}), etc.
    Put simply, these instabilities arise 
    from entropically unfavorable distributions of relative velocity.
    Historically, kinetic instabilities have been a key part of 
    transport theory in collisionless plasma~(\cite{drummond_micro_diffusion, yoon_anom}),  
    and kinetic simulations of instability have contributed to advances in dynamical sciences~(\cite{escande_contrib})
    and modeling for fluid closures~(\cite{conner1994survey}).
    The intrinsic complexity of phase space turbulence arising from kinetic instabilities has made nonlinear simulation
    key to the construction of more comprehensive models,
    and a huge potential remains for kinetic simulations to enrich plasma theory as advances
    in technique and computing power overcome the curse of dimensionality~(\cite{ham_pic, choi2021high}).

    We begin by stating the purpose of this work and its context in the existing literature, by clarifying commonly used terms here,
    and by listing which codes may benefit from the described techniques.
    This work reviews eigenfunction solutions to the linearized plasma kinetic equations,
    here also referred to as ``self-consistent plasma-field configurations'' or ``kinetic eigenfunctions'',
    and discusses how these kinetic eigenfunctions can be utilized to
    cleanly initialize kinetic instabilities simulated by the continuum kinetic method.
    Here ``continuum kinetic method'' means that the kinetic equation is solved by an Eulerian method in the phase space,
    for example the finite element method~(\cite{heath_dg}), in contrast to the particle-in-cell method.
    The phase space configuration corresponding to the kinetic eigenfunction will be referred to as a ``phase space eigenfunction''.
    Then, having reviewed the basic theory and discussed the benefits of eigenfunction initialization, the method is illustrated
    with model problems of streaming, pressure anisotropy, and loss-cone cyclotron instabilities.
    Commentary on the instability physics is made throughout and some novel results are noted.
    A summary is made of novel results in the conclusion.

    Historically, phase space eigenfunctions have been utilized in the space physics community for analysis of collisionless
    energy transfer, for example in the role played by kinetic slow modes in the solar wind~(\cite{dsp_kinetic_slow_modes1, dsp_kinetic_slow_modes2})
    or by phase space perturbations in Alfvenic turbulence~(\cite{dsp_kinetic_alfven_turbulence,dsp_alfven_energy_transfer}).
    In the realm of simulation,
    the major continuum fusion gyrokinetic codes (GENE, GS2, GYRO/CGYRO) all have a kinetic eigenfunction initialization capability
    regularly employed in studies of astrophysical and laboratory fusion plasmas~(\cite{gyrokinetics1, gyrokinetics2, gyrokinetics3, gyrokinetics4}).
    On the other hand, codes solving the unreduced kinetic models (Vlasov-Poisson or Vlasov-Maxwell) with continuum kinetic
    method appear, to the best of our knowledge, not to utilize kinetic eigenfunctions.
    The methodology advocated for in this work should be applicable and beneficial for all continuum kinetic codes, such as
    HVM~(\cite{hvm_valentini}), Vlasiator~(\cite{vlasiator_kempf1, vlasiator_kempf2}), Gkeyll~(\cite{gkeyll1,gkeyll2}), ViDA~(\cite{vida}),
    the Lawrence Livermore Vlasov-Poisson code~(\cite{LLNL_code}), the Ruhr University Vlasov code~(\cite{ruhr_university_code}),
    WARPXM~(\cite{SHUMLAK20111767, datta_kinetic}), and the Los Alamos spectral Vlasov solver~(\cite{los_alamos_spectral1, los_alamos_spectral2}).
    Eigenfunctions of the unreduced kinetic equations are analytically tractable in many situations of interest, for example
    the unmagnetized (streaming) or magnetized (loss-cone) electrostatic instabilities, and
    the electromagnetic unmagnetized (Weibel) or magnetized (field-parallel whistler) pressure anisotropy instabilities.
    Of these examples, the whistler is not treated in this work.
    
    
    
    
    Kinetic eigenfunctions consist of structure in both the field and the phase space self-consistently.
    It is fairly common practice to initialize simulations using only the field-part of the eigenfunctions
    but without the corresponding phase space configuration,
    with field sources obtained by spatial moments of a Maxwellian distribution~(\cite{hvm_clarify_spatial_mode, gkeyll_clarify_spatial_mode}).
    For example, to achieve density perturbations $n_1/n_0 = A\sin(kx)$
    the equilibrium distribution function $f_0$ is perturbed as $f_1(x,v) = A\sin(kx)f_0(v)$.
    Indeed, linear modes~(namely, self-consistent plasma-field configurations) 
    have a phase space structure more like $f_1(x,v) = (\zeta - v)^{-1}\partial_vf_0 e^{ikx}$ with $\zeta=\omega/k$ a complex
    phase velocity, and so a significant portion of the perturbation energy is channeled into the Landau-damped modes 
    and the model's energy trace begins by a transient energetic reorganization via Landau damping.
    The consequence is that spatial moment-based perturbations partition
    perturbation energy in a manner different from what the researcher may have intended.
    The partition of energy into non-eigenfunction perturbations is not harmful to the intended solution when 
    initial amplitudes are small, but at larger perturbation amplitudes 
    Landau damping may non-physically contribute to nonlinear phenomena as these modes are usually activated by thermal fluctuations.
    There is an additional numerical consideration of perturbations with both growing and damped components,
    in which case precise measurement of growth rates is obscured by Landau damping.  
    
    To recapitulate, the numerical and theoretical methods and results presented here
    should be beneficial for those conducting continuum kinetic simulations,
    including researchers utilizing the major codes listed above,
    those investigating new numerical methods for kinetic equations~(\cite{einkemmer2019performance}),
    and for problems considered with smaller targeted codes~(\cite{crews_kinetic, paul_and_sharma}).
    The methods presented here apply most directly to approaches
    where the perturbed distribution may be functionally specified.
    This is the case in most continuum kinetic numerical methods,
    such as in our work where we use a mixed Fourier spectral/finite element
    method discussed in Appendix~\ref{app:summary_of_numerical},
    with some variations noted throughout the applications.
    However, it seems to the authors that sophisticated
    particle-in-cell methods~(\cite{kraus_2017,glasser_qin_2020,benedikt_geometric,barnes_pic})
    can utilize these results as well.

    This article is organized as theory followed by simulation, treating progressively the unmagnetized and
    magnetized electrostatic and unmagnetized electromagnetic problems.
    Section~\ref{sec:electrostatic_theory} reviews unmagnetized electrostatic phase space eigenfunctions and Landau modes,
    including a historical summary,
    a review of the initial-value problem,
    and an energetic analysis.
    The difference between kinetic eigenfunctions and Landau damping modes is discussed, with only unstable perturbations
    resulting in genuine eigenfunctions.
    Section~\ref{subsec:quasilinear-theory} notes the importance of kinetic eigenfunctions in quasilinear theory (QLT).
    Section~\ref{sec:magnetized-electrostatic-plasma-modes} then discusses the electrostatic cyclotron modes, 
    making new connections to the theory of special functions through
    hypergeometric functions and the Laguerre polynomials in the dielectric tensor of loss cones,
    and explores the helical phase space structure of Bernstein modes.
    Section~\ref{sec:electromagnetic_theory} considers the vector eigenmodes of the Vlasov-Maxwell system
    by casting the dielectric tensor as an eigenvalue problem for a system of integral equations
    and presents a practical method to calculate them.
    The natural consistency of the plasma-field configuration resulting from this method is
    observed as a benefit, so that the initial condition automatically satisfies Poisson's equation, for example.
    The distinction between phase space eigenfunctions and Landau damping modes is generalized to the vector case,
    and the one-dimensional and two-dimensional Weibel instabilities are explored for the anisotropic
    Maxwellian distribution.
    
    Illustrative simulations are presented in the relevant sections and include the
    multidimensional multi-mode two-stream instability in Section~\ref{subsec:two_d_electrostatic_simulation},
    single-mode Dory-Guest-Harris instability in Section~\ref{subsec:two_case_studies},
    single-mode Weibel instability in Section~\ref{subsec:single-mode-saturation-one-d}, and multidimensional
    multi-mode Weibel problem in Section~\ref{subsec:two_dimensional_weibel_simulation}.
    Commentary is provided throughout on the physics of these problems as nonlinear structures
    evolve from the initialized linear eigenfunctions.
    The general numerical method is described in Appendix~\ref{app:summary_of_numerical} and problem-specific modifications
    are noted for each problem prior to initialization specifics.
    The magnetized electromagnetic problem is not treated here, but we mention that the analytic kinetic eigenfunctions
    are fairly simple, as parallel-field whistler modes, for example, involve only the first cyclotron harmonic.
    The linear theory for whistler emission can be found in~\cite{gurnett2017introduction}, page 415, and the
    eigenfunction is briefly described in Chapter 7 of~\cite{crews2022numerical}.

    \section{Electrostatic plasma modes with zero-order ballistic trajectories}\label{sec:electrostatic_theory}
    Perhaps the simplest problem in plasma kinetic theory is that of electrostatic modes in unmagnetized homogeneous plasma,
    meaning that the zero-order orbits are simply free-streaming ballistic motions.
    Here the problem is treated beginning with a historical discussion,
    followed by analysis of the initial-value problem, and an energetic analysis of response
    to eigenfunction and non-eigenfunction phase space perturbations. 
    Recall that there are two ways of considering the linearized dynamics:
    the eigenvalue problem ($t\in(-\infty,\infty)$) and the
    initial-value problem ($t\in[0,\infty)$).
    A somewhat subtle theoretical point is the distinction between eigenfunctions and Landau-damped modes;
    to clarify this distinction requires a review of the Case-van Kampen modes and
    the theory of linear Landau damping, which can be found in~\cite{crews2022numerical} and is omitted here for brevity.
    Nevertheless, in the following the distinction is reviewed at a qualitative level.

    \subsection{Historical summary and context}\label{subsec:historical_summary}
    The study of eigenfunctions of collisionless and collisional plasma kinetic equations has a long history.
    Vlasov was the first to suggest a method to estimate the plasma oscillation frequencies by prescribing that
    the principal value be taken at the resonant velocity in the dispersion function.
    Following this~\cite{bohmgross} side-stepped the resonant term by considering
    high-enough phase velocities that the distribution function was effectively zero at the pole.
    Famously~\cite{landau_damping} formally solved the linearized initial-value problem
    for Vlasov-Poisson dynamics using Laplace transformation, discovering a discrete set of solutions
    for the Maxwellian plasma, wherein the electric potential damped in time.
    This collisionless decay phenomenon is called Landau damping.
    For the collisionless plasma the decaying Landau modes are not eigenfunctions,
    meaning that such solutions cannot evolve independently.
    Yet Landau's analysis also found unstable modes for certain distributions $f_0$.
    Thus, despite both unstable and dissipative modes sharing a similar phase space
    structure like $(v-\zeta)^{-1}\partial_vf_{0}e^{ikx}$, unstable modes evolve as a single analytic function
    while dissipation spreads across the entire Landau spectrum.

    As the eigenvalue problem remained unsolved,~\cite{van1955theory} and~\cite{case1959plasma}
    formally solved the problem and determined the spectrum of the linearized kinetic equation
    to be continuous with a possibly discrete component.
    Discrete eigenvalues arise only for unstable modes where $\text{Im}(\omega) > 0$.
    The continuous part of the spectrum consists of ballistic modes
    in the form $\varepsilon(k,\zeta)\delta(v - \zeta)$,
    while the discrete part in the analytic form $(v-\zeta)^{-1}\partial_v f_0$.
    Here $\varepsilon(k,\zeta)$ is the dielectric function and $f_0(v)$ the homogeneous equilibrium.
    The discrete part of the Case-van Kampen spectrum is identical to Landau's unstable modes.
    Landau's damping modes are represented as an integral expansion over the continuous Case-van Kampen spectrum.
    As a complete orthogonal system the Case-van Kampen modes are a useful though under-utilized tool.
    An insightful application was accomplished by P.J.~Morrison and colleagues in
    constructing a linear integral transform, termed ``G transform,''
    to reduce the linearized Vlasov equation to an advection problem
    by utilizing the Case-van Kampen modes as a basis~(\cite{morrison1992dielectric, morrison2000hamiltonian, heninger_morrison_2018}).
    

    As the Landau damping modes are not collisionless eigenfunctions their status has remained somewhat obscure.
    Light is shed on this problem by considering weak dissipation in the Vlasov equation, such as the
    collision operator of~\cite{lenard}.
    The landmark studies of~\cite{numerical_ng} and~\cite{analytical_short} have shown that
    as dissipation tends to zero the dissipative eigenfunctions converge to the Landau damping modes,
    with the conclusion that dissipation is a singular perturbation of the collisionless dynamics.
    \cite{bratanov} numerically confirmed this limit for discrete systems and~\cite{ng_eigenmodes_collisional_2004}
    extended and formalized the results for weakly collisional plasma.
    However, the authors wish to highlight here that the Case-van Kampen modes still play a role in the dissipative picture.
    Namely, one can show~(\cite{bratanov,crews2022numerical}) that the propagator of the
    linearized kinetic equation with the Lenard-Bernstein operator limits to the Case-van Kampen modes as the
    dissipation $\nu\to 0^+$.
    This fact clarifies the relationship between the continuous Case-van Kampen spectrum and Landau damping modes,
    as even in the dissipative picture the Landau modes are represented as an integral over the diffusive propagator, and
    the collisionless Case-van Kampen modes indeed play the role of non-diffusive propagators~(\cite{balescu1997statistical}).
    To understand why Landau modes are easily identified in the initial-value problem, consider that Landau
    damping originates from phase mixing~(\cite{villani_2011}) and so the modes possess the peculiar
    property of decaying in both directions of time.
    Thus they arise by propagating initial data, or otherwise must be represented 
    as an interference of free-streaming modes un-mixing from
    $t\in (-\infty, 0)$ and re-mixing from $t\in (0, \infty)$.

    In summary, in collisionless plasma kinetic theory unstable modes are normal modes and evolve independently,
    while dissipative modes either occur as a summation of non-orthogonal transient modes or must be represented in the
    Case-van Kampen spectrum.
    The Case-van Kampen spectrum has found fruitful application as the basis of constructing an integral transform theory
    for linearized dynamics, most recently explored in~\cite{heninger_morrison_2018}.
    In weakly collisional dynamics dissipative modes are also eigenfunctions
    and limit to the collisionless Landau mode spectrum as dissipation tends to zero.
    As a consequence, it is possible to excite a single-mode instability by initializing its eigenfunction,
    but one may not excite a lone Landau-damping plane wave under collisionless dynamics.
    The following discussion illustrates this point.

    \subsection{Phase space linear response and the dielectric function}\label{subsec:linear_response}
    The electric susceptibility $\chi$ is the linear response function which relates in
    the spatiotemporal frequency domain $(\bm{x}, t) \to (\bm{k},\omega)$ the polarization
    $\bm{\widetilde{P}}(\bm{k},\omega)$ and electric field $\bm{\widetilde{E}}(\bm{k}, \omega)$ by the constitutive relation
    $\bm{\widetilde{P}}=\varepsilon_0\chi\bm{\widetilde{E}}$.
    In the scheme of electrostatic theory one aims to determine the susceptibility and consequently
    the dielectric permittivity $\varepsilon(\bm{k},\omega) = \varepsilon_0(1 + \chi)$ of a plasma with equilibrium
    distribution $f_0(\bm{v})$.
    The susceptibility is determined through the self-consistent particle response $f_1(\bm{k},\bm{v},\omega)$ such that
    $f = f_0 + f_1$,
    leading to the sought-after modal structures in the charge density.
    The distribution $f_1$ encodes a linear response of the charges to the electric potential
    $\varphi$ according to the relation $f_1 = h(\bm{k}, \bm{v}, \omega)\varphi$ for some response function $h$.
    For this reason, we mean by ``phase space linear response function'' the self-consistent particle response $f_1$
    to the potential $\varphi$.
    The permittivity is often referred to as simply the dielectric function
    because the permittivity tensor reduces to a scalar for isotropic equilibria.

    The following review of the initial-value problem is done in detail for the simplest case in order
    to build intuition for the electromagnetic and magnetized problems where propagation of the
    initial data is tedious\footnote{
    The following analysis applies quasi-analysis, which considers
    the eigenvalue problem and analytically continues the dispersion function from the upper-half to the lower-half
    complex frequency plane. The analysis does not consider damped modes to be eigenfunctions.}.
    Using the results, we then consider the evolution of perturbations in a plasma with Maxwellian $f_0$,
    namely a so-called Maxwellian perturbation (that is, where $f_1(x,v) = A\sin(kx)f_0(v)$)
    and a special pertubation describing a propagating, damping plane wave.
    When this plane wave is unstable the special pertubation grows as an eigenmode,~\textit{i.e.} a constant phase
    space structure with time-dependent amplitude, and when damped its structure evolves in phase space.

    Rather than the usual Laplace transform notation, we use a more standard notation for
    the one-sided temporal Fourier transform pair as
    \begin{align}
        f(\omega) &= \int_0^{\infty} f(t)e^{i\omega t}dt,\label{eq:one_sided}\\ %
        f(t) &= \frac{1}{2\pi}\int_{-\infty + is}^{\infty + is}f(\omega)e^{-i\omega t}d\omega\label{eq:inverse}%
    \end{align}
    where the factor $s$ keeps the contour above all poles of the integrand in order that Eq.~\ref{eq:one_sided}
    converges.
    The contour in Eq.~\ref{eq:inverse} is closed at infinity around the lower half-plane, and the inverse transform
    is then given by the residue theorem as
    \begin{equation}\label{eq:residue_theorem}
        f(t) = -i\sum\text{Res}(f(\omega))
    \end{equation}
    with the sum over all poles of the response function $f(\omega)$.
    The spatial Fourier transform $\bm{x}\to\bm{k}$ is defined as usual over all of space by
    $f(\bm{k}) = \int d\bm{x}e^{-i\bm{k}\cdot\bm{x}}f(\bm{x})$.
    We consider for simplicity the response of a single species of particle charge $q$ and mass $m$ amidst a uniform
    neutralizing Maxwellian background, so we do not write a species subscript.

    The Vlasov-Poisson system linearized by $f(\bm{x},\bm{v},t) = f_0 + f_1$ with $f_1\ll f_0$ is given by
    \begin{align}
        \frac{\partial f_{1}}{\partial t} &+ \bm{v}\cdot\nabla f_{1} - \frac{q}{m}(\nabla_x\varphi)\cdot\nabla_v
        f_{0} = 0,\label{eq:linear_transport}\\
        \nabla^2\varphi &= -\frac{qn_0}{\varepsilon_0}\int f_{1} d\bm{v}.
    \end{align}
    Fourier transforming for all of space and for time $t\in(0,\infty)$ as described, the phase space linear response is obtained
    as
    \begin{equation}\label{eq:linear_response_function}
        f_1(\bm{k}, \bm{v}, \omega) = i\frac{g(k,v)}{\omega - \bm{k}\cdot\bm{v}}
    - \frac{q}{m}\varphi\frac{\bm{k}\cdot\nabla_v f_{0}}{\omega - \bm{k}\cdot\bm{v}}.
    \end{equation}
    where $g(k,v)=f_1(t=0,k,v)$ is the initial condition.

\begin{figure}
        \centering
        \includegraphics[width=0.45\columnwidth]{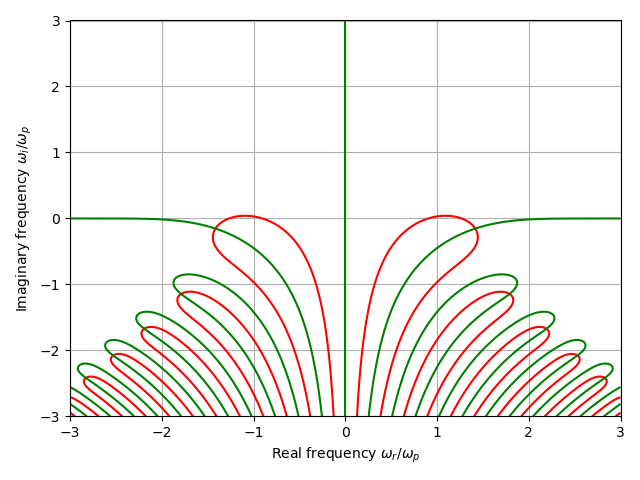}
        \caption{Locations of solutions to $\varepsilon(k,\omega)=0$ showing contours of $\text{Re}(\varepsilon)=0$ in
        red and $\text{Im}(\varepsilon)=0$ in green for a Maxwellian
        $f_0(v)$ at short wavelength, $k\lambda_D=0.5$.
        Solutions to $\varepsilon(k,\omega)=0$ occur at the intersection of the real and imaginary zero-contours.
        These damped modes represent the transient Landau spectrum which accompanies
        any simulation of kinetic instability not initialized with an eigenfunction.}\label{fig:kinetic_damping_modes}
    \end{figure}

    The following result is as described in~\cite{landau_damping} with some change in notation.
    Define the Cauchy integral
    $g^*(k,\zeta) \equiv \int_{\mathcal{C}}\frac{g(k,v)}{\zeta - v}dv$, with $\mathcal{C}$ Landau's contour (that is,
    analytically-continued into the lower-half $\zeta$-plane).
    Choose the coordinates such that one axis aligns with the wavevector $\bm{k}$, so that when computing the zeroth
    velocity moment of $f_1$, two of the velocity coordinates integrate out.
    The potential is found to be
    \begin{equation}\label{eq:potential_response}
      \varphi(k, \zeta) = i\frac{\sigma}{k^3}\frac{g^*(k,\zeta)}{\varepsilon(k,\zeta)}
    \end{equation}
    where $\sigma = q/|q|$ and
    $\varepsilon(k, \zeta) = 1 + k^{-2}\int_{\mathcal{C}}\frac{1}{\zeta-v}\frac{\partial f_0}{\partial v}dv$
    is the electrostatic dielectric function.
    Here the wavenumber $k$ is normalized to the Debye length.
    If each root $\zeta_j$ of $\varepsilon(k,\zeta)=0$ is simple and $g^*(k,\zeta)$ has \textit{no poles}
    then inverse transforming in time gives
    \begin{equation}\label{eq:potential_solution}
        \varphi(k, t) = \frac{\sigma}{k^3}\sum_j\frac{g^*(k,\zeta_j)}{\varepsilon_\omega(k,\zeta_j)}e^{-i\omega_jt}
    \end{equation}
    with $\varepsilon_\omega \equiv \frac{\partial\varepsilon}{\partial\omega}$.
    Figure~\ref{fig:kinetic_damping_modes} shows the typical locations of the infinite set of zeros $\varepsilon(k,\omega)=0$
    in the lower-half frequency plane making up the Landau damping mode spectrum.
    The amplitude of each mode $\zeta_j$ is given by the Cauchy integral of the initial perturbation weighted
    by that mode's factor $\varepsilon_\omega^{-1}$.
    For example, a typical perturbation used in numerical simulations is Maxwellian in velocity space,
    namely $g(v) = (\sqrt{\pi}v_t)^{-1}e^{-v^2/v_t^2}$.
    In this case our Cauchy integral is $g^*(\zeta)= v_t^{-1}Z(\zeta/v_t)$ with $Z(\zeta)$
    the plasma dispersion function defined by~\cite{fried_plasma}.
    Normalizing phase velocity to $v_t$, the self-consistent electric potential response for the Maxwellian
    perturbation is
    \begin{equation}\label{eq:maxwellian_potential}
      \varphi(k,t) = -\sigma\sum_{j}\frac{Z(\zeta_j)}{Z''(\zeta_j)}e^{-i\omega_j t}
    \end{equation}
    as $\varepsilon(\zeta)=1-k^{-2}Z'(\zeta)$ and  $\varepsilon'(\zeta)= -k^{-2}Z''(\zeta)$.

    In the theory of continuous dielectrics~(\cite{landau1946electrodynamics,nicholson1983introduction}),
    wave energy density consists of electric field energy density
    multiplied by the so-called Brillouin factor $\partial_\omega(\omega\varepsilon_r)$.
    The denominator of Eq.~\ref{eq:potential_solution} evokes the Brillouin factor
    because for each root $\omega_n=k\zeta_n$, we have
    $\partial_\omega(\omega\varepsilon) = \omega_n\varepsilon_\omega(\omega_n)$.
    This energy factor $\varepsilon_\omega(k, \zeta_j)$ is monotonically increasing towards the higher Landau modes.
    For this reason we speculate that in Landau damping the lowest-energy state is also the least-damped mode,
    and that the lower-energy states are also of greater amplitude in general perturbations.
    
    \subsection{Electrostatic eigenfunctions and transient responses}\label{subsec:electrostatic_and_transient}
    Equation~\ref{eq:potential_solution} is the response when the initial data $g^*(k,\zeta)$ is an
    entire function of $\zeta$.
    However, the phase space linear response Eq.~\ref{eq:linear_response_function} itself has a simple pole.
    Since the initial data supplied to the kinetic equation is supposed to simulate the self-consistent response of the
    plasma to some perturbation, it follows that appropriate initial data may also have a pole.
    Consider the initial condition to be the linear response,
    \begin{equation}\label{eq:special}
        g(k, v) = -\frac{q}{m}\frac{1}{k^2}\frac{1}{\zeta_n-v}\frac{\partial f_0}{\partial v}
    \end{equation}
    where $\zeta_n$ is a root of the dielectric function, $\varepsilon(k, \zeta_n) = 0$.
    The Cauchy transform of Eq.~\ref{eq:special} is the dielectric function with an added residue for $\text{Im}(\zeta)<0$,
    \begin{equation}\label{eq:transform_of_special}
        g^*(k,\zeta) = \frac{\sigma}{\zeta-\zeta_n}\Big(\varepsilon(k,\zeta) +
        \begin{cases}
            0, & \text{Im}(\zeta_n) > 0\\
            \frac{2\pi i}{k^2}\frac{\partial f_0}{\partial v}\Big|_{v=\zeta_n}, & \text{Im}(\zeta_n) < 0
        \end{cases}\Big)
    \end{equation}
    Consider first $\text{Im}(\zeta)> 0$.
    Combining Eqs.~\ref{eq:potential_response} and~\ref{eq:transform_of_special}
    and performing an inverse Fourier transform with Eq.~\ref{eq:inverse} gives the potential
    $\varphi=k^{-2}e^{i(kx-\omega_nt)}.$
    The phase space structure is given by Eq.~\ref{eq:linear_response_function},
    and again observing the form of Eq.~\ref{eq:special} and combining with the expression for the potential gives
    \begin{equation}\label{eq:distribution_special_sol}
        f_1(x,v,t) = g(k,v)e^{i(kx - \omega_n t)}.
    \end{equation}
    Therefore Eq.~\ref{eq:distribution_special_sol} is a linear eigenfunction of the Vlasov equation.
    The phase space fluctuation grows in time and there is no phase mixing.
    On the other hand, in the case of $\text{Im}(\zeta)<0$ the spectral potential contains a residue,
    \begin{equation}\label{eq:transform_of_damping_ic}
        \varphi(k,\zeta) = \frac{i}{k^3}\frac{1}{\zeta-\zeta_n} -
    \frac{2\pi}{k^3}\frac{1}{(\zeta-\zeta_n)\varepsilon(k, \zeta)}\frac{1}{k^2}\frac{\partial f_0}{\partial v}\Big|_{v=\zeta_n}.
    \end{equation}
    Equation~\ref{eq:transform_of_damping_ic} has a double pole at $\zeta=\zeta_n$ in the second term,
    and a simple pole at all other roots $\varepsilon(k, \zeta_j)=0$ with $j\neq n$.
    Inverting the solution obtains the expression
    \begin{equation}\label{eq:potential_damping_time}
    \begin{split}
        \varphi(k,t) &= \Big\{1 +
    \frac{i\pi}{k^2}\frac{\varepsilon_{\omega\omega}(k,\zeta_n)}{\varepsilon_\omega^2(k,\zeta_n)}
    \frac{\partial f_0}{\partial v}\Big|_{v=\zeta_n}\Big\}\frac{e^{-i\omega_n t}}{k^2}\\
        &\quad\quad + \sum_{j\neq n}\Big(\frac{1}{(\omega_n-\omega_j)\varepsilon_\omega(k,\zeta_j)}
    \frac{2\pi}{k^2}\frac{\partial f_0}{\partial v}\Big|_{v=\zeta_n}\Big)\frac{e^{-i\omega_jt}}{k^2}.
        \end{split}
    \end{equation}
    Although the kinetic mode with frequency $\omega_n$ is preferentially excited by this perturbation,
    it is evident that all Landau modes are also necessarily involved.
    The form of Eq.~\ref{eq:potential_damping_time}
    demonstrates phase mixing and decay at the Landau damping frequencies.
    However, at long wavelength the mode propagates decoupled from the others to $\mathcal{O}(k^{-2})$.
    
\begin{figure}
        \centering
        \includegraphics[width=\columnwidth]{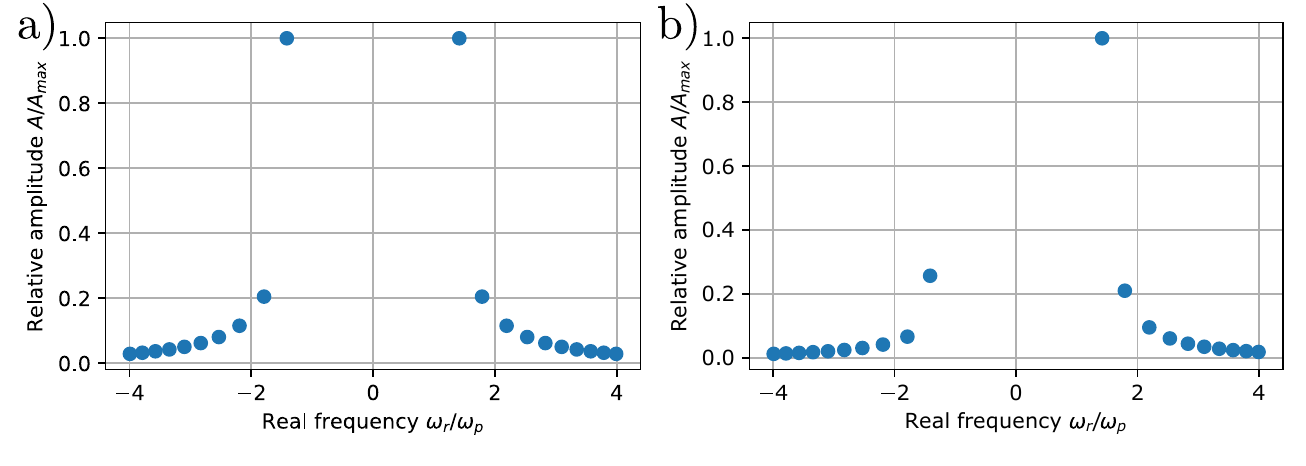}
        \caption{Relative potential amplitudes (normalized to the highest magnitude) of the Landau damping spectrum
          for the first $\pm 10$ frequencies $\omega_j$
          given a Maxwellian equilibrium distribution $f_0(v)$ at $k\lambda_D=0.5$
          for a) the Maxwellian perturbation with $g^*(\zeta)=Z(\zeta)$, and
          b) initial data as Eq.~\ref{eq:special} with a single complex pole.
          The modes are ordered by the real part of their frequency.
          Higher modes are rapidly damped, as seen in Fig.~\ref{fig:kinetic_damping_modes} which is computed for
            the same equilibrium Maxwellian distribution.
          Case b) shows that while Eq.~\ref{eq:special} can excite a propagating, damping plane wave it is necessarily
          accompanied by transient oscillations in its phase space structure with initial amplitudes up to 20\% of
            the primary oscillation.
        }\label{fig:modal_amplitudes_figure}
    \end{figure}

    So far the results are independent of the specific form of $f_0(v)$ other than its spatial uniformity.
    Now, to illustrate the partition of energy into the damped modes,
    the relative response amplitudes to (a) a Maxwellian perturbation and (b) one with a single pole,
    namely Eq.~\ref{eq:special},
    are computed for a Maxwellian equilibrium distribution $f_0(v)$.
    In case (a) the Maxwellian perturbation evolves with the potential
    \begin{equation}\label{eq:maxwellian_potential_response}
        \varphi(k, t) = -2\sum_j\frac{Z(\widetilde{\zeta}_j)}{Z''(\widetilde{\zeta}_j)}e^{-i\omega_j t}
    \end{equation}
    where $\widetilde{\zeta}_j = \zeta_j/v_t$, while case (b) evolves with the potential
    \begin{equation}\label{eq:linear_response_potential_form}
        \varphi(k, t) = \frac{1}{k^2}\Big\{\Big(1 - 2\pi i \frac{Z'''(\widetilde{\zeta}_n)}{(Z''(\widetilde{\zeta}_n))^2}
    \frac{\partial f_0}{\partial v}\Big)e^{-i\omega_n t}
    - 4\sqrt{2}\pi\frac{\partial f_0}{\partial v}\sum_{j\neq n}\frac{e^{-i\omega_jt}}{(\zeta_n-\zeta_j)Z''(\widetilde{\zeta}_j)}\Big\}.
    \end{equation}
    Figure~\ref{fig:modal_amplitudes_figure} compares these relative potential amplitudes
    for the two cases (a) and (b), where (b) represents a rightward-propagating Langmuir wave.
    While mostly the primary plasma oscillation mode is excited, the higher modes make a substantial contribution.
    
    In summary, Landau damping modes are not eigenfunctions of the Vlasov equation.
    If they are initialized and time is run either forward or backward they damp through phase mixing in either
    direction of time.
    However, their phase space structure is essentially the same as that of the unstable eigenfunctions, namely
    the plasma part of the plasma-field configuration occurring in a plasma wave.
    On the other hand, unstable modes are true eigenfunctions whose phase space structures do not change in time.

    \subsection{Visualizing the electrostatic phase space eigenfunctions}\label{subsec:eigenfunctions}
    Let us visually explore the phase space structure just discussed mathematically.
    Given a solution $\zeta_n$ to $\varepsilon(\zeta_n, k_n)=0$ for particular $k_n$ (for instability,
    $\text{Im}(k_n\zeta_n) > 0$) the perturbed distribution is given by Eq.~\ref{eq:distribution_special_sol}
    in complex-conjugate pair.
    Examining the real part gives
    \begin{align}
      f_1(x, v, t_0) &= \alpha\text{Re}(\psi)\frac{\partial f_0}{\partial v},\label{eq:11}\\
       \psi(x, v) &\equiv \frac{e^{ik_n (x - \zeta_n t_0)}}{\zeta_n - v}\label{eq:22}
    \end{align}
    where $\alpha$ is the perturbation amplitude.
    Provided that no solution to $\varepsilon(\zeta, k)=0$ has $\text{Im}(\zeta) = 0$, the denominator of $\psi$ does not
    vanish and the function is well-defined.
    The complex function $\psi$ is defined for convenience to account for all phase information in the perturbation.
    The real part is chosen arbitrarily as the linear modes come in conjugate pairs.
    Note that one can think of the mode as an instantaneous Cauchy transform of the distribution gradient.

\begin{figure}
        \centering
            \includegraphics[width=0.618\columnwidth]{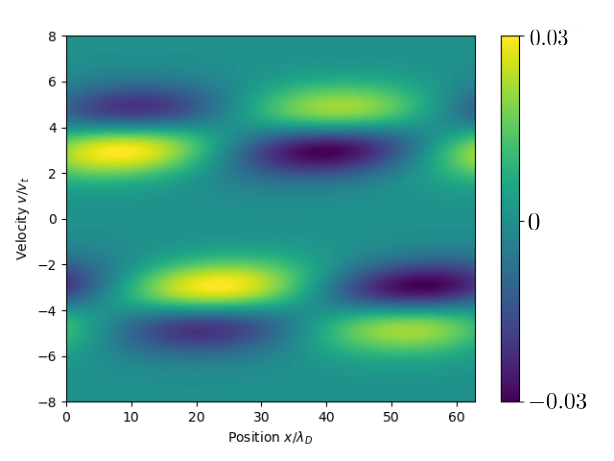}
        \caption{
          Depicted is the phase space eigenfunction of an unstable two-stream mode on two drifting Maxwellians
          of drift velocities $v_d = \pm 4v_t$, which appears as coupled plasma waves on each drifting distribution
          in the specific sense that the phase space structure on each beam is visually and mathematically similar to a
            Landau-damped mode of a thermal Maxwellian (i.e. the phase space structure of Eq.~\ref{eq:special}).
          However, these plasma wave structures only occur as a normal mode, or eigenfunction, when unstable.
            The mode has structure in both $x$ and $v$, but its zeroth velocity moment is a pure sine wave and its first
            moment is zero.
        }\label{fig:2}
    \end{figure}

    For example, consider the unstable two-stream mode of two drifting Maxwellians of drift velocities
    $v_d = \pm 4 v_t$ with wave-number $k_0\lambda_D = 0.1$,
    giving a growth rate $\omega_i \approx 0.28\omega_p$.
    Figure~\ref{fig:2} visualizes the phase space of the corresponding mode given by Eqs.~\ref{eq:11}-\ref{eq:22}.
    The perturbation is a two-dimensional oscillatory structure in phase space, yet the zeroth moment is a pure sine wave
    resulting in an initial electric potential $\varphi(x) = \varphi_0\sin(x)$.
    The non-separability of the perturbation is evident in Fig.~\ref{fig:2}.

    \subsection{Kinetic eigenfunctions applied to nonlinear initial-value problems}\label{subsec:two_stream}
    The physically correct initial data is usually discussed in the context of
    the thermal fluctuation spectrum~(\cite{ichimaru1992statistical}).
    Linear eigenfunctions grow from spontaneous thermal fluctuations until
    nonlinear saturation at some significant fraction of the thermal energy~(\cite{yoon2007spontaneous, crews_kinetic}).
    The Vlasov model does not resolve thermal fluctuations, but this is acceptable for typical plasmas as the magnitude
    of such fluctuations is much less than the thermal energy.

    The eigenfunction part of a general perturbation amplifies its energy while the
    non-eigenfunction part decays with the same timescale $\omega_p^{-1}$.
    Clearly, with sufficiently small initial amplitude the non-eigenfunction part of a general perturbation
    does not participate in nonlinear saturation,  
    so that sufficiently low-amplitude general perturbations are physically correct.
    Yet in the same way, an eigenfunction perturbation of initially large amplitude compared to the thermal fluctuation level
    is also physically correct.
    For this reason, eigenfunction perturbations yield a physically meaningful computational cost-savings
    when initialized with amplitudes just below nonlinear levels, 
    while high-amplitude general perturbations introduce nonlinear Landau damping.
    Initialization at high amplitude translates to considerable computational savings for high-dimensional,
    computationally intesive continuum kinetic problems. 


    We mention an application of eigenfunction perturbations to small-amplitude perturbation problems.
    Sometimes linear instability growth rates are measured for verification of
    model implementation~(\cite{ho2018physics,einkemmer2019performance}). 
    Small-amplitude eigenfunction perturbations allow linear instability growth rates to be deduced from
    data with basically arbitrary precision
    because there is a complete absence of Landau damping.

    \subsubsection{Phase space eigenfunctions applied to the two-stream instability problem}\label{subsec:the-one-dimensional-two-stream-instability}
    Here we refer to perturbations as ``separable'' when they factor as $g(x,v) = h(x)f_0(v)$ with $h(x)$
    representing the desired density perturbation.
    Given the preceding discussion, it is illustrative to compare energy traces of fully nonlinear Vlasov-Poisson simulations
    initialized both with general separable perturbations and eigenfunction perturbations.
    Consider, for example, the two-stream unstable distribution
    \begin{equation}\label{eq:two_stream_distribution}
        f_0(v) = \frac{1}{2\sqrt{2\pi}v_t}\Big(\exp\Big\{-\frac{(v-v_d)^2}{2v_t^2}\Big\}
    + \exp\Big\{-\frac{(v+v_d)^2}{2v_t^2}\Big\}\Big).
    \end{equation}
    Initialization with a separable Maxwellian perturbation, namely
    $g(x,v) = \alpha f_0(v)e^{ikx}$ with $\alpha$ a scalar amplitude, leads to the linear solution 
    \begin{equation}\label{eq:potential_response_two_stream}
        \varphi(k, t) = -2\alpha \sum_j
    \frac{Z(\zeta_{+,j}) + Z(\zeta_{-,j})}{Z''(\zeta_{+,j}) + Z''(\zeta_{-,j})}e^{-i\omega_j t}
    \end{equation}
    where $\zeta_{\pm,j} \equiv (\zeta_j \pm v_d)/\sqrt{2}v_t$ are the beam-shifted phase velocities.
    Table~\ref{tab:relative_amplitudes_table} lists the greatest amplitudes of Eq.~\ref{eq:potential_response_two_stream} 
    for drifts $v_d=5v_t$ at wavenumber $k\lambda_D=0.126$, and shows that the unstable
    mode is only the fifth-largest amplitude.
    Langmuir waves influence dynamics 
    by masking the growing mode or through nonlinear Landau damping.

    \begin{table}
        \begin{center}
          \begin{tabular}{lcccccc}
        Mode      & 1 & 2 & 3 & 4 & 5 & 6\\
        Frequency, $\omega_r/\omega_p$             & $\pm$ 1.42                 & $\pm$ 0.0157 & 0     & $\pm$ 1.10   & $\pm$ 1.20   & $\pm$ 1.29   \\ 
        Growth rate, $\gamma/\omega_p$             & $-3.2\times 10^{-7}$ & -0.341 & 0.335 & -0.228 & -0.377 & -0.488 \\ 
        Amplitude, $|\varphi_j|/\text{max}(|\varphi|)$ & 1       & 0.710  & 0.335 & 0.0701 & 0.0581 & 0.0466 \\ 
        \end{tabular}
        \caption{Relative amplitudes of electric potential for the first six linear mode pairs in a separable perturbation
        of two counterstreaming thermal plasmas of drift velocities $v_d= 5v_t$ and at wavenumber $k\lambda_D=0.126$. 
        In this case, for each non-zero frequency $\omega$ there is a pair of solutions. Due to these mode pairs the unstable mode
            (here of growth rate $\gamma/\omega_p=0.335$)
            has only the fifth-largest relative amplitude of electric potential in the initial condition
            (counting the modes twice, as they come in pairs).
            The significance is that the unstable mode contains only a small fraction of the initialized energy,
            as all modes occur at the same wavelength.
            The amplitude and growth rate matching is here only a coincidence.
          }\label{tab:relative_amplitudes_table}
        \end{center}
    \end{table}

\begin{figure}
        \centering
        \includegraphics[width=\columnwidth]{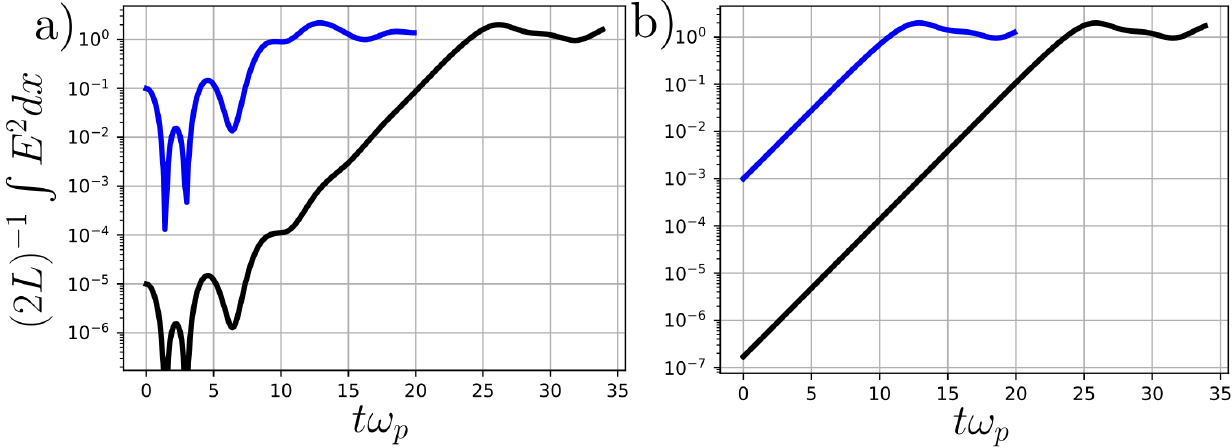}
        \caption{
          Comparison of energy traces for nonlinear simulations of a single wavelength two-stream instability
          of drift velocities $v_d = 5v_t$
            at wavenumber $k\lambda_D=0.126$ showing (a) separable perturbation
            and (b) eigenfunction perturbation.
            Two subcases are shown. The black trace represents a small-amplitude perturbation
            and the blue trace a large-amplitude one.
            The large-amplitude Maxwellian perturbation introduces
            nonlinearity, polluting the solution,
            while the large-amplitude eigenmode saturates equivalently as if seeded from small amplitude.
            In any case solutions initialized by eigenfunction perturbation undergo pure growth
            in the linear phase.
        }\label{fig:two_stream_comparison_figure}
    \end{figure}

    Figure~\ref{fig:two_stream_comparison_figure} compares the energy traces of nonlinear simulations of the two-stream
    instability initialized by both separable and eigenfunction perturbations.
    When initialized at small amplitude the type of perturbation does not make a difference to saturation dynamics.
    On the other hand, a perturbation of large amplitude reaches saturation much faster.
    The large amplitude Maxwellian perturbation in Fig.~\ref{fig:two_stream_comparison_figure}a
    introduces nonlinearities and changes the energy trace from the desired evolution.
    Observe that the simulations initialized with the eigenfunction perturbations undergo a pure growth.

    \subsection{Multidimensional dispersion function for the two-stream instability}
    Electrostatic turbulence at the Debye length scale generated by streaming instability of electron beams
    is a ubiquitous plasma phenomenon~(\cite{rudakov1978strong, che2016electron}),
    and is inherently three dimensional. 
    This section considers the two-stream instability in the computationally tractable
    two-dimensional configuration space as an example of
    the methodology used to compute electrostatic phase space eigenfunctions in multiple dimensions.
    Recall that the electrostatic dielectric function is determined by
    \begin{equation}\label{eq:multi_d_electrostatic}
        \varepsilon(\omega, \bm{k}) = 1 + \frac{\omega_p^2}{k^2}\int \frac{\bm{k}\cdot\nabla_v f_0}{\omega - \bm{k}\cdot\bm{v}}d\bm{v} = 0.
    \end{equation}
    Now consider a thermal two-stream distribution with equal temperatures on each beam,
    \begin{equation}\label{eq:two_stream_dist}
    \begin{split}
      f_0(u,v,w) = \frac{e^{-(v^2+w^2)/(2u_t^2)}}{2(2\pi)^{3/2}u_t^3}
        \Big(\exp\Big(-\frac{(u-u_d)^2}{2u_t^2}\Big) + \exp\Big(-\frac{(u+u_d)^2}{2u_t^2}\Big)\Big),
    \end{split}
    \end{equation}
    which differs from Eq.~\ref{eq:two_stream_distribution} in retaining three components of velocity.
    Having assumed an isotropic thermal velocity $u_t$ greatly simplifies analysis; otherwise complications arise due
    to the elliptical level-sets of $f_0$.
    Let the wavevector lie in the $(x,y)$-plane and consider Eq.~\ref{eq:multi_d_electrostatic}.
    The $w$-component integrates out immediately, while the $(x,y)$-directed velocities must be rotated into the frame
    of the wavevector.
    Rotating through an angle $\varphi$ to coordinates $(v_\parallel, v_\perp)$, the distribution
    function is $f_0(v_\parallel, v_\perp) = (f_+ + f_-)/2$ with
    \begin{equation}\label{eq:rotated_two_stream}
    \begin{split}
        f_{\pm} \equiv \frac{1}{2\pi u_t^2}\exp\Big(-\frac{(v_\parallel \mp \cos(\varphi)u_d)^2}{2u_t^2}\Big)
        \exp\Big(-\frac{(v_\perp \pm \sin(\varphi)u_d)^2}{2u_t^2}\Big)
    \end{split}
    \end{equation}
    Evaluating the integral for each drifting component $f_\pm$ as $-Z'(\zeta_\pm)/2u_t^2$ as in~\cite{Skoutnev_2019}
    with drift velocity-shifted phase velocity $\zeta_\pm \equiv (\zeta \mp \cos\varphi u_d)/\sqrt{2}u_t$ gives
    \begin{equation}\label{eq:electrostatic_general_angle}
        \varepsilon(k, \zeta, \varphi) = 1 - \frac{1}{(k\lambda_D)^2}\frac{Z'(\zeta_+) + Z'(\zeta_-)}{2} = 0.
    \end{equation}
    The wavevector having transverse components to the drift axis decreases the effective drift by the cosine of $\varphi$,
    leading to maximum growth rate of longitudinal waves parallel to the streaming velocity.
    However, the growth of these transverse-axis components seeds a multi-dimensional turbulence,
    depending on the configuration space dimensionality.

    \subsection{Nonlinear simulation of two-stream instability in two spatial dimensions}\label{subsec:two_d_electrostatic_simulation}
\begin{figure}
        \centering
        \includegraphics[width=0.718\columnwidth]{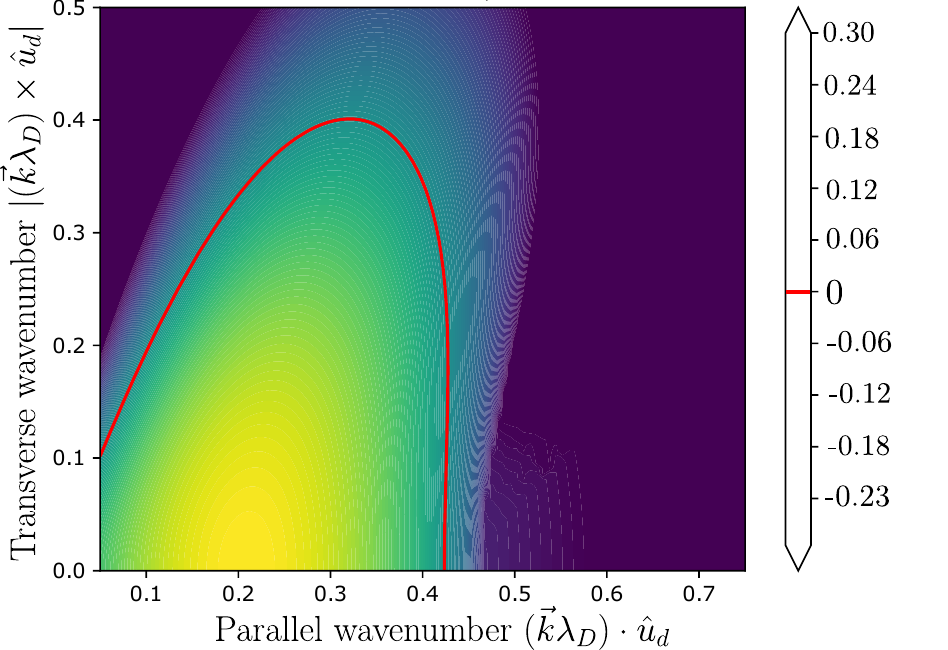}
        \caption{Growth rates of the electron two-stream instability as $\text{Im}(\omega)/\omega_{pe}$ for
        two counter-streaming Maxwellians with drift speeds relative to the thermal speed of $u_d/u_t = \pm 3$,
            in terms of parallel and perpendicular wavenumbers relative to the beam axis vector $\hat{u}_d$.
        The red contour indicates the line of marginal stability where $\text{Im}(\omega) = 0$.
        While the fastest growing mode is aligned with the beam axis (occuring at $(k\lambda_D)\cdot\hat{u}_d\approx 0.2$),
            modes of comparable growth rates occur with
            transverse wavelengths roughly five-to-ten times the unstable wavelength on-axis.}\label{fig:electrostatic_growth_rates}
    \end{figure}

    Although the fastest growing mode of the two-stream problem has a beam-axis-aligned wavevector,
    eigenmodes with a long-wavelength transverse part $k_\perp \ll k_\parallel$ grow comparably to the
    fastest mode, as illustrated in Fig.~\ref{fig:electrostatic_growth_rates} by the dielectric function $\varepsilon(k_\parallel, k_\perp)$ computed using
    Eq.~\ref{eq:electrostatic_general_angle}.
    In this way, streaming in unmagnetized plasma generically produces multi-dimensional
    Langmuir turbulence.  
    In practice this is a three-dimensional phenomenon, but for computational tractability 
    a streaming simulation is presented here with two space and two velocity dimensions (2D2V).
    We speculate that 2D2V nonlinear phase mixing is quite similar to 3D3V because the ``total dimensionality'' of the phase space turbulence
    is greater than two.
    On the other hand, nonlinear spatial dynamics of the saturated state are likely quite different in 2D2V versus full 3D3V.
    A dedicated study of this issue with sufficient compute capability would be welcome.
    
    \subsubsection{Initialization of the two-dimensional two-stream simulations}
    Our numerical method is summarized in Appendix~\ref{app:summary_of_numerical}.
    The domain used is periodic and set by fundamental wavenumbers $k_x\lambda_D=0.05$ and $k_y\lambda_D=0.02$.
    The $x$-axis is divided into forty evenly-spaced collocation nodes and the $y$-axis into fifty nodes.
    Velocity space is truncated at $v_{\max} = \pm 11.5 v_t$, and each axis divided into fourteen finite elements
    each of a seventh-order Legendre-Gauss-Lobatto polynomial basis.
    A non-uniform velocity grid is used; ten elements are linearly clustered between $v\in (-7.5, 7.5)v_t$ and two elements into $v\in \pm (7.5, 11.5)v_t$.
    The drift velocity in Eq.~\ref{eq:two_stream_dist} is set to $u_d=3v_t$.
    Finally, a spatial hyperviscosity $\nu \nabla_x^4 f$ with $\nu=10$ is added to the kinetic equation to mitigate spectral
    blocking with this low spatial resolution, as in~\cite{crews_kinetic}.
    Many modes of comparable growth rates are initialized using the eigenfunction perturbations
    \begin{equation}\label{eq:vlasov_2d_perturbations}
        f_1(x, y, u, v) = \frac{\alpha}{k}\text{Re}\Big(
        \frac{k_x \partial_u f + k_y\partial_v f}{\omega(k_x, k_y) - k_x u - k_y v} \exp(i(k_x x + k_y y + \theta))\Big)
    \end{equation}
    with $\alpha$ an amplitude scalar, $\omega(k_x, k_y)$ solution to the dispersion relation
    for the mode $(k_x, k_y)$, and of random phases $\theta$.
    A total of thirty-three modes are excited, each with the amplitude $\alpha=0.01$;
    for each harmonic of the $x$-fundamental $k_{n,x} = n k_{1,x}$
    with $n=2$, $3$, and $4$, eleven harmonics of the $y$-fundamental are excited with $k_{m,y} = m k_{1,y}$ with $m = 0$,
    $\pm 1$, $\pm 2$, $\pm 3$, $\pm 4$, and $\pm 5$.
    There is no symmetry in the $y$-direction as different phases $\theta$ are used for the modes $\pm m$.
    Mode $n=1$ has small growth rate and is not initialized.

    \subsubsection{Two-dimensional nonlinear simulations of the streaming instability}
\begin{figure}
        \centering
        \includegraphics[width=0.5\columnwidth]{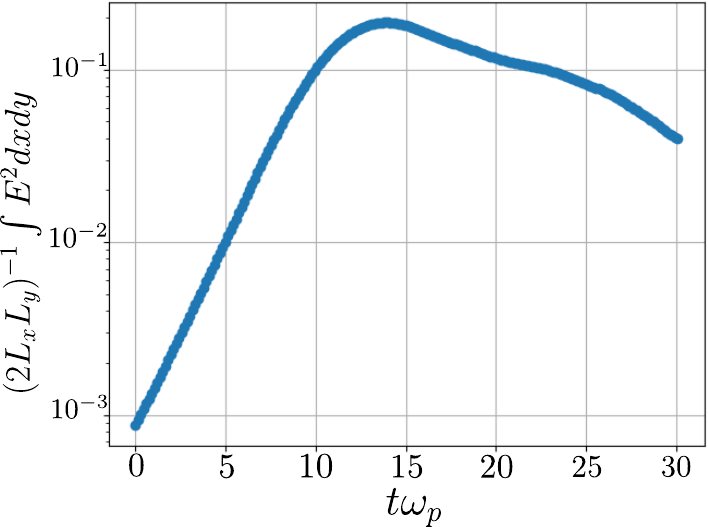}
        \caption{Electric energy trace of the 2D2V two-stream instability
          with thirty-three excited domain modes.
          Initialization with eigenmodes enables a perturbation just two decades below saturation energy. 
          The aggressive spatial hyperviscosity of $\nu=10$ leads to an energy loss of
        $\mathcal{O}(10^{-3})$ by the stop time of this simulation.
        This artificial dissipation would not be necessary with greater spatial resolution,
        but the loss is well beneath the electric energy at the stop time.
        }\label{fig:electrostatic_energy}
    \end{figure}

    The simulation is run to a stop time of $t\omega_p=30$.
    Figure~\ref{fig:electrostatic_energy} shows the simulation's electric field energy trace.
    Hyperviscosity with this spatial resolution leads to a domain energy loss of $\mathcal{O}(10^{-3})$ while electric
    energy saturates at $\mathcal{O}(10^{-1})$.
    Due to the use of eigenfunction perturbations there are no oscillations in the electric field energy trace.
    Therefore, the simulation was initialized with a perturbation energy just two orders below saturation.
    In this case, this saves approximately $10\omega_p^{-1}$ of simulation time compared to, for example,
    a perturbation of initial energy $10^{-6}$.

    \begin{figure}
        \centering
        \includegraphics[width=0.95\columnwidth]{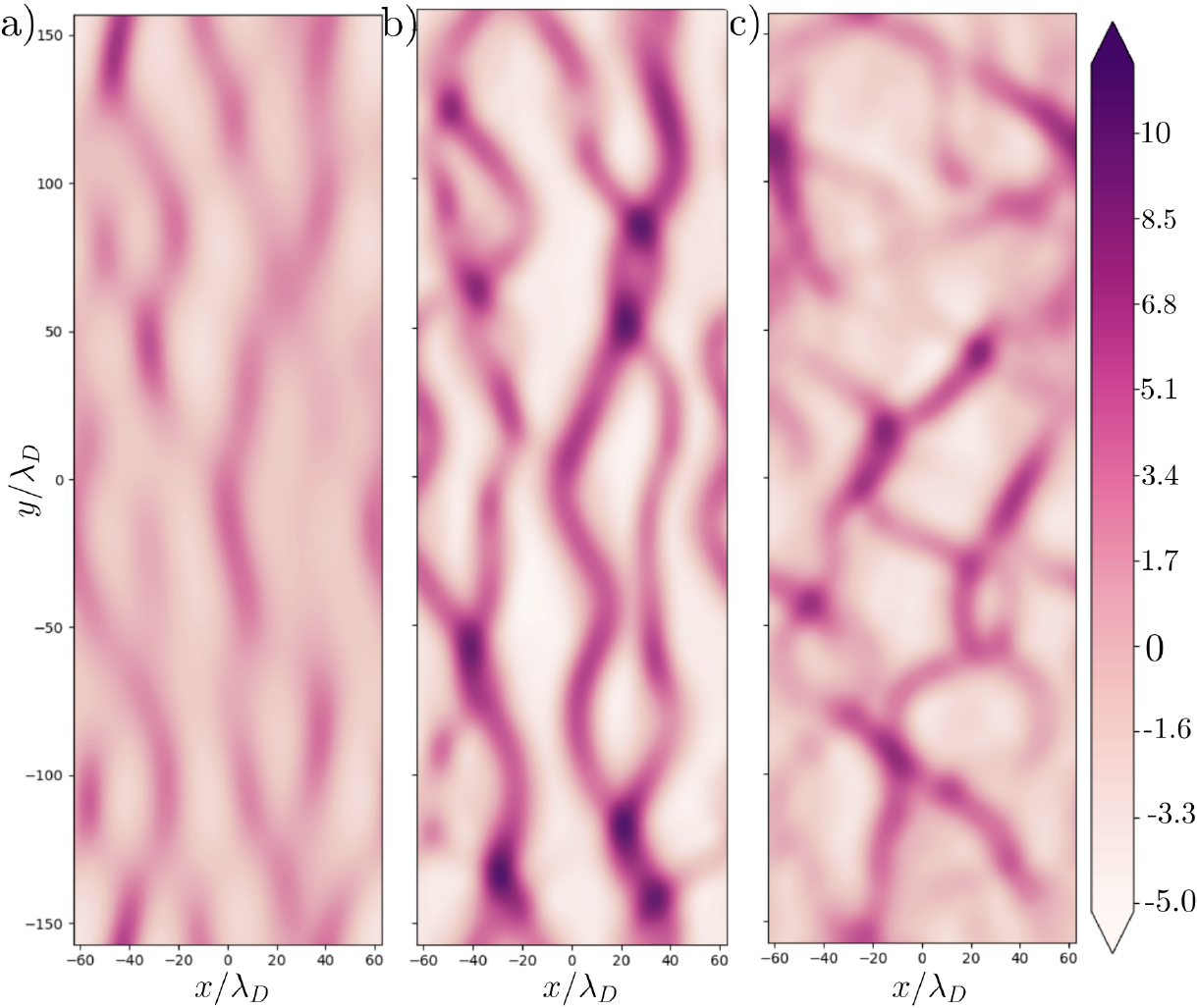}
        \caption{Electric potential $\varphi(x,y)$ of the 2D2V electron two-stream instability
        for three times: a) $t\omega_p=8$ (linear phase);  b)
        $t\omega_p=18$ (nonlinear phase); c) $t\omega_p=27$ (isotropizing).
        The evolution begins with the formation of one-dimensional electron holes, or phase space vortex lines,
            transverse to the streaming axis.
        These vortex lines then break up after saturation into two-dimensional hole structures
        with more complex orbits which maintain connection to one another
        by the Vlasov-dynamical conservation of phase space circulation.
        The lines of potential in this simulation may be understood as a projection of these
        phase space vortex tubes.}\label{fig:electrostatic_two_dimensional_potential}
    \end{figure}

    \begin{figure}
        \centering
        \includegraphics[width=\columnwidth]{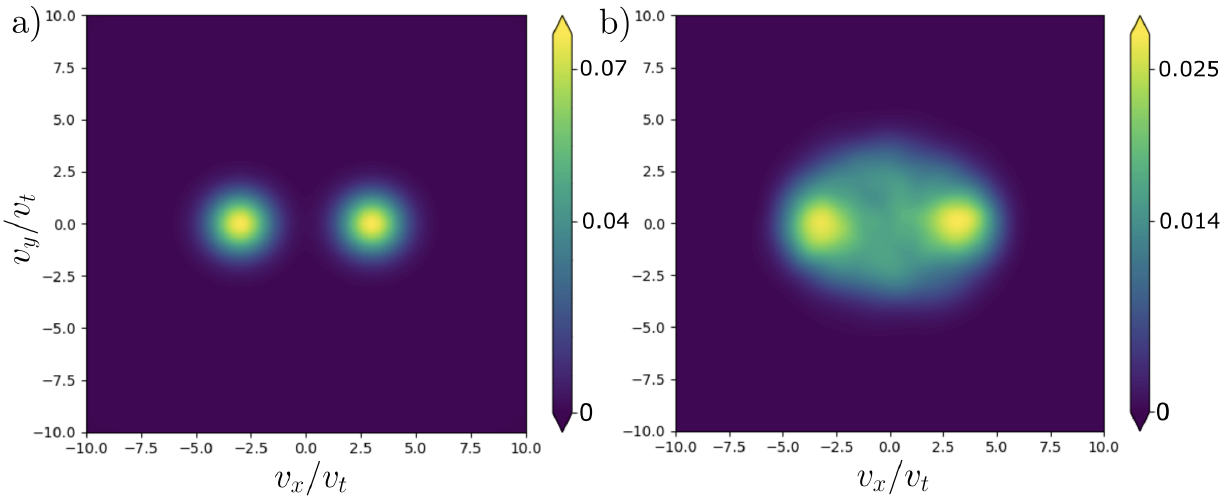}
        \caption{Domain-averaged (coarse-grained) distribution $\langle f\rangle_{(x,y)}(v_x,v_y)$ showing relaxation to a
        Penrose-stable distribution on averaged scales due to multidimensional Langmuir turbulence, for two times: (a) $t\omega_p=0$, the unstable
        initial condition; (b) $t\omega_p=30$, the stop time of the simulation.
        At the end of the simulation the average distribution is double-humped, 
        having heated significantly in the direction transverse to the beam axis compared to the initial state.}\label{fig:electrostatic_two_dimensional_distribution}
    \end{figure}
    
    Figure~\ref{fig:electrostatic_two_dimensional_potential} plots evolution of electric potential.
    Beam-axis wave energy initially predominates as electron holes form with wavenumber $k_x\gg k_y$,
    yet the transverse perturbations with comparable growth rates to the maximal beam-axis part
    lead to two-dimensional structures.
    Following nonlinear saturation, wave energy significantly increases in the transverse direction as the
    holes tilt, consolidate, and isotropize.
    Figure~\ref{fig:electrostatic_two_dimensional_distribution} visualizes 
    the domain-averaged (coarse-grained) distribution and demonstrates that isotropization 
    associated with beam-driven electrostatic turbulence distributes coarse-grained energy
    into the beam-transverse directions.
    That is, beam-transverse temperature increases significantly in the resulting
    marginally stable double-humped distribution~(\cite{penrose}),
    as heat transfers from coarse-grained wave energy to the coarse-grained distribution~(\cite{nicholson1983introduction}).
    These results support the notion that the continuum of electron hole solutions is key to strongly driven plasma transport~(\cite{schamel2023pattern}).

    \subsection{Quasilinear kinetic simulation with phase space eigenfunctions}\label{subsec:quasilinear-theory}
    Quasilinear theory (QLT) is the name given to the simplest closure in the hierarchy of equations resulting
    from separating the variables of a turbulent system into fluctuating and mean components~(\cite{vedenov1963quasi}).
    The scheme of the theory is as follows: a suitable method of averaging is defined, typically temporal, spatial, or ensemble
    averaging; the dynamical equation is averaged and the mean subtracted from the original equation to obtain the mean and
    fluctuating components of the system; lastly a closure hypothesis is made by neglecting the ``second fluctuation''.
    Under this procedure, known as ``classic QLT'',
    the equation for the fluctuation becomes quasilinear and can be solved by spectral methods~(\cite{crews_kinetic}).
    Substitution into the equation for the mean gives a diffusion equation in velocity space.
    Beyond classic QLT, a formulation more applicable to inhomogeneous plasmas is presented in~\cite{dodin2022quasilinear}.
    
    Diffusion equations are numerically stiff when diffusivity is large.
    A drawback of QLT posed as a diffusion problem is that the diffusivity is asymptotically singular in the relaxed
    state of an unstable system.
    The singularity arises as $\text{Im}(\zeta)\to 0$ around the purely real frequencies of the relaxed state~(\cite{crews_kinetic}).
    This singularity is side-stepped by solving the equations of QLT as an initial-value problem for a system of
    first-order equations, resolving the linear kinetic eigenfunctions, linear Landau damping,
    and the asymptotic ($t\to\infty$) saturation of the distribution function.
    There is no dimensionality reduction like in the diffusion theory,
    but significant advantage remains over the fully nonlinear
    theory because the turbulent nonlinear cascade does not form and only the unstable scales need be resolved.
    Further, as a first-order system there is no need to solve the dielectric function in a quasilinear simulation.
    To prevent spurious Landau damping it is wise to utilize kinetic eigenfunction perturbations
    as in Section~\ref{subsec:two_stream}.

    To demonstrate we consider the kinetic equation for electrons in a neutralizing background.
    Splitting the distribution function $f = \langle f\rangle_L + \delta f$ where $\langle\cdot\rangle_L$ is a spatial average,
    the quasilinear system in normalized units is~(\cite{crews_kinetic})
    \begin{align}
        \frac{\partial\langle f\rangle_L}{\partial t} &= \frac{\partial}{\partial v}\langle E\delta f\rangle_L\label{eq:qlt1}\\
        \frac{d(\delta f)}{dt} &= \frac{\partial}{\partial v}E\langle f\rangle_L\label{eq:qlt2}
    \end{align}
    with $\frac{d}{dt}=\partial_t + v\partial_x$ the change along a zero-order trajectory.
    Consider a finite periodic domain $x\in (0,L)$ and the expansion of the distribution in Fourier series,
    \begin{equation}
        f(x,v,t) = f_0 + \sum_{n=1}^{\infty}\big(f_n e^{ik_n x} + f_n^*e^{-ik_{n}x}\big).\label{eq:fourier_vlasov1}
    \end{equation}
    The $n=0$ component of Eq.~\ref{eq:fourier_vlasov1} is the average distribution, $f_0=\langle f\rangle_L$,
    and the remaining Fourier coefficients make up the Fourier spectrum of the fluctuation with $f_n^* = -f_n$.
    Thus dropping the symbols $\delta(\cdot)$ and $\langle\cdot\rangle_L$, Eqs.~\ref{eq:qlt1} and~\ref{eq:qlt2} are
    \begin{align}
        \frac{\partial f_0}{\partial t} &= \frac{\partial}{\partial v}\langle Ef\rangle_L,\label{eq:fourier_qlt1}\\
        \frac{\partial f_n}{\partial t} &= -ik_{n}vf_n + \frac{\partial}{\partial v}\big(E_n f_0\big),\quad n\geq 1\label{eq:fourier_qlt2}\\
        E_n &= ik_n^{-1}\int_{-\infty}^{\infty} f_{n}dv,\quad n\geq 1\label{eq:gauss_fourier_qlt}
    \end{align}
    where Eq.~\ref{eq:gauss_fourier_qlt} is obtained from Gauss's law in the Fourier basis.

    \subsubsection{Numerical method for the initial-value problem for the quasilinear equations}\label{sec:num_qlt}
    Equations~\ref{eq:fourier_qlt1} and~\ref{eq:fourier_qlt2} are discretized in velocity space
    and Eq.~\ref{eq:gauss_fourier_qlt} is applied as a constraint.
    First, the Fourier series is truncated at a chosen mode number (Galerkin projection) to resolve the range of instability,
    with the corresponding spatial grid identified as the evenly spaced collocation nodes of the frequency range
    in a standard manner through the fast Fourier transform.
    The velocity axis is to be discretized by discontinuous Galerkin (DG) method similarly to the method in Appendix~\ref{app:summary_of_numerical}.
    We consider the two velocity fluxes in Eqs.~\ref{eq:fourier_qlt1} and~\ref{eq:fourier_qlt2},
    \begin{align}
        \mathcal{T}(v) &\equiv \langle Ef\rangle_L = L^{-1}\int_0^L E\delta f(x,v) dx,\\
        \mathcal{M}_n(v) &\equiv E_n f_0 = \Big(ik_n^{-1}\int_{-\infty}^\infty f_n(v')dv'\Big)f_0(v).
    \end{align}
    We evaluate the flux $\mathcal{T}(v)$ by the trapezoidal rule because of its ideal trigonometric quadrature properties~(\cite{boyd2001chebyshev})
    using the inverse FFT of the spectra $E_n$ and $f_n(v)$.
    On the other hand, the fluxes $\mathcal{M}_n(v)$ depend only on the local field mode $E_n$ and the mean distribution $f_0(v)$,
    so this quantity is simply computed by quadrature in $v$.
    Both fluxes $\mathcal{T}$ and $\mathcal{M}_n$ are then utilized in the DG method as a nonlinear flux.
    The linear translation operator $ik_n vf_n$ is discretized by quadrature
    and 
    the system is integrated in time semi-implicitly by Strang splitting, as outlined in Appendix~\ref{app:summary_of_numerical}.

    \subsubsection{Simulation of the bump-on-tail instability using phase space eigenfunctions}
\begin{figure}
        \centering
        \includegraphics[width=\columnwidth]{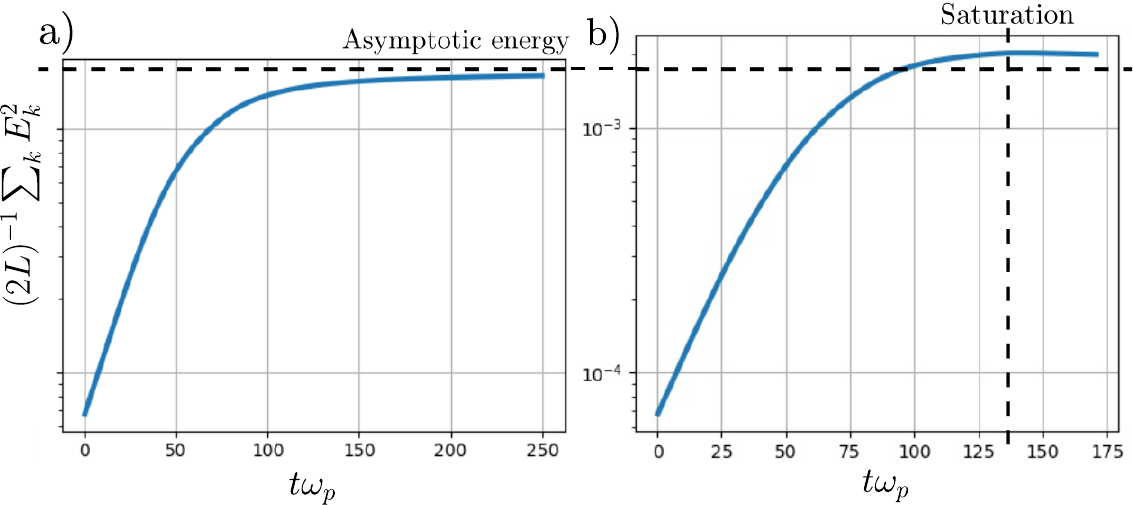}
        \caption{Electric energy traces are compared for (a) the quasilinear evolution,
          and (b) the fully nonlinear Vlasov-Poisson simulation (from the data of~\cite{crews_kinetic})
            from identical initial conditions and perturbations.
            The quasilinear system approaches the marginally stable state asymptotically, while the fully nonlinear
            system saturates in a finite time due to nonlinear particle trapping.
            Both simulations are initialized at high though sub-nonlinear amplitude without subsequent Landau damping
            oscillations because they utilize kinetic eigenfunction perturbations,
            while non-eigenfunction perturbations would oscillate significantly.}\label{fig:qlt_energies}
    \end{figure}

\begin{figure}
        \centering
        \includegraphics[width=\columnwidth]{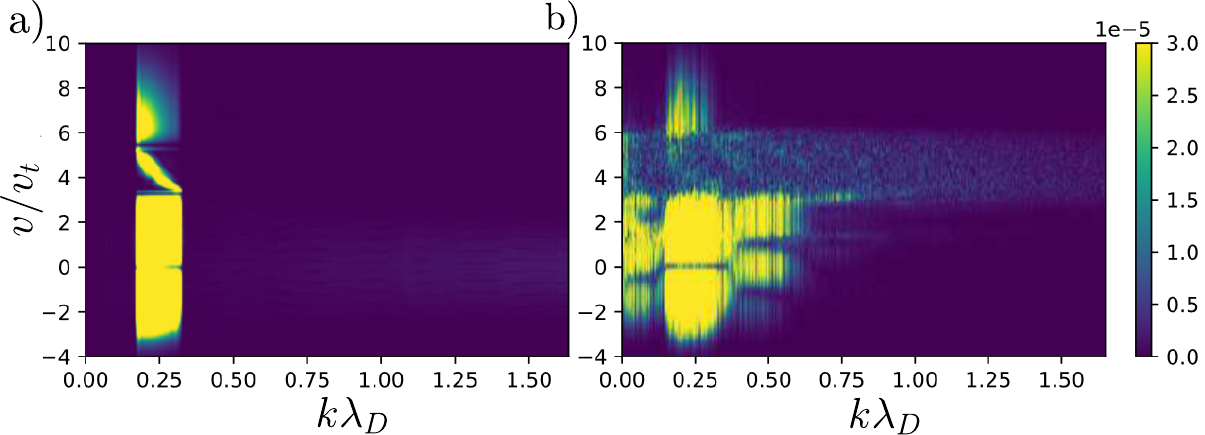}
        \caption{Spectral $(k\lambda_D, v/v_t)$ phase space is compared at saturation $t=150\omega_p^{-1}$
            between (a) the quasilinear system, and (b) the corresponding fully nonlinear Vlasov-Poisson simulation.
            The spectral phase of the quasilinear system is grossly overresolved in $k\lambda_D$ space only for the
            purpose of comparison; energy remains spatially localized in QLT
            because the system is unable to cascade through turbulent mixing of phase space eddies, a fully nonlinear phenomenon.
            The phase space structures in a simulation of QLT always remain linear eigenfunctions of the form of Eq.~\ref{eq:special}.
            In this way sub-Debye length scales need not be resolved in simulation of QLT.}\label{fig:qlt_phase_space}
    \end{figure}

    We repeat the calculation of~\cite{crews_kinetic} for the nonlinear and quasilinear evolutions of
    the bump-on-tail instability, with the difference that here QLT is solved
    as an initial-value problem in phase space instead of as a diffusion problem for
    the reduced distribution, and using the numerical method described in Section~\ref{sec:num_qlt}.
    See~\cite{crews_kinetic} for the details of initialization including the domain and the zero-order distribution.
    In~\cite{crews_kinetic} the perturbation is constructed using the phase space eigenfunctions given by Eq.~\ref{eq:special}.
    Figure~\ref{fig:qlt_energies} compares the energy traces of the nonlinear and quasilinear simulations where
    QLT is solved as an initial-value problem,
    while Fig.~\ref{fig:qlt_phase_space} compares the spectral phase space at saturation
    and demonstrates the absence of phase space cascade in QLT. 
    In addition, there is an absence of Landau damping as the perturbation is constructed from eigenfunctions.

    \section{Electrostatic eigenmodes with zero-order cyclotron motion}\label{sec:magnetized-electrostatic-plasma-modes}
    Here the electrostatic kinetic eigenfunctions 
    of strongly magnetized plasma are illustrated by first reviewing the linearized theory~(see~\cite{gurnett2017introduction} page 382 for step-by-step derivation)
    and then applying the linearized theory to the particular case of ring distributions
    of the Dory-Guest-Harris or $\chi$-distribution type~(\cite{dory1965unstable})
    that are of special interest in space plasmas and magnetic traps.
    By strongly magnetized plasma we mean that the zero-order thermal magnetic force $q v_{th}B_0$ exceeds the first-order
    perturbation force $q E_1$ such that the zero-order trajectories are cyclotrons.
    Strongly magnetized electrostatic modes are longitudinal oscillations characterized by
    two fundamental frequencies, the plasma frequency $\omega_p$ and the cyclotron frequency $\omega_c$.
    Defining the obliquity through $\bm{k}\cdot\bm{B} = kB\cos(\phi) = k_\parallel B$, the wavevector decomposes as
    $
        \bm{k} = k_\parallel\hat{e}_\parallel + k_\perp\hat{e}_\perp
    $.
    The spectrum is determined by the dielectric function roots $\varepsilon(\omega, k_\perp, k_\parallel) = 0$,
    and depends strongly on the angle between the wavevector $\bm{k}$ and magnetic field $\bm{B}$.
    In contrast to the unmagnetized case there are undamped waves perpendicular to $\bm{B}$,
    including the well-known Bernstein modes of Maxwellian plasma~(\cite{bernstein1958waves}).
    We will see that electrostatic kinetic eigenfunctions have helical structure in the perpendicular velocity
    phase space.

    \subsection{Review of the Harris dispersion relation}\label{subsec:harris_dispersion}
    Our treatment here follows~\cite{gurnett2017introduction} Section 10.2.1
    with a particular focus on the phase space perturbation as a kinetic eigenfunction.
    Coordinates are chosen such that $\bm{B} = B\hat{z}$, $\bm{k} = k_\perp\hat{x} + k_\parallel\hat{z}$
    and the velocity space is expressed in cylindrical coordinates as $\bm{v} = v_\perp\cos(\phi)\hat{v}_x
    + v_\perp\sin(\phi)\hat{v}_y + v_\parallel\hat{v}_z$.
    In these coordinates the Vlasov equation linearizes around $f_0(v_\perp, v_\parallel)$ as, with $\omega_c=qB/m$,
    \begin{equation}\label{eq:linearized_vlasov}
        \frac{\partial f_1}{\partial t} + v_{\perp}\cos\phi\frac{\partial f_1}{\partial x} +
        v_{\parallel}\frac{\partial f_1}{\partial z} - \omega_c\frac{\partial f_1}{\partial\phi} =
    \frac{q}{m}\Big(E_x\cos\phi\frac{\partial f_0}{\partial v_\perp} + E_z\frac{\partial f_0}{\partial v_\parallel}\Big).
    \end{equation}
    The linearized equation is then Fourier transformed $(x,z,t)\to (k_\perp,k_\parallel,\omega)$ to yield
    \begin{equation}\label{eq:fourier_lin_vlasov}
        \frac{df_1}{d\phi} + i\frac{\omega - k_\perp v_\perp \cos\phi - k_\parallel v_\parallel}{\omega_c}f_1 =
        -i\frac{q\varphi}{m}\Big(k_\perp\cos\phi \frac{\partial f_0}{\partial v_\perp} +
    k_\parallel\frac{\partial f_0}{\partial v_\parallel}\Big).
    \end{equation}
    Equation~\ref{eq:fourier_lin_vlasov} is a first order inhomogeneous equation in the cylindrical velocity-space angle
    $\phi$ and can be solved by the usual methods.
    Integrating the inhomogeneous term along the solution of the homogeneous equation yields
    \begin{equation}\label{eq:inhomogeneous_solution}
        f_1(v_\perp, \phi, v_\parallel) = -\frac{q\varphi(k)}{m}\exp(ik_\perp v_y/\omega_c)\Big[
            \frac{1}{v_\perp}\frac{\partial f_0}{\partial v_\perp}\Upsilon_1 +
            k_\parallel\frac{\partial f_0}{\partial v_\parallel}\Lambda_1\Big]
    \end{equation}
    where the terms $\Upsilon_1$ and $\Lambda_1$ are auxiliary wavefunctions defined as the polar Fourier series
    \begin{align}
        \Upsilon_1 &= \sum_{n=-\infty}^{\infty}\frac{n\omega_c}{\omega-k_\parallel v_\parallel - n\omega_c}
        J_n(k_\perp v_\perp/\omega_c)e^{in\phi},\label{eq:polar1}\\
        \Lambda_1 &= \sum_{n=-\infty}^{\infty}\frac{\omega_c}{\omega-k_\parallel v_\parallel - n\omega_c}
        J_n(k_\perp v_\perp/\omega_c)e^{in\phi},\label{eq:polar2}
    \end{align}
    with $J_n(z)$ a Bessel function of the first kind.
    The terms of $\Upsilon_1$ associated with perpendicular propagation decay one order slower in $n$ than
    $\Lambda_1$, meaning that high-order resonances are more important for perpendicular propagation.
    Equation~\ref{eq:inhomogeneous_solution} is the phase space linear response for electrostatic fluctuations
    in a strongly magnetized plasma.

    The self-consistent spectrum consists of all pairs $(\omega, k_\perp, k_\parallel)$
    such that the zeroth moment of $f_1(\omega, k_\perp, k_\parallel, v_\perp, \phi, v_\parallel)$
    results in an electric potential mode of wavenumber $(k_\perp, k_\parallel)$.
    Integration of the phase space fluctuation gives the density fluctuation as
    \begin{equation}\label{eq:zero_moment_solution}
    n_1(\omega, k_\perp, k_\parallel) = -i\frac{q\varphi}{m}\int_{-\infty}^{\infty}\int_0^\infty \Big[
            \frac{1}{v_\perp}\frac{\partial f_0}{\partial v_\perp}\Upsilon_2 +
            k_\parallel\frac{\partial f_0}{\partial v_\parallel}\Lambda_2\Big]2\pi v_\perp dv_{\perp}dv_{\parallel}
    \end{equation}
    where a set of additional series, analogs of Eqs.~\ref{eq:polar1} and~\ref{eq:polar2}, are defined as
    \begin{align}
        \Upsilon_2 &= \sum_{n=-\infty}^{\infty}\frac{n\omega_c}{\omega - k_\parallel v_\parallel - n\omega_c}
        J_n^2(k_\perp v_\perp/\omega_c),\label{eq:upsilon2}\\
        \Lambda_2 &= \sum_{n=-\infty}^{\infty}\frac{\omega_c}{\omega - k_\parallel v_\parallel - n\omega_c}
        J_n^2(k_\perp v_\perp/\omega_c)\label{eq:lambda2}.
    \end{align}
    Substitution of $n_1$ into Poisson's equation gives Harris's dispersion relation
    \begin{equation}\label{eq:harris_dispersion}
        \varepsilon(\omega, k_\perp, k_\parallel) \equiv 1 - \Big(\frac{\omega_p}{\omega_c}\Big)^2\frac{1}{(k\lambda_D)^2}
        \int_{-\infty}^{\infty}\Big(\mathbb{V}_\perp f_\parallel + k_\parallel \mathbb{V}_\parallel
    \frac{\partial f_\parallel}{\partial v_\parallel}\Big) dv_{\parallel} = 0
    \end{equation}
    where the integration over perpendicular velocities is broken out into the two quantities
    \begin{align}
        \mathbb{V}_{\perp} &= \int_0^\infty\frac{1}{v_\perp}\frac{\partial f_\perp}{\partial v_\perp}\Upsilon_2(v_\perp,
        v_\parallel)2\pi v_{\perp}dv_{\perp},\label{eq:v_perp}\\
        \mathbb{V}_{\parallel} &= \int_0^\infty f_\perp \Lambda_2(v_\perp,
        v_\parallel)2\pi v_{\perp}dv_{\perp},\label{eq:v_para}
    \end{align}
    and separability of the background $f_0(v_\perp, v_\parallel) = f_\perp(v_\perp)f_\parallel(v_\parallel)$ has been assumed.

    \subsection{Amplitude limitation of linearization around zero-order cyclotron orbits}\label{subsec:linearization_validity}
    Given a zero-order spatially uniform magnetic field 
    $\bm{B} = \bm{B}_0$ and first-order
    electric field perturbation $\bm{E} = \bm{E}_1$, the zero- and first-order kinetic equations are
    \begin{align}
        (\bm{v}\times\bm{B}_0)\cdot\nabla_{v}f_0 &= 0,\label{eq:gyrotropy}\\
        \partial_{t}f_1 + \bm{v}\cdot\nabla_{x}f_1 + \frac{q}{m}(\bm{v}\times\bm{B}_0)\cdot\nabla_{v}f_1 +
    \frac{q}{m}\bm{E}_1\cdot\nabla_{v}f_0 &= 0\label{eq:first_order}
    \end{align}
    assuming a homogeneous zero-order distribution $f_0 = f_0(\bm{v})$.
    Equation~\ref{eq:gyrotropy} indicates gyrotropy of $f_0$. 
    This ordering is valid when the zero-order cyclotron acceleration is much greater
    than the electrostatic acceleration of a typical particle.  
    Validity translates to an amplitude restriction on electric potential
    and the density fluctuation.
    Assuming $k_\parallel=0$ and comparing terms proportional to $\nabla_{v}f_0$ in
    Eqs.~\ref{eq:gyrotropy} and~\ref{eq:first_order} for a thermal particle gives $E_1 \ll v_{th}B_0$.
    Estimating the field $E_1$ of wavenumber $k$ by Gauss's law gives $E_1 = e\delta n / k\varepsilon_0$ for density fluctuation
    $\delta n$.
    Combining these estimates results in equivalent conditions on amplitude as measured by $\delta n$ or $\varphi$,
    \begin{align}\label{eq:estimate}
        \frac{\delta n}{n_0} &\ll \Big(\frac{\omega_c}{\omega_p}\Big)^2(kr_L),\\ 
        \frac{e\varphi}{kT} &\ll \frac{1}{kr_L},
    \end{align}
    for Larmor radius $r_L = v_{th}/\omega_c$. 
    Amplitudes which exceed these inequalities are subject to electrostatic Landau damping as
    the dielectric function of Eq.~\ref{eq:harris_dispersion} is not valid.
    Typical cyclotron instabilities have $k_\perp r_L \approx 1$ which limits the amplitudes of the linear modes  
    considered in this section to amplitudes $\delta n/n_0 \ll (\omega_c/\omega_p)^2$ and $e\varphi\ll kT$.

    \subsection{The dielectric function for ring ($\chi$-) distributions}\label{subsec:dielectric_for_losscone}
    The linear mode spectrum depends on the background distribution~(the zero-order equilibrium).
    Plasma theory textbooks consider Maxwellian plasmas by expansion in the cyclotron harmonics~(\cite{gurnett2017introduction} Section 10.2.3).
    Of course, in ideal collisionless plasmas with plasma parameter $\Lambda\to\infty$ distributions are expected
    to be observed only close enough to Maxwellian such that the Penrose criterion is satisfied.

    Recent analytical work on non-Maxwellian distributions focuses on the kappa distributions~(\cite{mace2009new})
    to model observations in space plasma~(\cite{pierrard2010kappa}).
    Kappa ($\kappa$-) distributions, also called q-Gaussians, are
    motivated by recent advances in entropy methods~(\cite{livadiotis2023entropy, zhdankin2023dimensional}).
    It is thought that such entropy methods may facilitate the extension of maximum entropy principles to the prediction of
    metastable equilibria such as non-Maxwellian velocity distributions
    or self-organized equilibria in magnetic confinement including tokamaks~(\cite{dyabilin2015thermodynamic})
    and Z pinches~(\cite{crews2023kadomtsev}).
    \cite{ewart_2022} is a significant recent advance with a lucid description of collisionless relaxation.
    
    Spatially uniform strongly magnetized plasmas must have zero-order gyrotropy so
    that non-Maxwellian features in perpendicular velocity space are typically ring-shaped.
    Ring distributions commonly arise from the loss cone mechanism of magnetic traps or planetary magnetospheres.
    Early identifications of velocity-space instability in ring-distributed plasmas were made by~\cite{dory1965unstable}.
    Dory's ring distribution, known in the mathematics literature as a $\chi$-distribution, is a type of maximum entropy
    distribution subject to two constraints on variance. 
    The studies of~\cite{tataronis1970_cyclotron1} and~\cite{tataronis1970_cyclotron2}
    extended the theory to oblique propagation, showing maximal growth rates for near-perpendicular propagation,
    though analytical work was performed only with singular ring distributions.
    Around the turn of the millennium $q$-analogs of Dory's analytic ring distributions were introduced by~\cite{leubner2001general}
    and extended in~\cite{pokhotelov2002linear}, motivated by the successful use of $\kappa$-distributions
    as $q$-deformations of Maxwell-Boltzmann statistics.
    Dory's $\chi$-distribution is the $q\to 1$ limit of Leubner's $\kappa$-like ring distributions in the same way that
    the Maxwell-Boltzmann distribution is the $q\to 1$ limit of the $q$-Gaussian (or $\kappa$-) distributions.

    For this reason, in this section we focus on Dory's ring distribution 
    and analyze the dielectric function for such rings assuming separability of the zero-order distribution as
    \begin{equation}\label{eq:product_distribution}
        f_0(v_\parallel, v_\perp) = f_{\parallel}(v_\parallel)f_{\gamma}(v_\perp)
    \end{equation}
    where $f_\parallel(v_\parallel)$ is a Maxwellian of thermal velocity $v_t$ and $f_\gamma(v_\perp)$ is
    Dory's ring function
    \begin{equation}\label{eq:ring_distribution}
        f_{\gamma}(v) = \frac{1}{\pi\alpha^2}\frac{1}{\Gamma(\gamma+1)}\Big(\frac{v^2}{\alpha^2}\Big)^{\gamma}\exp\Big(-\frac{v^2}{\alpha^2}\Big)
    \end{equation}
    of thermal velocity $\alpha$ and parameter $\gamma$, where $\Gamma(z)$ is the Gamma function.
    The function $f_\gamma$ is a polar distribution,  
    \begin{equation}\label{eq:normalization}
        \int_0^\infty f_\gamma(v)2\pi vdv = 1,\quad\text{ for } \text{Re}(\gamma) > -1
    \end{equation}
    yet is bounded at zero only for $\gamma \geq 0$.
    Equation~\ref{eq:ring_distribution} is also known as a $\chi$-distribution and is ``two-temperature''.
    That is,
    the ring distribution $f_\gamma(v_\perp)$ is the maximum entropy distribution for $v_\perp\in[0,\infty)$
    subject to the \textit{two} constraints $\langle v_\perp^2\rangle = \big(\gamma+1\big)\alpha^2$ and
    $\big\langle(v_\perp - \langle v_\perp\rangle)^2\big\rangle = \alpha^2\Big(1 + \gamma -
    \big(\frac{\Gamma(\gamma+3/2)}{\Gamma(\gamma+1)}\big)^2\Big)$ or
    $\big\langle(v_\perp - \langle v_\perp\rangle)^2\big\rangle = \frac{1}{4}\alpha^2 + \mathcal{O}(\gamma^{-1})$
    such that the temperature in the gyrating frame is independent of $\gamma$ as $\gamma\to\infty$.
    Thus the physical meaning of $\alpha$ is the thermal velocity in the gyrating frame, while $\gamma$
    is the boost to thermal energy in the laboratory frame (and need not be an integer).
    In this sense the distribution has two temperatures.

    The ring distribution satisfies the recurrence (and $f_{-1}=0$),
    \begin{equation}\label{eq:recurrence_relation}
        \frac{1}{v}\frac{df_\gamma}{dv} = \frac{1}{\alpha^2}\frac{f_{\gamma-1} - f_\gamma}{2}.
    \end{equation}
    In the case of ring distributions of the form of Eq.~\ref{eq:ring_distribution}, the integrals in both quantities
    $\mathbb{V}_\perp$ and $\mathbb{V}_\parallel$ involve only $f_\gamma$ due to the recurrence
    Eq.~\ref{eq:recurrence_relation}, and suggests defining
    \begin{equation}\label{eq:equation_with_parameters}
        \mathbb{F}_{n,\gamma}(k) \equiv \int_0^{\infty}f_{\gamma}(v)J_n^2(kv)2\pi vdv
    \end{equation}
    which, as shown in Appendix~\ref{sec:integral_appendix},
    is a type-$_{2}F_2$ hypergeometric function with series
    representation~(\cite{gradshteyn})
    \begin{equation}\label{eq:hypergeometric}
    \begin{split}
        \mathbb{F}_{n,\gamma}(x) = \frac{1}{\Gamma(\gamma+1)}
    \sum_{\ell=0}^{\infty}&\frac{\Gamma(2n+2\gamma+2\ell+1)}
    {\Gamma^2(n+\ell+1)\Gamma(2n+\ell+1)}\frac{(-1)^\ell}{\ell!}x^{2(\ell+n)}.
    \end{split}
    \end{equation}
    We now proceed with integrating this power series over the parallel velocities.
    First, we can make a note on an alternative possibility.
    Rather than integrating the auxiliaries Eqs.~\ref{eq:upsilon2} and~\ref{eq:lambda2} in their summation form, it is
    possible to first close the summation with the Lerche-Newberger summation theorem~(\cite{newberger1982new}), and
    to determine the perpendicular velocity integrals in Eqs.~\ref{eq:v_perp} and~\ref{eq:v_para} in closed form as
    hypergeometric functions, as in Appendix~\ref{sec:perp_app}, for arbitrary $k_\parallel$.
    However, the integration over parallel velocities must then proceed by a series expansion around the poles of these
    hypergeometric functions, making a power series approach inevitable.
    On the other hand, the power series developed in this section maintains separability of terms containing the parallel
    velocity.

    Proceeding to the integration over parallel velocities, with the parallel distribution $f_{\parallel}(v_\parallel)$
    taken as a Maxwellian distribution, the two integrals are
    \begin{align}
        \int_{-\infty}^\infty \frac{n\omega_c}{\omega - k_\parallel v_\parallel - n\omega_c}f_\parallel dv_\parallel &=
        -\frac{n}{k_\parallel}Z(\zeta_n)\\
        \int_{-\infty}^{\infty} \frac{\omega_c}{\omega - k_\parallel v_\parallel - n\omega_c}
        \frac{\partial f_\parallel}{\partial v_\parallel}dv_{\parallel}
        &= \frac{1}{2k_\parallel}Z'(\zeta_n)
    \end{align}
    where $Z(\zeta)$ is the plasma dispersion function and the cyclotron harmonic-shifted phase velocity is defined as
    $\zeta_n \equiv \frac{\omega-n\omega_c}{\sqrt{2}k_\parallel v_t}$.
    Thus the dielectric function for loss-cones is
    \begin{equation}\label{eq:ring_dispersion}
        \begin{split}
            \varepsilon(\omega, k) &= 1 - \Big(\frac{\omega_p}{\omega_c}\Big)^2\frac{1}{(kr_L)^2}
            \sum_{n=-\infty}^{\infty}\Big[Z'(\zeta_n)\mathbb{F}_{n,\gamma}(k_\perp r_L)
            \\
            &\quad\quad\quad+\frac{n}{k_\parallel r_L}Z(\zeta_n)
            \Big\{\mathbb{F}_{n,\gamma-1}(k_\perp r_L) - \mathbb{F}_{n,\gamma}(k_\perp r_L)\Big\}\Big].
        \end{split}
    \end{equation}
    Equation~\ref{eq:ring_dispersion} reduces to the standard series for a Maxwellian ($\gamma=0$) by the identity
        $\mathbb{F}_{n,0}(k) = e^{-k^2}I_n(k^2)$
    where $I_n(k)$ is the modified Bessel function of the first kind.

    \subsection{Perpendicular propagation in ring distributions}\label{subsec:perpendicular_propagation}
    In the limit of perpendicular propagation, $k_\parallel\to 0$, the series Eq.~\ref{eq:ring_dispersion} in the
    cyclotron harmonics simplifies as the terms proportional to $Z(\zeta_n)$ vanish.
    Further, by use of the Lerche-Newberger summation theorem two alternatives to the power series
    for the perpendicular cyclotron wave dielectric function may be developed which incorporate the contributions
    from the cyclotron harmonics to all orders.
    The derivation of these expressions may be found in Appendix~\ref{sec:perp_app}.
    The first is a closed form, a hypergeometric function with complex poles at the cyclotron resonances,
    \begin{equation}\label{eq:closed_form}
        \begin{split}
        \varepsilon(\omega, k_\perp) = 1 + \Big(\frac{\omega_p}{\omega_c}\Big)^2\frac{1}{(k_\perp r_L)^2}
        \Big\{
            &\pFq{2}{2}{\frac{1}{2},\quad\quad \gamma+1}{1+\omega/\omega_c, 1-\omega/\omega_c}(-2(k_\perp r_L)^2)\\
            &-\pFq{2}{2}{\frac{1}{2},\quad\quad \gamma}{1+\omega/\omega_c,1-\omega/\omega_c}(-2(k_\perp r_L)^2)
        \Big\} = 0
        \end{split}
    \end{equation}
    and the second form is a representation of Eq.~\ref{eq:closed_form} as a trigonometric integral
    generalizing that used in~\cite{tataronis1970_cyclotron1, vogman2014dory, datta2021electromagnetic},
    \begin{equation}\label{eq:trig_form}
        \varepsilon(\omega, k) = 1 + \Big(\frac{\omega_p}{\omega_c}\Big)^2\csc(\pi\omega/\omega_c)
        \int_0^{\pi}\sin(\theta)\sin(\theta\omega/\omega_c)L_\gamma(\beta)\exp(-\beta)d\theta
    \end{equation}
    with $\beta\equiv 2\cos^2\big(\theta/2\big)(k_\perp r_L)^2$ and $L_\gamma(\beta)$ the Laguerre
    polynomial of order $\gamma$.
    Equation~\ref{eq:trig_form} is particularly suited to numerical calculation by quadrature.
    Hypergeometrics similar to Eq.~\ref{eq:closed_form} have been obtained for the Maxwellian plasma and for
    $\kappa$-distributions~(\cite{mace2004generalized, mace2009new}),
    and reduce to the Maxwellian result for the parameter $\gamma= 0$.
    Observe that Eqs.~\ref{eq:closed_form} and~\ref{eq:trig_form} are functions of frequency and not
    of phase velocity as there is no ballistic contribution to the zero-order motion.
    It is expected that Eqs.~\ref{eq:ring_dispersion}--\ref{eq:trig_form} will be the basic dispersion functions
    from which to build the q-analogs for $\chi$-distributions proposed by~\cite{leubner2001general} 

    \subsection{Visualization of the dispersion function and phase space eigenfunctions}\label{subsec:visualization_and_eigenmodes}
\begin{figure}
        \centering
        \includegraphics[width=\columnwidth]{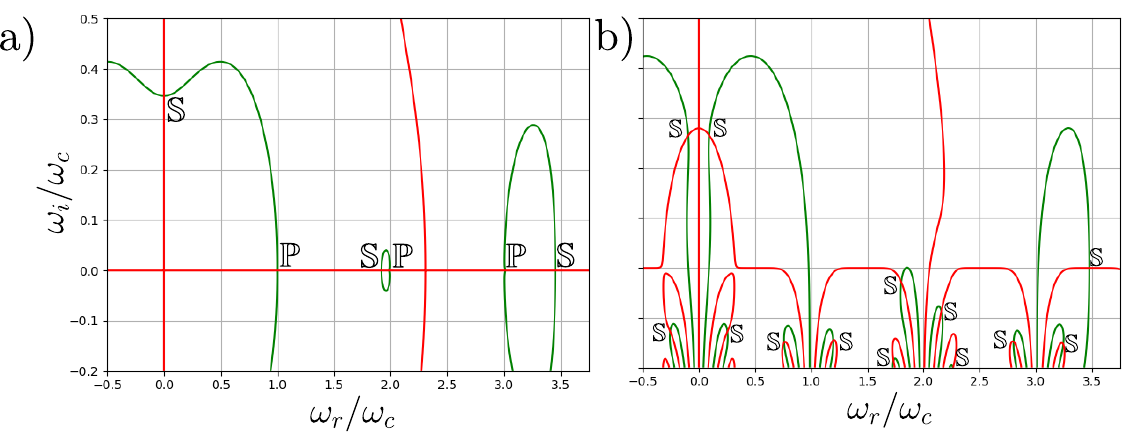}
        \caption{Contours of $\text{Re}(\varepsilon)=0$ (green) and $\text{Im}(\varepsilon)=0$ (red)
            for the electrostatic dispersion function of a loss cone-distributed plasma with parameters
            $\omega_p = 10\omega_c$, $\gamma=6$, and
            $k_\perp r_L=0.9$ for case a) $90^\circ$ propagation and case b)
            $85^\circ$ propagation.
            Either a solution or a pole occurs at an intersection of these two curve families.
            Solutions to the dispersion relation $\varepsilon(\omega, k)=0$ are labelled $\mathbb{S}$ while simple poles
            are labelled $\mathbb{P}$. There are no poles for oblique $(< 90^\circ)$ propagation.
            Solutions correspond to the first few electron Bernstein modes.
            The dispersion function shows Gaussian-like responses in the lower-half plane for the oblique wave, an
            indicator of dissipative wave-particle resonance near the cyclotron harmonics.}\label{fig:oblique0}
    \end{figure}

    Figure~\ref{fig:oblique0} plots the electrostatic dispersion function in the complex frequency plane for
    a ring distribution of $\gamma=6$ for magnetization $\omega_c/\omega_p=0.1$ at wavenumber $kr_L=0.9$.
    A plethora of solutions to the complex dispersion function $\varepsilon=0$ is illustrated by the intersection
    of the zeros of the real and imaginary parts, which take place at both solutions and poles.
    Complex poles at the cyclotron harmonics occur only for the case $k_\parallel\to 0$, as evident from the cosecant
    function $\csc(\pi\omega/\omega_c)$ in Eq.~\ref{eq:trig_form}.
    In this way solutions and simple poles can be clearly distinguished.
    It is clear that, given zero-order cyclotron orbits, oblique modes are Landau damped but perpendicular modes are not.
    When wave amplitudes violate the inequality of Eq.~\ref{eq:estimate} nonlinear phenomena occur, and
    perpendicular waves are also Landau damped.
    In this situation one may see streaming instabilities in numerical experiments in the magnetization transition
    regime $\omega_c \approx \omega_p$.

\begin{figure}
        \centering
        \includegraphics[width=0.618\columnwidth]{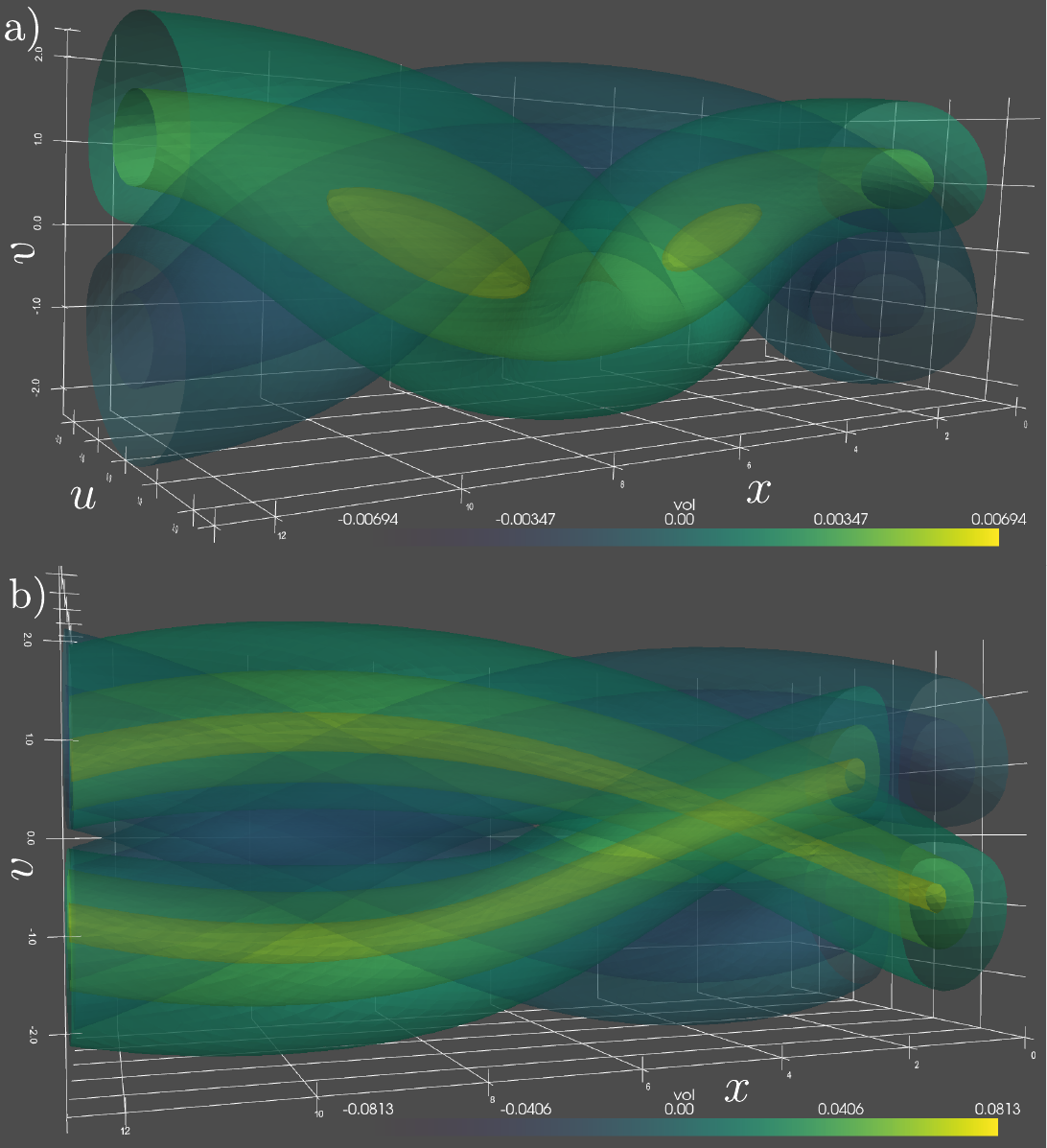}
    \caption{Phase space eigenfunctions of (a) the first and (b) the second cyclotron-harmonic
    electron Bernstein waves propagating orthogonal to $\bm{B}_0$, visualized in the 
    phase space $(x, u, v)$.
    The order of the cyclotron harmonic determines the number of helical strands of a given sign.
    The phase space eigenfunction propagates in the lab frame, 
    but when the velocity space is observed at a fixed spatial coordinate the eigenfunction
        rigidly rotates in the perpendicular velocity space.
    Balanced counterpropagating modes produce a stationary wave.}\label{fig:bernstein_eigenfunctions}
    \end{figure}

    Therefore perpendicularly propagating linear modes do not experience Landau resonance so that
    all such modes are true eigenfunctions.
    The phase space structure associated with these cyclotron waves (that is, Eq.~\ref{eq:inhomogeneous_solution})
    consists of helical modes in the perpendicular velocity space,
    since the primary phase component is $\exp(i(kx + n\phi - \omega t))$ for a mode with frequency $\omega \approx n\omega_c$,
    with $\phi$ the cylindrical velocity space angle.
    The simplest example of such helical phase space modes are the electron Bernstein modes.
    Figure~\ref{fig:bernstein_eigenfunctions} visualizes the eigenfunctions of the first and second cyclotron harmonics
    for a Maxwellian background distribution $f_0(v_\perp)$ in the phase space $(x, u, v)$ with $(u, v)$ the perpendicular
    velocity space and $x$ the propagation coordinate perpendicular to the background magnetic field.
    With non-zero real frequency these helical modes propagate through phase space.

    \subsection{Simulation of perpendicular electron cyclotron loss-cone instability}\label{subsec:two_case_studies}
    Here kinetic eigenfunction initialization is illustrated for
    the instability of an electron loss cone to perpendicular cyclotron waves
    in a neutralizing background, as studied in~\cite{vogman2014dory}.
    Normalizing to the Debye length, plasma frequency,
    and thermal velocity, and the fields by $E_0 = v_{th}B_0$, the Vlasov-Poisson equations are
    \begin{equation}
        \partial_{t}f + F^j\partial_j f = 0,\label{eq:vlasov_cyclotron}
    \end{equation}
    \begin{equation}
        \frac{d^2\varphi}{dx^2} = \int_{-\infty}^\infty\int_{-\infty}^\infty f(x,u,v) dudv - 1,\label{eq:poisson_cyclotron}
    \end{equation}
    \begin{equation}\label{eq:doesnt_need_one}
        F =
        \begin{bmatrix}
            u,&
            \frac{d\varphi}{dx} - vB_{\text{ext}},&
            uB_{\text{ext}}
                \end{bmatrix}^T
    \end{equation}
    with coordinates $(x,u,v)$.
    The external magnetic field $B_{\text{ext}}$ is set such that the magnetization parameter
    $\omega_c / \omega_p = B_\text{ext}$ in normalized units.
    Two single-mode simulations termed $A$ and $B$ are performed for Eqs.~\ref{eq:vlasov_cyclotron}--\ref{eq:doesnt_need_one}
    in the highly unstable over-dense parameter regime $\omega_p = 10\omega_c$
    using as eigenvalues two solutions to $\varepsilon(\omega,k)=0$, namely
    $(k_A\lambda_D,\omega_A /\omega_c)\approx (0.886, 0.349i)$ and
    $(k_B\lambda_D, \omega_B/\omega_c)\approx (1.4, 1.182 + 0.131i)$
    found using the integral form of Eq.~\ref{eq:trig_form}
    with fifty point Gauss-Legendre quadrature.

    \begin{figure}
        \centering
        \includegraphics[width=\columnwidth]{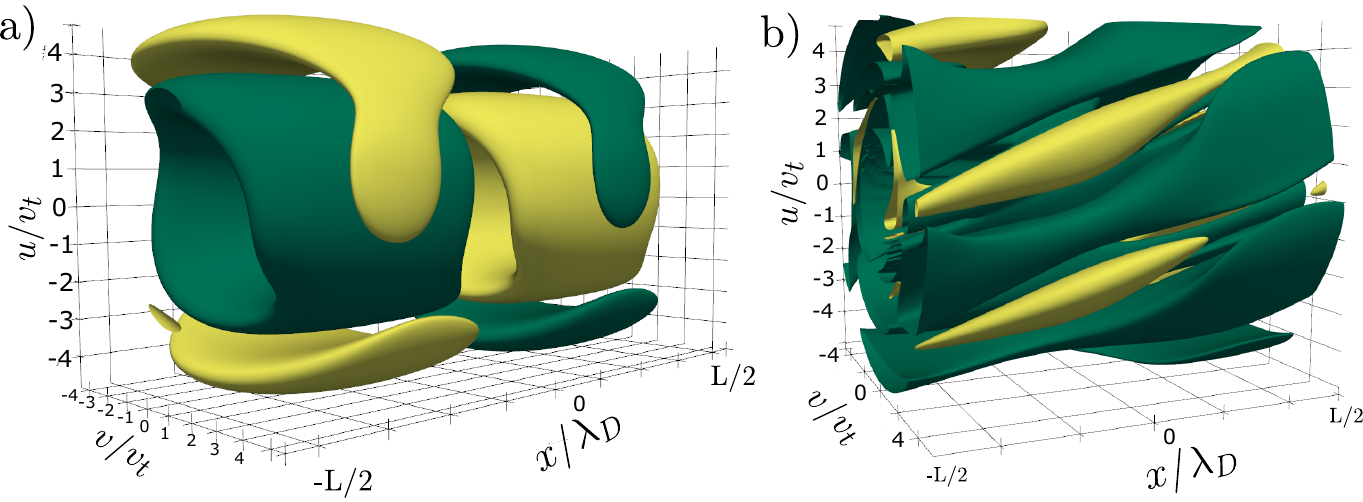}
        \caption{Shown here are the perturbations $f_1 \equiv f - f_0$ at time $t=0$ for a) simulation $A$
            with $\omega_r = 0$ and b) simulation $B$ with $\omega_r\neq 0$,
            each with iso-surfaces at $30\%$ of the minimum (green) and maximum (yellow).
            The considered eigenfunctions $f_1(x,u,v)$ consist of twisting islands in the phase space,
            capturing the combined physics of translation, electric acceleration, and magnetic gyration.
            Translating modes ($\omega_r\neq 0$) are associated with a helical structure.}\label{fig:perturbation}
    \end{figure}

    In this situation, case $A$ corresponds to a stationary mode with wavelength long compared to the thermal Larmor radius,
    similar to the two-stream instability studied in the unmagnetized case,
    while case $B$ corresponds to a destabilized propagating Bernstein-like mode at the first cyclotron harmonic with
    more significant finite Larmor radius effect.
    The spatial domain is set to $L = 2\pi / k$ in each case and the velocity boundaries
    to $u_{\text{max}}, v_{\text{max}}= \pm 8.5$.
    We perturb these nonlinear simulations using the phase space eigenfunctions corresponding to the eigenvalue
    pairs $(k_A, \omega_A)$, $(k_B, \omega_B)$.

    Figure~\ref{fig:perturbation} shows isocontours of the phase space eigenfunctions used as the
    initial perturbations.
    Case $A$ has a phase space structure similar to the basic plasma wave seen in the two-stream instability,
    while case $B$ has a helical structure as a cyclotron mode with $\omega_r\approx \omega_c$.
    We reiterate here that the eigenfunction perturbation allows arbitrary perturbation amplitude and still produces
    the same nonlinear phenomena, namely the saturated state or mode coupling/conversion.
    However, with zero-order $\bm{B}_0$ the initial amplitude must not exceed the limit of
    Eq.~\ref{eq:estimate} or nonlinear phenomena will develop as the perturbation electric force is not first order
    compared to the thermal magnetic force.

    \begin{figure}
        \centering
        \includegraphics[width=0.618\columnwidth]{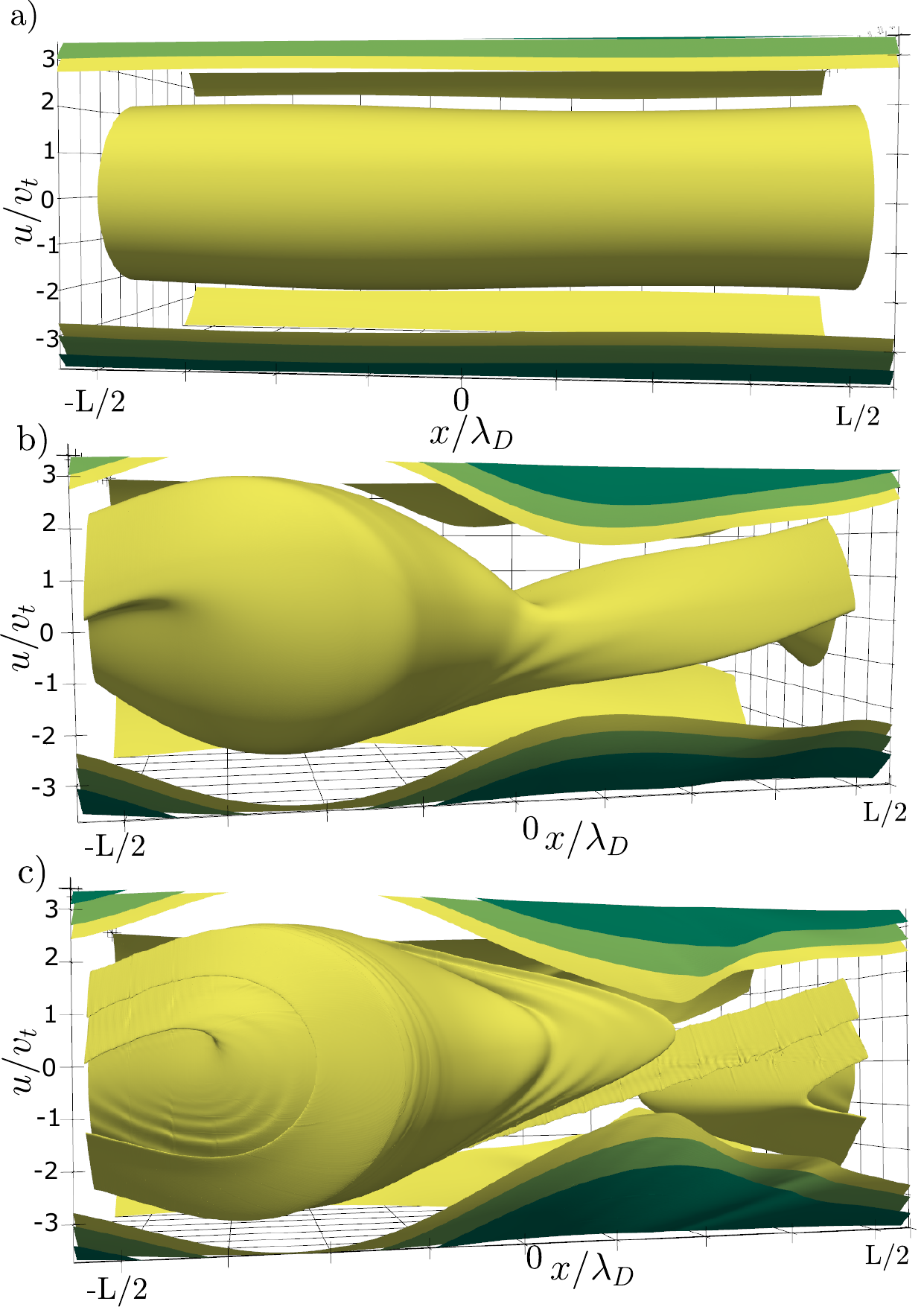}
        \caption{Phase space view $(x,u,v)$ of simulation $A$ focused on $(x,u)$-plane as
        iso-contours at $15\%$ of $\text{max}(f)$ shown in yellow,
        at times a) $t=0$, b) $t=80$, and c) $t=120$.
        The domain within the original ring is shown with $(u,v)\in (-3.5, 3.5)$ to focus on the trapping dynamics.
        The trapping structure consists of a ribbon winding around
        a separatrix, while the outer bulk ring distribution maintains passing trajectories.
        This saturation geometry is typical for electrostatic potentials in a magnetic field as the
        magnetic force depends on the sign of the transverse velocity.}\label{fig:simulation_a_evolution}
    \end{figure}

    \subsubsection{Numerical method for loss-cone simulations}\label{sec:num_loss_cone}
    The problem is evolved numerically using the discontinuous Galerkin (DG) method described in
    Appendix~\ref{app:summary_of_numerical} and in~\cite{crews_kinetic}, with the difference that the spatial
    coordinate is not Fourier transformed but also discretized by DG method.
    We use an element resolution $(N_x, N_u, N_v) = (25, 50, 50)$ and nodal basis of $n=8$ LGL nodes per dimension,
    while the Shu-Osher SSPRK3 method is used to integrate the semi-discrete equation in time.
    These instabilities grow on a slow time-scale relative to the plasma frequency;
    that is, they grow at a fraction of the cyclotron time-scale $\omega_c^{-1}$, while time $t$ is
    normalized to the plasma frequency $\omega_p^{-1}$.
    Thus these instabilities take many plasma periods to reach nonlinear saturation beginning from amplitudes below
    the limit of Eq.~\ref{eq:estimate}.
    Simulation $A$ reaches saturation around $t=100$ and runs to $t=175$ while simulation $B$ saturates at around
    $t=175$ and stops at $t=200$.

\begin{figure}
        \centering
        \includegraphics[width=\columnwidth]{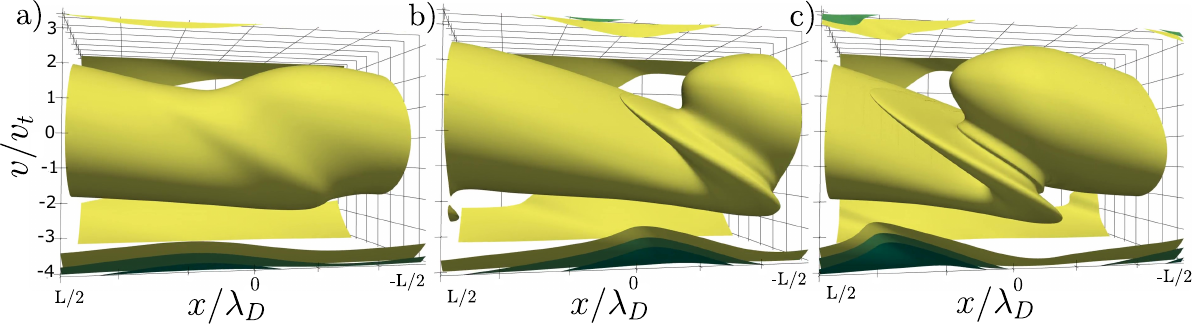}
        \caption{Phase space view looking on $(-x,v)$-plane of simulation B at $15\%$ iso-contours of $\text{max}(f)$
            (yellow), with a) the nonlinear mode developing at $t=100$, b) the developed vortex at $t=160$,
            and c) the saturated vortex translating at $t=180$.
            The mode is seen to be a growing, translating electrostatic potential $\varphi(x)$ of positive phase
            velocity $v_{\varphi} = \omega_r / k$ with an underlying phase space vortex structure centered at $(u,v)=0$.
            The vortex shape is explained by considering the trajectory of a test particle in the wave.
            That is, particles with a velocity close to that of the wave see a stationary potential and are accelerated to a
            high $u$-velocity.             
            They then translate towards positive $x$ while their velocity vector is rotated by the Lorentz force
            to $-u$ at a rate close to the wave frequency (as $\omega_r \approx 1.2\omega_c$) and repeats the cycle.}
        \label{fig:simulation_b_evolution}
    \end{figure}

    Three-dimensional isosurface plots were produced using PyVista, a Python package for VTK.
    To prepare the data, an average is taken of nodal values lying on element boundaries for smoothness, and the
    $8$-nodes per element are resampled to $25$ linearly-spaced points per axis and per element onto the basis functions
    of the DG method.
    These iso-contours are shown for simulations $A$ and $B$ in Figs.~\ref{fig:simulation_a_evolution} and~\ref{fig:simulation_b_evolution}
    respectively.

    \subsubsection{Fully nonlinear single-mode loss-cone simulation results}
\begin{figure}
        \centering
        \includegraphics[width=\columnwidth]{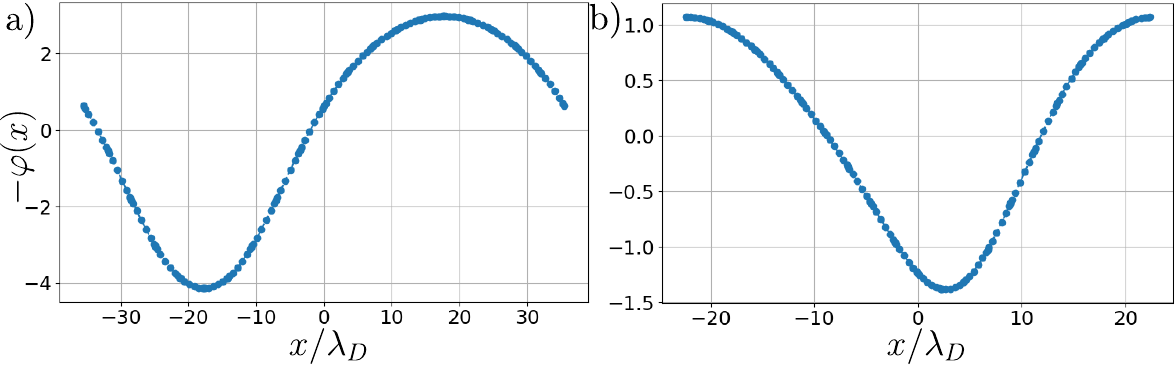}
        \caption{Electric potentials $\varphi(x)$ at saturation of the two studied cases, for (a) simulation A at $t=120$ and
            (b) simulation B at $t=180$. The potential of A is stationary while that of B is translating to the right.
        The negative of potential $-\varphi(x)$ is shown in order to account for the electron's negative charge.
        In both cases electron holes develop in the potential wells of $-\varphi(x)$.}\label{fig:potentials}
    \end{figure}

    Figure~\ref{fig:potentials} shows the electric potentials $\varphi(x)$ in simulations $A$ and $B$,
    computed using the numerical method described in Section~\ref{sec:num_loss_cone}.
    In simulation $A$ the wave potential $\varphi(x)$ is stationary with a weakly fluctuating boundary,
    so that part of the density $f(x,v)$ within the potential well executes trapped orbits.
    This results in a trapping structure with orbits tracing a nonlinear potential similar to the
    characteristic pendulum-like cats-eye separatrix of the single-mode electrostatic two-stream instability.
    In this case the electrons are magnetized and execute zero-order cyclotron motion so that the trapping separatrix in
    simulation $A$ is instead in the form shown by the isosurfaces.
    The saturated state of simulation A is a lattice of one-dimensional strongly magnetized electron holes.
    The continuum of such hole solutions is key to strongly driven plasma transport physics~(\cite{schamel2023pattern}).
    Two-dimensional axisymmetric magnetized holes are the focus of analytical work
    in~\cite{hutchinson2020particle} and~\cite{hutchinson2021synthetic}.  

\begin{figure}
        \centering
        \includegraphics[width=\columnwidth]{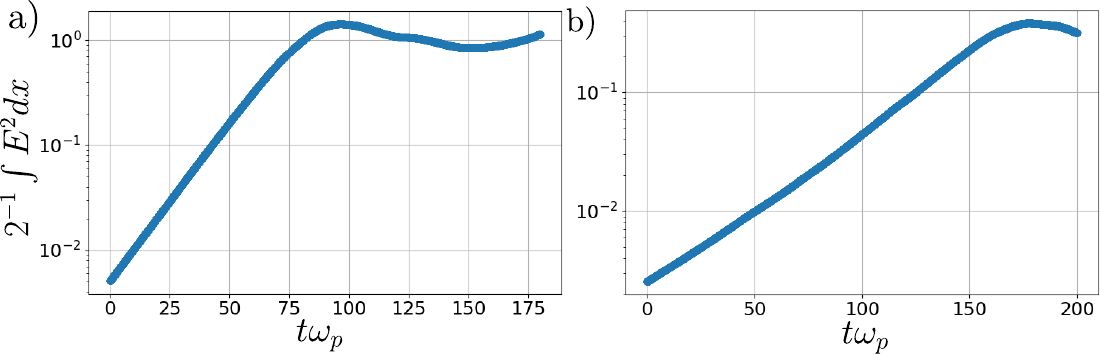}
        \caption{Domain-integrated electric field energy traces for simulations (a) A and (b) B.
        The thermal energy of the zero-order distribution is $0.25$ per unit length with $\gamma=6$ and $\alpha = 1$.
        In simulation A this corresponds to a domain-integrated thermal energy of $E_A\approx 17.8$ and in simulation B
        to $E_B\approx 11.2$. Therefore in both cases the instability saturates with an electric energy a few percent of the
            thermal energy, approximately $6\%$ in A and $2.5\%$ in B.}\label{fig:cyclotron_energy}
    \end{figure}

    The saturated wave potential of simulation $B$, on the other hand, translates with positive phase velocity
    $\zeta \approx \omega_r / k$.
    The region of particle interaction translates along with the wave potential and forms a vortex structure in the
    phase space density $f(x, u, v)$.
    The center of this kink continues to tighten as the simulation progresses, leading to progressively
    finer structures just as in simulation $A$.
    This effect is in agreement with the filamentation phenomenon and introduces simulation error as the structures
    lead to large gradients on the grid scale where discreteness produces dispersion error.
    For this reason the simulation is stopped at $t = 200$.
    This solution is perhaps somewhat artificial as it is obtained by single-mode initialization via its
    phase space eigenfunction, thereby not disturbing modes of greater growth rate.
    This is demonstrated through the energy traces in Fig.~\ref{fig:cyclotron_energy}, showing that simulation $A$
    saturates with a greater proportion of the plasma thermal energy than simulation $B$.
    The solution of simulation B has the significance of a propagating train of nonlinear electrostatic cyclotron waves
    with associated electron holes, driven unstable by the free energy of a ring distribution.

    \subsubsection{Experimental consequences of cyclotron loss-cone instabilities}
    Magnetic mirror trapping requires the maintenance of a loss-cone distribution in the confined plasma.
    Simulations such as these, and quasilinear theories, maintain that kinetic instabilities lead to a relaxation of
    the distribution function on macroscopic scales towards near-Maxwellian distributions, along with
    long-lived vortical structures in the phase space.
    By examination of the dispersion functions one finds that phase space instability may be suppressed when
    $\omega_p \ll \omega_c$.
    Assuming equal electron and ion temperatures and densities, we may write for the plasma beta
    \begin{equation}\label{eq:beta_as_f_of_omegap}
        \beta = \Big(\frac{v_{ti}}{c}\Big)^2\Big(\frac{\omega_{pi}}{\omega_{ci}}\Big)^2 = \Big(\frac{r_{Li}}{\lambda_i}\Big)^2
    \end{equation}
    with $\lambda_i = c/\omega_{pi}$ the ion skin depth.
    Thus, non-Maxwellian features such as ring distributions, as $\chi$-distributions or their $\kappa$-analogs
    (\cite{pokhotelov2002linear, leubner2004fundamental}), are expected to persist in very low--$\beta$ plasma,
    in much the same way in which for weakly magnetized plasma the $\kappa$-distributions persist in the collisionless regime
    when the plasma parameter $\Lambda\gg 1$.
    Further discussion on the consequences of mirror instability in high-$\beta$ space plasma
    can be found in~\cite{pokhotelov2004mirror}.
    
    \section{Electromagnetic eigenmodes with zero-order ballistic trajectories}\label{sec:electromagnetic_theory}
    We return to weakly magnetized plasma where the zero-order magnetic field is weak enough
    compared to perturbations such that zero-order motion is ballistic.  
    We consider the electromagnetic linear response, determine the plasma and field configurations 
    of the kinetic eigenfunctions, and utilize them to initialize one- and two-dimensional nonlinear
    simulations of collisionless electromagnetic instability.
    We then study the magnetic-trapping electron holes resulting from electromagnetic instability.
    Purely electromagnetic instability arises from pressure anisotropy,
    in which case the linear eigenfunctions are known as Weibel instabilities~(\cite{weibel_orig}),
    although streaming instability may still be determined as electrostatic theory is contained in the limit $\zeta\ll c$.

    Linearization of the Vlasov-Maxwell system
    around a weakly magnetized spatially uniform equilibrium $f_0(\bm{v})$ with no mean drift
    yields the system
    \begin{align}
        \partial_t f_1 &+ \bm{v}\cdot\nabla_x f_1 +
        \frac{q}{m}(\bm{E} + \bm{v}\times\bm{B})\cdot\nabla_v f_0 = 0,\\
        \partial_t\bm{B} &= -\nabla\times\bm{E},\\
        c^{-2}\partial_t\bm{E} &= -\mu_0\bm{j}_1 + \nabla\times\bm{B}\label{eq:ampere_law2}.
    \end{align}
    Faraday's equation gives $\bm{B} = \omega^{-1}\bm{k}\times\bm{E}$
    such that the spectral Lorentz force is
    \begin{equation}\label{eq:fourier_lorentz}
    \begin{split}
        \bm{E} + \bm{v}\times\bm{B} 
        &= \omega^{-1}\big((\omega - \bm{k}\cdot\bm{v})\bm{E} + (\bm{v}\cdot\bm{E})\bm{k}\big),
    \end{split}
    \end{equation}
    and a two-sided-in-time quasi-analysis produces the Vlasov linear response as
    \begin{equation}\label{eq:linear_response_electromagnetic}
        i\omega f_1 = \frac{q}{m}\bm{E}\cdot\nabla_{v}f_0 +
    \frac{q}{m}(\bm{v}\cdot\bm{E})\frac{\bm{k}\cdot\nabla_v f_0}{\omega - \bm{k}\cdot\bm{v}}.
    \end{equation}
    The spectral time-derivative of the perturbed current follows as
    \begin{equation}\label{eq:perturbed_current}
        \mu_0(-i\omega\bm{j}_1) = \frac{\omega_p^2}{c^2}\Big(\bm{E} -
    \int\bm{v}(\bm{v}\cdot\bm{E})\frac{\bm{k}\cdot\nabla_v f_0}{\omega - \bm{k}\cdot\bm{v}}d\bm{v}\Big).
    \end{equation}
    The first term of the phase space linear response has no resonant denominator
    and thus yields a non-thermal perturbed current independent of the zero-order distribution function,
    while the second term encodes resonance between the plasma and its wavefield.

    \subsection{Tensor components for arbitrary Cartesian coordinates}
    In Cartesian coordinates $(x,y,z)$ with wavevector
    $\bm{k} = k_x\hat{x} + k_y\hat{y} + k_z\hat{z}$, the dielectric tensor~(\cite{Skoutnev_2019})
    is obtained from combination of Eqs.~\ref{eq:perturbed_current} and~\ref{eq:ampere_law2} as
    \begin{equation}\label{eq:general_tensor}
        \begin{split}
        \varepsilon_{ij} E_j &= 0,\\
            \varepsilon_{ij} &= \delta_{ij} - \frac{c^2k^2}{\omega^2}\Big(\delta_{ij} - \frac{k_ik_j}{k^2}\Big) -
        \frac{\omega_p^2}{\omega^2}\Big(\delta_{ij} - \int v_i v_j
        \frac{\bm{k}\cdot\nabla_v f_0}{\omega - \bm{k}\cdot\bm{v}}d\bm{v}\Big).
        \end{split}
    \end{equation}
    In 
    the formal initial-value problem an initial-value vector results in the system $\varepsilon_{ij}E_j=g_i$.
    Typically in Cartesian form 
    the moment integrals are inseparable because
    of the resonant denominator $\omega - \bm{k}\cdot\bm{v}$,  
    yet are separable when the frame is chosen with one coordinate aligned with the wave-vector  
    such that the resonant denominator appears as $\omega-kv_\parallel$.

    \subsection{Electromagnetic susceptibility and the eigenvalue problem}\label{subsec:susceptibility_and_eigenvalue}
    Reformulation of Eq.~\ref{eq:general_tensor} as an eigenvalue problem 
    for the phase velocity $\zeta$
    allows calculation of electric field eigenfunctions $\bm{E}$ 
    naturally consistent with the corresponding
    phase space eigenfunction $f_{1}$ given by Eq.~\ref{eq:linear_response_electromagnetic}.
    This facilitates construction of initial conditions for 
    simulation of Vlasov-Maxwell instabilities.
    Multiply by $\omega^2$ and pull out the diagonal tensor $\omega^2 \delta_{ij}$. 
    Define the integrals encoding resonant wave-particle interaction into the self-consistent perturbation current as
    \begin{equation}\label{eq:integral_source}
        I_{ij} \equiv \int_{\mathcal{C}} v_i v_j\frac{\bm{k}\cdot\nabla_v f_0}{\omega - \bm{k}\cdot\bm{v}}d\bm{v}
    \end{equation}
    where the integral is evaluated on the Landau contour $\mathcal{C}$, \textit{i.e.}~analytically continued to the
    lower-half complex $\omega$-plane.
    The dielectric tensor system may then be rewritten as
    \begin{equation}\label{eq:eigenvectors_part1}
        \Big\{\delta_{ij} - I_{ij} + (k\lambda_c)^2\Big(\delta_{ij} - \frac{k_ik_j}{k^2}\Big)\Big\}E_j =
        \Big(\frac{\omega}{\omega_p}\Big)^2 E_i
    \end{equation}
    where $\lambda_c = c/\omega_p$ is the inertial length.
    The resonant integrals are naturally functions of the phase velocity $\zeta=\omega/k$, so one can also express
    the system as
    \begin{equation}\label{eq:eigenvectors_part2}
    \frac{1}{(k\lambda_D)^2}\Big\{\delta_{ij} - I_{ij} + (k\lambda_c)^2\Big(\delta_{ij} - \frac{k_ik_j}{k^2}\Big)\Big\}E_j =
    \Big(\frac{\zeta}{v_t}\Big)^2 E_i.
    \end{equation}
    Equation~\ref{eq:eigenvectors_part2} casts the problem in eigenvalue form as the system of integral equations
    \begin{equation}\label{eq:eigenvectors_part3}
        \chi(k, \zeta)\vec{E} = \zeta^2\vec{E}.
    \end{equation}
    As in the scalar Poisson problem there is a spectrum of solutions $\zeta$ for a given $k$,
    obtained by determining a root of the characteristic function $D(k, \zeta) \equiv$ det$(\chi - \zeta^2 I) = 0$.
    As in the scalar problem only unstable solutions satisfying $\text{Im}(\zeta) > 0$ constitute normal modes of
    oscillation.
    In unmagnetized spatially uniform plasma these modes are either streaming instabilities~(two-stream, Buneman,
    ion-acoustic) or generalized Weibel instabilities.
    In either case their effect is thermalizing on coarse-grained scales by reducing relative velocities far from equilibrium.
    Casting the dielectric tensor for zero-order cyclotron motion~(\cite{gurnett2017introduction} page 408)
    into an eigenvalue problem proceeds in the same manner;
    the main difference is that the integrals $I_{ij}$ are sums over Doppler-shifted cyclotron
    resonances and their calculation, though methodical, is lengthy.
    
    Having determined a particular eigenvalue $\zeta_n^2$ such that $D(k, \zeta_n)=0$,
    the corresponding electric field eigenfunction $\bm{E}_n$ is found by solving for the eigenvector of the matrix
    $\chi_n \equiv \chi(\zeta_n, k)$ with eigenvalue $\zeta_n^2$.
    The other two eigenvalues of the matrix $\chi_n$ are spurious as they do not correspond to solutions of
    Eq.~\ref{eq:eigenvectors_part3}.
    The magnetic field eigenfunction is obtained through $\bm{B}_n = \zeta_n^{-1}\hat{\bm{k}}\times\bm{E}_n$,
    and the phase space eigenfunction $f_{1,n}$ from Eq.~\ref{eq:linear_response_electromagnetic}.

    \subsection{Dielectric tensor components for the anisotropic Maxwellian}\label{subsec:dielectric_bimaxwellian}
    In order to illustrate the Weibel instability due to anisotropy
    in a plasma with zero-order ballistic trajectories it is useful to consider the anisotropic Maxwellian,
    or multiple-temperature, zero-order distribution~(\cite{davidson1972nonlinear}).
    Anisotropy is a key element of kinetic equilibrium in flowing magnetized plasmas~(\cite{mahajan2000sheared}),
    and the anisotropic Maxwellian is the simplest anisotropic model distribution to investigate pure Weibel instability.
    Let $(x,y,z)$ and $(u,v,w)$ be Cartesian coordinates in configuration space and velocity space respectively.
    Consider a three-temperature Maxwellian distribution given by 
    \begin{equation}\label{eq:anisotropic_maxwellian}
        f_0 = \frac{1}{(2\pi)^{3/2}\theta_u\theta_v\theta_w}\exp\Big\{-\frac{1}{2}\Big(\frac{u^2}{\theta_u^2} +
    \frac{v^2}{\theta_v^2} + \frac{w^2}{\theta_w^2}\Big)\Big\}.
    \end{equation}
    For purpose of illustration we obtain from Eq.~\ref{eq:anisotropic_maxwellian} one- and two-dimensional model
    problems of Weibel instability by letting the wavevector lie in the $(x,y)$-plane and the $\hat{z}$-direction
    be the wave binormal.
    These model problems have the advantage of zero-order electrostatic stability such that all eigenfunctions
    are fully electromagnetic.

\begin{figure}
        \centering
        \includegraphics[width=0.45\textwidth]{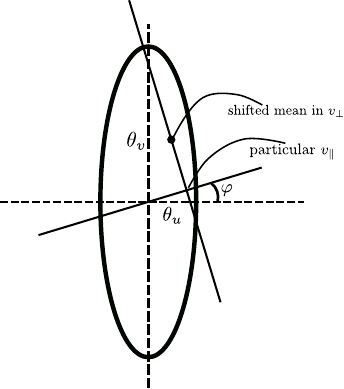}
        \caption{
        In the frame of reference of a particle propagating in a direction not aligned with the principal axes of an
        anisotropic distribution there is a mean drift velocity in the direction transverse to the particle's motion
        which couples together the electromagnetic wave's longitudinal and transverse components.
        In other words, in a frame where the velocity coordinates $(u,v)$ are rotated through an angle $\varphi$ into coordinates
            $(v_\parallel, v_\perp)$,
            for a particular value of $v_\parallel$ there is a non-zero mean value of $v_\perp$
            such that the off-diagonal integrals $I_{ij}\neq 0$ of Eq.~\ref{eq:integral_source}.
        }\label{fig:rotated_anisotropic}
    \end{figure}

    With wavevector in the $(x,y)$-plane the off-diagonal integrals $I_{13}$, $I_{23}$ 
    vanish as $\langle w\rangle=0$, so that the $\hat{z}$-directed perturbation does not contribute to the $(x,y)$-plane's perturbation current,
    decoupling the binormal from the longitudinal and transverse components~(\cite{sharma1976wave,datta2021electromagnetic}).
    A fully general wavevector would couple all three components of the perturbation.
    Equation~\ref{eq:general_tensor} simplifies to 
    \begin{equation}\label{eq:simplified_dielectric_tensor}
        \begin{Bmatrix}
            \omega^2 - \omega_p^2(1 - I_{11}) & \omega_p^2 I_{12} & 0\\
            \omega_p^2 I_{12} & \omega^2 - c^{2}k^2 - \omega_p^2(1 - I_{22}) & 0\\
            0 & 0 & \omega^2 - c^2 k ^2 - \omega_p^2 (1 - I_{33})
        \end{Bmatrix}
        \begin{Bmatrix}
            E_1\\ E_2\\ E_3
        \end{Bmatrix} = 0.
    \end{equation}
    Focusing on the $(x,y)$-plane we consider a reduced distribution, the bi-Maxwellian
    $f_0(u,v)=\int f_0(u,v,w)dw$
    whose level sets form ellipses in the $(u,v)$ velocity plane.
    Take $\theta_v > \theta_u$ so that the semi-major axis of each ellipse is $v$-directed and the 
    characteristic eccentricity is $e^2=1 - \theta_u^2/\theta_v^2$.
    To ensure separability of the resonant integrals $I_{ij}$, the $(x,y)$-plane is rotated through an angle $\varphi$,
    transforming velocities $(u,v)\to(v_\parallel,v_\perp)$ such that the resonant denominator is $\omega - k v_\parallel$.
    By completing the square on $v_\perp$, the anisotropic Maxwellian of Eq.~\ref{eq:anisotropic_maxwellian}
    in the reduced coordinates
    $(u,v)\to(v_\parallel,v_\perp)$ 
    is 
    \begin{equation}\label{eq:simplified_rotated_distribution}
        f_0(v_\parallel, v_\perp) = \frac{1}{2\pi\theta_\parallel\theta_\perp}
        \exp\Big(-\frac{v_\parallel^2}{2\theta_\parallel^2}\Big)\exp\Big(-\frac{(v_\perp - \alpha v_\parallel)^2}{2\theta_\perp^2}\Big),
    \end{equation}
    where the rotated thermal and mean velocities are defined as
    \begin{equation}\label{eq:rotated_parameters}
        \theta_{\parallel}^2 \equiv \frac{(1 - e^2)\theta_u^2}{1-e^2\cos^2(\varphi)},\quad
        \theta_{\perp}^2 \equiv \frac{\theta_u^2}{1 - e^2\cos^2(\varphi)},\quad
        \alpha \equiv \frac{e^2\sin(\varphi)\cos(\varphi)}{1-e^2\cos^2(\varphi)}.
    \end{equation}
    Equation~\ref{eq:simplified_rotated_distribution} shows that in the rotated frame the distribution is a shifted Maxwellian.
    That is, in the frame of a resonant particle 
    the distribution has a mean velocity $\alpha v_\parallel$ transverse to the wavevector, 
    as illustrated in Fig.~\ref{fig:rotated_anisotropic}.
    The off-diagonal integrals 
    are non-zero, $I_{ij}\neq 0$, for anisotropic distributions as the apparent transverse current in the resonant 
    particle frame 
    couples the longitudinal and transverse wave components for propagation not aligned with the principal axes,
    even though there is no net current in the lab frame.
    In general all three components are coupled for non-principal propagation.
    This coupling is characteristic of anisotropy and contrasts with isotropic distributions for which the longitudinal plasma wave
    and the two transverse electromagnetic wave components have fully independent dispersion relations.

    To evaluate the integrals $I_{11}$, $I_{12}$, and $I_{22}$ for the distribution in Eq.~\ref{eq:simplified_rotated_distribution}
    in the rotated coordinates, note the following integrals related to the plasma dispersion function $Z(\zeta)$,
    \begin{align}
        \frac{1}{\sqrt{\pi}}\int_{-\infty}^{\infty}\frac{x}{\zeta-x}e^{-x^2/2a^2}dx
        &= \frac{a}{2}Z'(\widetilde{\zeta}),\\
        \frac{1}{\sqrt{\pi}}\int_{-\infty}^{\infty}\frac{x^2}{\zeta-x}e^{-x^2/2a^2}dx
        &= -\frac{a^2}{2}(Z''(\widetilde{\zeta}) + 2Z(\widetilde{\zeta})),\\
        \frac{1}{\sqrt{\pi}}\int_{-\infty}^{\infty}\frac{x^3}{\zeta-x}e^{-x^2/2a^2}dx
        &= \frac{a^3}{2^{3/2}}(Z'''(\widetilde{\zeta}) + 6Z'(\widetilde{\zeta})),
    \end{align}
    where $\widetilde{\zeta}=\zeta / \sqrt{2} a$.
    These identities are found through integration by parts using
    the Hermite relation $\psi_n(x)=\frac{d^n}{dx^n}e^{-x^2/2}$
    and the identity $Z'(\zeta)=-2(1+\zeta Z(\zeta))$.
    With the gradient 
    $\frac{\partial f}{\partial v_\parallel} = \big(-v_\parallel/\theta_\parallel^2 +
    \alpha(v_\perp-\alpha v_\parallel)/\theta_\perp^2\big)f$
    the integrals $I_{ij}$ in Eq.~\ref{eq:simplified_dielectric_tensor} work out to
    \begin{align}
        I_{11} &= -\frac{1}{4}(Z'''(\widetilde{\zeta}) + 6Z'(\widetilde{\zeta})),\\
        I_{12} &= -\frac{\alpha}{4}(Z'''(\widetilde{\zeta}) + 6Z'(\widetilde{\zeta})),\\
        I_{22} &= -\frac{\alpha^2}{4}(Z'''(\widetilde{\zeta}) + 2Z'(\widetilde{\zeta})) -
    \frac{\theta_\perp^2}{\theta_\parallel^2}\frac{Z'(\widetilde{\zeta})}{2},\\
        I_{33} &= -\frac{\theta_w^2}{\theta_\parallel^2}\frac{Z'(\widetilde{\zeta})}{2},\label{eq:I33}
    \end{align}
    where $\widetilde{\zeta}\equiv \zeta/\sqrt{2}\theta_\parallel$.
    Let the angular eccentricity be $\gamma = \arcsin(e)$.
    The characteristic anisotropies are then $A_1\equiv \theta_\perp^2/\theta_\parallel^2-1=\frac{e^2}{1-e^2}=\tan^2(\gamma)$
    for the in-plane anisotropy and
    $A_2\equiv \theta_w^2/\theta_\parallel^2-1$ 
    for the binormal anisotropy. 
    These two anisotropy parameters induce electromagnetic Weibel instability to relax the anisotropy of their
        respective dimensions.
    For propagation along the principal axes, $\varphi\to 0, \pi/2$ the parameter $\alpha\to 0$,
    decoupling the $\hat{x}$- and $\hat{y}$-components as $I_{12}\to 0$
    and $I_{22} \to -\frac{\theta_v^2}{\theta_u^2}\frac{Z'(\zeta/\sqrt{2}\theta_u)}{2}$ as in Eq.~\ref{eq:I33}.

    \subsubsection{Dispersion function for the classic Weibel instability}
    Focusing on the decoupled component in Eq.~\ref{eq:simplified_dielectric_tensor}, namely $\varepsilon_{33}E_3=0$,
    leads to
    \begin{equation}\label{eq:weibel_bimaxwellian_with_anisotropy}
        1 - \Big(\frac{\zeta}{c}\Big)^2 + \frac{1}{(k\lambda_c)^2}\Big(1 +
    \frac{Z'(\widetilde{\zeta})}{2} + A_2\frac{Z'(\widetilde{\zeta})}{2}\Big) = 0
    \end{equation}
    with $A_2$ the bi-normal anisotropy.
    Equation~\ref{eq:weibel_bimaxwellian_with_anisotropy} is valid for all propagation angles.
    For principal propagation at $\varphi=0$ (the $\hat{x}$-direction) the principal anisotropies $A_1=A_2$
    such that Eq.~\ref{eq:weibel_bimaxwellian_with_anisotropy} describes
    both the transverse and binormal components.
    It is instructive to observe that isotropy reduces Eq.~\ref{eq:weibel_bimaxwellian_with_anisotropy} to the
    ordinary wave kinetic dispersion relation.
    The coupled branch of solutions to Eq.~\ref{eq:simplified_dielectric_tensor} is described by
    \begin{equation}\label{eq:dispersion_function_restricted_twod_weibel1}
        \Big(-\Big(\frac{\zeta}{c}\Big)^2 + \frac{1}{(k\lambda_c)^2}(1 - I_{11})\Big)
    \Big(1 - \Big(\frac{\zeta}{c}\Big)^2 + \frac{1}{(k\lambda_c)^2}(1 - I_{22})\Big) - \frac{1}{(k\lambda_c)^4}I_{12}^2 = 0
    \end{equation} 
    with transverse magnetic field out-of-plane and two components of electric field in-plane.

    \subsection{Single-mode saturation of Weibel instability in one spatial dimension}\label{subsec:single-mode-saturation-one-d}
    Just as electrostatic instability saturates by nonlinear trapping of near-resonant particles in an
    electric potential energy well, electromagnetic instability saturates by magnetic trapping.
    As a supplement to this section, Appendix~\ref{sec:magnetic_well_app} builds an analytic model of the phase portraits
    associated with ideal magnetic trapping and a conceptual model of magnetic trapping as a magnetic potential momentum well.
    Simulation of single-mode Weibel instability by initialization with kinetic eigenfunctions allows one to self-consistently visualize
    the saturated phase space structures.
    The foundational simulations of electromagnetic instability due to anisotropy considered Weibel instability 
    evolving from the anisotropic Maxwellian using particle-in-cell method~(\cite{davidson1972nonlinear}).
    Later,~\cite{califano1997spatial} conducted one- and two-dimensional simulations in an inhomogeneous plasma
    using anisotropy induced by streaming beams. 
    Building on this,~\cite{cagas2017nonlinear} simulated single-mode Weibel saturation with continuum-kinetic method 
    using a zero-order counter-streaming electron beam distribution.
    In~\cite{cagas2017nonlinear} electrostatic streaming instability was proposed to explain the growth of beam-axis directed electric field
    close to nonlinear saturation.

    Zero-order beam distributions are often used because Weibel instability is induced in the laboratory by colliding
    high velocity plasmas~(\cite{hill2005beam,fox2013filamentation,huntington2015observation,shukla2018conditions}).
    While zero-order beam distributions are inherently anisotropic they are also possibly unstable to electrostatic
    streaming instability.
    The possible introduction of electrostatic streaming instability can confuse and complicate an attempt to isolate
    Weibel instability.
    On the other hand, there is no possibility of streaming instability when the Weibel instability is induced by
    a zero-order anisotropic Maxwellian distribution.
    For this reason the anisotropic Maxwellian is considered here using the continuum kinetic method,
    as originally considered with particle-in-cell method by~\cite{davidson1972nonlinear}.

    \subsubsection{Numerical method for single-mode Weibel instability simulation}\label{sec:num_single_weibel}
    We use the mixed spectral-DG method
    presented in Appendix~\ref{app:summary_of_numerical} and~\cite{crews_kinetic}, where the space coordinate is represented using
    Fourier modes and the two velocity dimensions are discretized with discontinuous Galerkin method.
    The field equations are chosen as follows: Ampere's law and Faraday's law are used to evolve the transverse
    electrodynamic field, and Gauss's law is used to constrain the electric field along the axis of the wavevector.
    When considering only a single spatial coordinate three field equations can be chosen in this way
    (one component each of the electrodynamic equations and Gauss's law) as there are only three field components.

    The geometry is established by aligning the hot direction of the bi-Maxwellian with the $y$-coordinate, the growing
    magnetic field with the $z$-direction, and perturbing the distribution function with wavenumber in the $x$-direction.
    This necessitates two dimensions of velocity, $v_x$ in the $x$-direction and $v_y$ in the $y$-direction, for a
    1D2V phase space geometry.
    Phase space is discretized using $N_x=100$ evenly spaced collocation nodes in the $x$-direction, and
    $N_{v_x}=N_{v_y}=22$ finite elements in velocity, each of $11$th polynomial order.
    Fourteen elements are evenly spaced between the velocity intervals $[-7, 7] v_t$, and four elements are evenly spaced
    within each interval $[-15,-7]v_t$ and $[7,15]v_t$. 
    A spatial hyperviscosity $10^{-4}\nabla_x^{4} f$ is used to prevent spectral blocking by the turbulent cascade
    saturation.
    The field equations are discretized by a standard Fourier spectral method.

    \subsubsection{Initialization with field and phase space eigenfunctions}
    The characteristic parameters are chosen such that $v_t/c=0.3$ and anisotropy $A = 3$
    (or ratio $\theta_{v}/\theta_{u}=2$) with the zero-order distribution given by Eq.~\ref{eq:anisotropic_maxwellian}
    and the direction of propagation set to $\varphi=0$.
    This is equivalent to using Eq.~\ref{eq:weibel_bimaxwellian_with_anisotropy} for the dispersion function.
    Velocities are normalized to $\theta_{u}$, time to $\omega_p^{-1}$, and lengths to $\lambda_D=\theta_{u}/\omega_p$.
    The domain length is then specified by the chosen wavenumber $k_x\lambda_D = 0.1$.
    Solution of Eq.~\ref{eq:weibel_bimaxwellian_with_anisotropy} gives the eigenvalue of the problem as the phase velocity
    $\zeta/v_t = 1.23i$.
    The phase space perturbation is constructed using Eq.~\ref{eq:linear_response_electromagnetic}
    for the phase space eigenfunction in the form
    \begin{equation}\label{eq:bi_maxwellian_perturbation}
        f_1(x, v_x, v_y) = \text{Re}\Big\{
    i\frac{\mathcal{A}}{k}\Big(\frac{v_y}{\zeta - v_x}\frac{\partial f_0}{\partial v_x} + \frac{\partial f_0}{\partial v_y}\Big)
        \exp(ikx)\Big\},
    \end{equation}
    longitudinal field $E_{x1}=0$, and the initial transverse electrodynamic fields by
    \begin{align}
        B_{z1} &= \text{Re}\Big(\mathcal{A} \exp(ikx)\Big),\\
        E_{y1} &= \text{Re}\Big(\zeta \mathcal{A} \exp(ikx)\Big),
    \end{align}
    where the amplitude is set to $\mathcal{A} = 10^{-3}$.
    This initial condition is consistent in the sense that the charge density is zero
    and the current density satisfies Ampere's law.

    \subsubsection{Fully nonlinear simulation results for single-mode Weibel saturation}\label{subsec:nonlinear_single_weibel}
\begin{figure}
        \centering
        \includegraphics[width=\columnwidth]{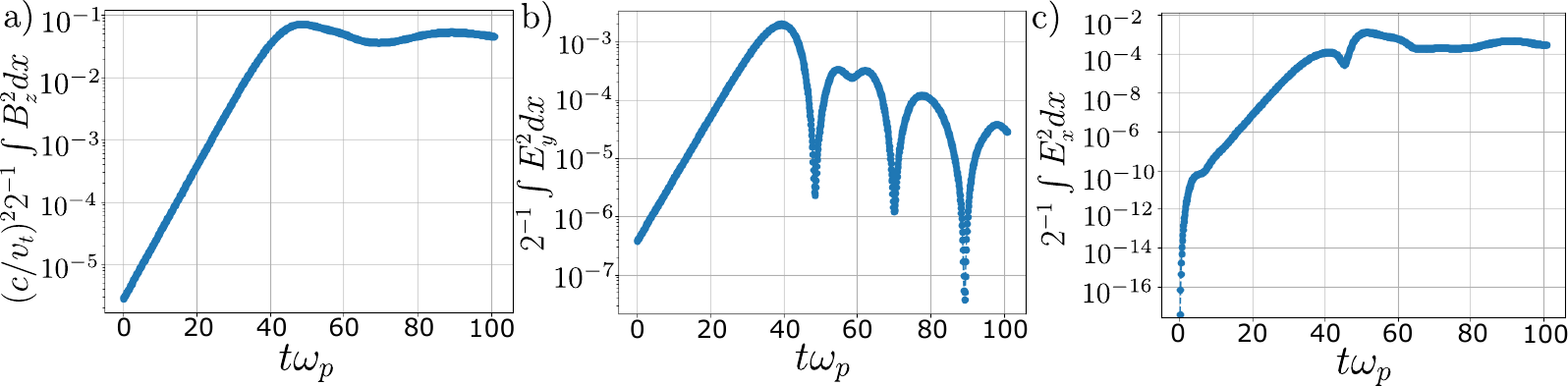}
        \caption{Growth of domain-integrated wave energies towards saturation of Weibel instability in one spatial
            dimension, namely (a) the magnetic energy, (b) the transverse electric energy of $E_y$, and (c) the longitudinal
        electric energy of $E_x$, or energy along the axis of the wavevector.
        A nonlinear phase is reached at $t\omega_p=40$, with peak magnetic energy around $t\omega_p=50$.
        Longitudinal electric energy is observed to grow with a similar trend to the magnetic energy.}\label{fig:1d_weibel_energy}
    \end{figure}

    The simulation is run until $t\omega_p=100$ using the numerical method of Section~\ref{sec:num_single_weibel}
    with time-step $\Delta t = 2.0\times 10^{-3}$.
    The change in domain-integrated wave energies is shown in Fig.~\ref{fig:1d_weibel_energy}.
    Of note is the oscillating transverse electric field at saturation, and how longitudinal electric energy
    corrects from zero to trend with magnetic energy.

\begin{figure}
        \centering
        \includegraphics[width=\columnwidth]{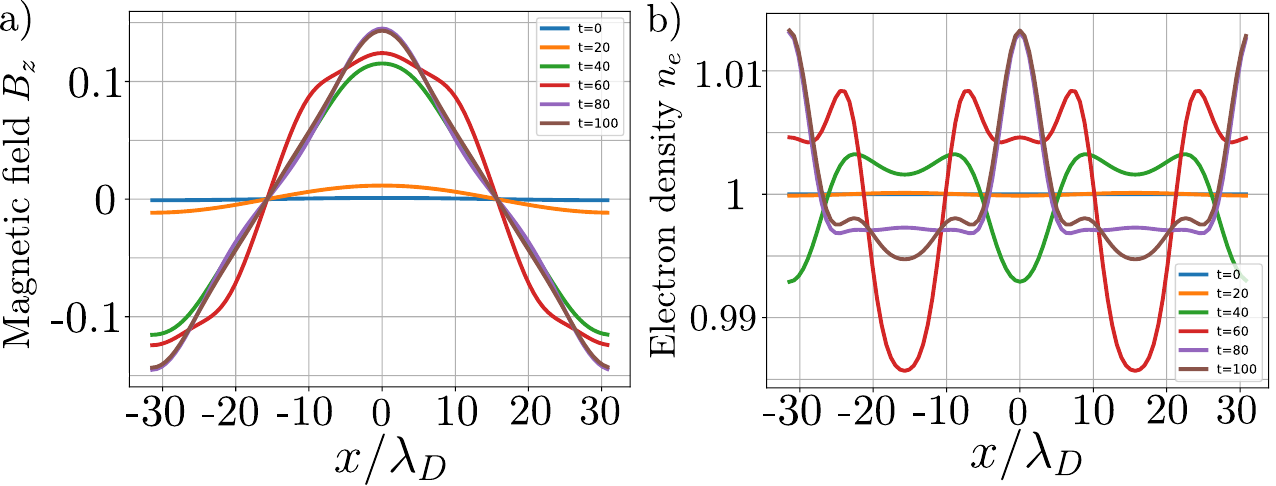}
        \caption{Evolution of (a) magnetic field $B_z(x)$ and (b) electron density $n_e(x)$, to nonlinear saturation of a single
        unstable Weibel mode. The mode saturates around $t\omega_p\approx 45$.
        Prior to saturation the magnetic field has a spectrum consisting of only even mode numbers
        with an apparent power law in logarithmic amplitudes. Since the function is clearly analytic this spectrum
        is consistent with an elliptic cosine function until nonlinear saturation.
        It is interesting that this higher-order phenomenon arises even from a single-mode eigenfunction perturbation.
        The electron holes at saturation are bounded by the maxima of magnetic energy $B^2/2\mu_0$, or equivalently
        bounded by the potential wells $A_y(x)$.}\label{fig:1d_weibel_field_density}
    \end{figure}

\begin{figure}
        \centering
        \includegraphics[width=\columnwidth]{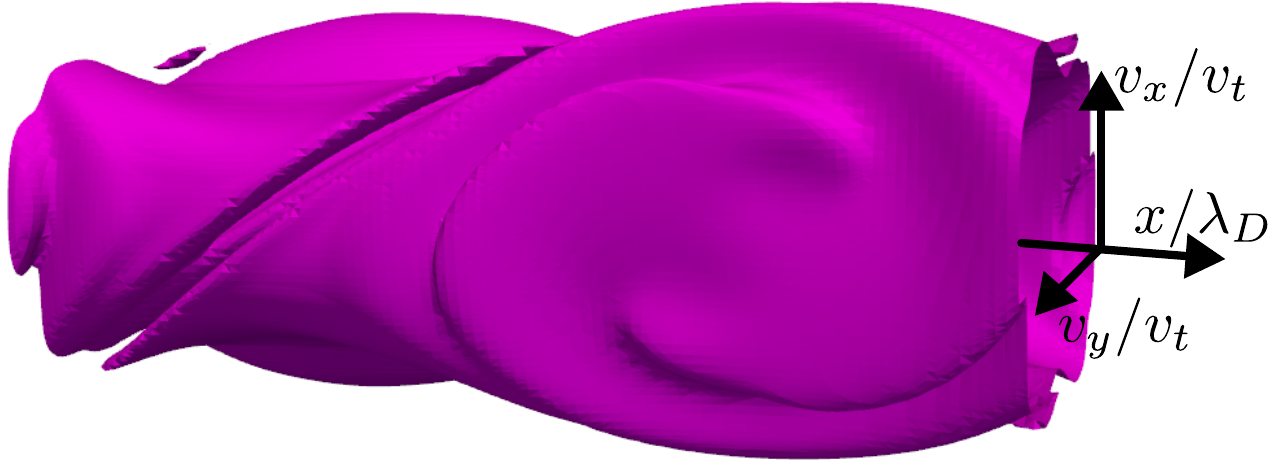}
        \caption{Phase space vortex shown by contour at $0.1\text{max}(f)$ in the coordinates $(x, v_x, v_y)$ at saturation
        of a single Weibel mode. The mid-domain $x$-coordinate corresponds to $x=0$.
        The magnetic trapping phase space vortex is distinguished from the electrostatic vortex as velocity-dependent
            since the momentum $p_y$ in Eq.~\ref{eq:electron_trajectory} is linear in $v_y$.
        In this way the single-mode phase space vortex is antisymmetric, with trapped and passing orbits swapping under
            $v_y\to -v_y$.
        For this reason there is no magnetic trapping along the $v_y=0$ plane.}\label{fig:1d_weibel_phase space}
    \end{figure}

    Figure~\ref{fig:1d_weibel_field_density} shows the time evolution of magnetic field and electron density in increments
    of $t\omega_p=20$.
    As magnetic energy grows, electrons are progressively magnetically trapped as described analytically in Appendix~\ref{sec:magnetic_well_app}.
    With the initial free energy released, the distribution function attempts to evolve toward a function of the constants of motion,
    namely the energy $H = \frac{1}{2}m(v_x^2 + v_y^2) - e\Phi(x)$ and canonical momenta
    $P_y = mv_y - e A_y(x)$, $P_x = mv_x$ where $\Phi(x)$ is the electric potential and
    $\bm{A}=A_y(x)\hat{y}$ is the magnetic vector potential.
    Specifically, trapped and passing phase space trajectories are determined by the equation~(\cite{morse1971numerical})
    \begin{equation}\label{eq:electron_trajectory}
        mv_x = \sqrt{2m(H + e\varphi(x)) - (P_y + eA_y(x))^2}
    \end{equation}
    for a particle's energy $H=\frac{1}{2}m(v_{x0}^2 + v_{y0}^2) - e\varphi(x_0)$ and momentum $P_y = mv_{y0} - eA_{y0}(x_0)$ constants.
    Figure~\ref{fig:1d_weibel_phase space} visualizes the phase space structure at nonlinear saturation
    by an isocontour of the distribution at $10\%$ of its maximum.
    Trapped and passing trajectories are seen at the right and left of the figure, respectively, for $v_y>0$.
    Trapped trajectories circulate within a phase space vortex while passing trajectories
    execute motion in the vicinity of the separatrix. 
    Upon inversion of the transverse velocity, $v_y\to -v_y$, the positions of the trapped trajectories and
    passing trajectories are inverted, $x\to -x$.

    Linear electrostatic instability is not possible due to the zero-order anisotropic Maxwellian, so here we explain 
    the growth of electrostatic energy observed both here and in~\cite{cagas2017nonlinear}
    as a second-order phenomenon arising from space-charge-generating
    filamentation (that is, the magnetic-trapping electron holes of Fig.~\ref{fig:1d_weibel_field_density}).
    While saturated filaments can be understood intuitively as electron holes,
    the progressive development of longitudinal electric energy in the linear phase can be understood as
    mode coupling of the longitudinal field to the transverse field at second order in the transverse dynamic field~(\cite{taggart1972second}).
    Appendix~\ref{sec:magnetic_well_app} presents a visual description of the density filamentation phenomenon
    as resulting from a bifurcation in phase space topology as the thermal trapping parameter $mv_{th}/eA_\text{max}$ passes through unity.
    When $eA_\text{max} \ll mv_{th}$ the primary phase space vortex structures are anti-symmetric across the $v_y$-plane, producing 
    a perturbation in current density without a perturbation in charge density. 
    Once $eA_\text{max} \approx mv_{th}$ particle orbits with low $P_y$ become more symmetric across the $v_y$-plane and
    produce a coherent perturbation in the density.

    Thus, both electric and magnetic field trapping are associated with local variations in the electron
    density which manifests as space charge.
    In the case of magnetic trapping the charge density is a higher-order effect, and the first-order effect
    is production of electric currents to sustain the magnetic mode.
    The development of space charge from magnetic pressure is anticipated in~\cite{morse1971numerical},
    and the longitudinal field is explained in~\cite{taggart1972second} to arise
    at second-order from the coupling of two magnetic modes.
    Dynamic space charge, or filamentation, has been observed in modeling to modify growth rates, in both
    early and more recent studies~(\cite{taggart1972second,tzoufras2006space}), pointing to the importance of
    higher-order effects prior to saturation.
    Since the fraction of trapped electrons is proportional to the magnetic energy, with saturation when the characteristic
    magnetic bounce frequency reaches the growth rate~(\cite{davidson1972nonlinear}), it follows 
    that longitudinal electric field should trend nonlinearly with the magnetic field.
    Finally, the theory of Appendix~\ref{sec:magnetic_well_app} interprets the density perturbation as arising from particles with
    $v_y\approx 0$ and energies $H<\frac{(eA_\text{max})^2}{2m}$.

    \subsection{Saturation of many unstable Weibel modes in two spatial dimensions}\label{subsec:two_dimensional_weibel_simulation}
\begin{figure}
        \centering
        \includegraphics[width=0.75\columnwidth]{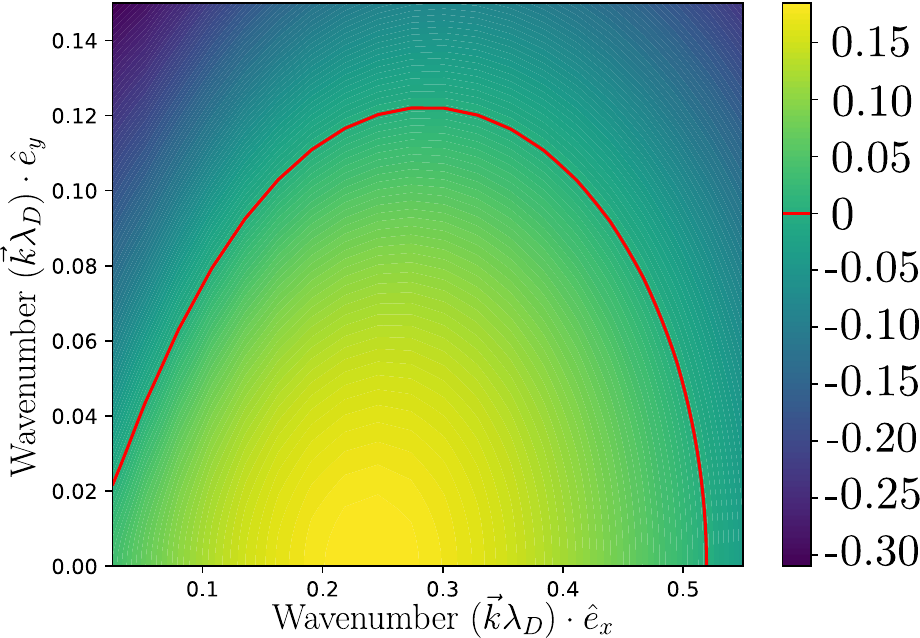}
        \caption{Growth rates of multidimensional Weibel instability as $\text{Im}(\omega)/\omega_{pe}$ for
        an anisotropic bi-Maxwellian of anisotropy $A=3$ where the electric field is assumed to lie
        in the $(x,y)$-plane and the first-order transverse magnetic field to be out-of-plane.
        The higher-temperature direction is assumed to lie in the $\hat{y}$-direction assuming $v_t/c=0.3$.
        The red line identifies marginal stability.}\label{fig:weibel_growth_rates}
    \end{figure}

    Weibel instability of a homogeneous unmagnetized plasma is inherently multidimensional as wavevectors
    oblique to the principal anisotropy axes have comparable growth rates to the principal axes
    (similar to the multidimensional streaming instability studied in Section~\ref{subsec:two_d_electrostatic_simulation}).
    We consider for example the branch of the dispersion function for the anisotropic Maxwellian
    given by Eq.~\ref{eq:dispersion_function_restricted_twod_weibel1} 
    with transverse magnetic field out-of-plane and the longitudinal and transverse electric fields in-plane.
    The out-of-plane magnetic field allows a reduced 2D2V phase space geometry for tractable continuum-kinetic
    simulation~(\cite{Skoutnev_2019}).
    It should be kept in mind that 
    the instability dynamics are truly three-dimensional just as in the multidimensional Langmuir turbulence
    simulation of Section~\ref{subsec:two_d_electrostatic_simulation}.
    Figure~\ref{fig:weibel_growth_rates} plots the growth rate as a function of the wavevector
    $\bm{k}=k_x\hat{x} + k_y\hat{y}$
    of the unstable branch of Eq.~\ref{eq:dispersion_function_restricted_twod_weibel1}
    for a zero-order bi-Maxwellian of anisotropy $A=3$ and with $v_t/c=0.3$.
    If the destabilized ordinary wave studied in Section~\ref{subsec:single-mode-saturation-one-d}
    were included in the simulation, the magnetic field would also take in-plane components, necessitating a
    third velocity dimension.
    We remind the same caveats for this simulation as for the two-dimensional two-stream simulation;
    the nonlinear phase space dynamics are likely similar to the three-dimensional problem, but the nonlinear
    dynamics of the saturated state (that is, the turbulence statistics) will likely be different.
    The essence of the problem lies in restricting the cylindrical symmetry of the near-maximal-growth-rate modes
    about the principal anisotropy axis to a plane.
    Fully three-dimensional simulations are necessary to clarify these issues.

    \subsubsection{Numerical methods for the two-dimensional multi-mode Weibel instability}\label{sec:num_twod_weibel}
    The simulations are conducted with the methods of Appendix~\ref{app:summary_of_numerical} and~\cite{crews_kinetic}
    with a few key differences.
    Specifically, Ampere's and Faraday's laws are used for the field equations which are time-integrated in Fourier spectral
    space with the same third-order Adams-Bashforth method as the kinetic equation.
    There is also an important difference in initialization of the unstable modes.
    In the Vlasov-Poisson system the field part of the kinetic eigenfunctions is described by the scalar potential, yet
    in the multidimensional Vlasov-Maxwell system the kinetic eigenfunctions consist of a vector
    self-consistent field-plasma configuration.
    Initialization with kinetic eigenfunctions in an electrodynamic problem 
    necessitates solving the eigenvalue problem $\chi\bm{E} = \zeta^2\bm{E}$ in the wavevector frame,
    as discussed in Section~\ref{subsec:susceptibility_and_eigenvalue}.
    In our implementation, the eigenfunctions are computed in the rotated wavevector frame
    and then anti-rotated back into the $(x,y)$-plane with components $\bm{E} = E_x\hat{x} + E_y\hat{y}$.
    Thus for each desired pair of wavenumbers $(k_x, k_y)$ a perturbation is applied as
    \begin{equation}\label{eq:weibel_2d_perturbation}
        f_1 = \mathcal{A}\text{Re}\Big[\frac{q}{i\omega}\Big\{E_x\Big(\frac{\partial f_0}{\partial u} +
        \frac{v_x(\bm{k}\cdot\nabla_v f_0)}{\omega - \bm{k}\cdot\bm{v}}\Big) +
        E_y\Big(\frac{\partial f_0}{\partial v} + \frac{v_y (\bm{k}\cdot\nabla_v f_0)}{\omega - \bm{k}\cdot\bm{v}}\Big) \Big\}
        e^{i(\bm{k}\cdot\bm{x} + \widetilde{\varphi})}\Big]
    \end{equation}
    with $\widetilde{\varphi}$ a randomly chosen phase shift per wavevector and the amplitude $\mathcal{A}=2\times 10^{-3}$.
    The magnetic field is then initialized as $B_z = \zeta^{-1}\hat{k}\times\bm{E}$ where $\zeta$ is the eigenvalue.
    Higher amplitude than the one-dimensional problem is chosen to reduce time to saturation.

    \subsubsection{Initial conditions for the two-dimensional Weibel instability simulations}\label{sec:ic_twod_weibel}
    The domain is specified by fundamental wavenumbers $k_x\lambda_D = 0.125$ and $k_y\lambda_D = 0.01$.
    Physical space is represented with $N_x = 32$ and $N_y = 128$ evenly-spaced collocation points, while velocity space
    is represented as a Cartesian tensor product of linear finite elements with eleven finite elements per 
    velocity axis, each linear element of seventh-order polynomial basis for $(v_x, v_y) \in [-11, 11]\theta_{u}$.
    As in Section~\ref{subsec:two_d_electrostatic_simulation} many modes are excited;
    for each of the first two harmonics of the fundamental wavenumber,
    namely $nk_x\lambda_D$ with $n=1,2$, five transverse wavenumbers $\pm mk_y\lambda_D$ are excited with $m=0,1,2$.
    The normalized thermal velocity $v_t/c=0.3$ and the anisotropy is $A=3$, as in the one-dimensional problem.
    The simulation is run to $t\omega_p=40$ using a time-step of $\Delta t=8\times 10^{-3}$ with an added
    hyperviscosity $\nu \nabla_x^4 f$ with $\nu=1$ to prevent spectral blocking.
    Due to the hyperviscosity total energy is conserved only to $\mathcal{O}(10^{-4})$ by the simulation stop time.

    \subsubsection{Nonlinear simulations of the two-dimensional Weibel instability}
\begin{figure}
        \centering
        \includegraphics[width=\columnwidth]{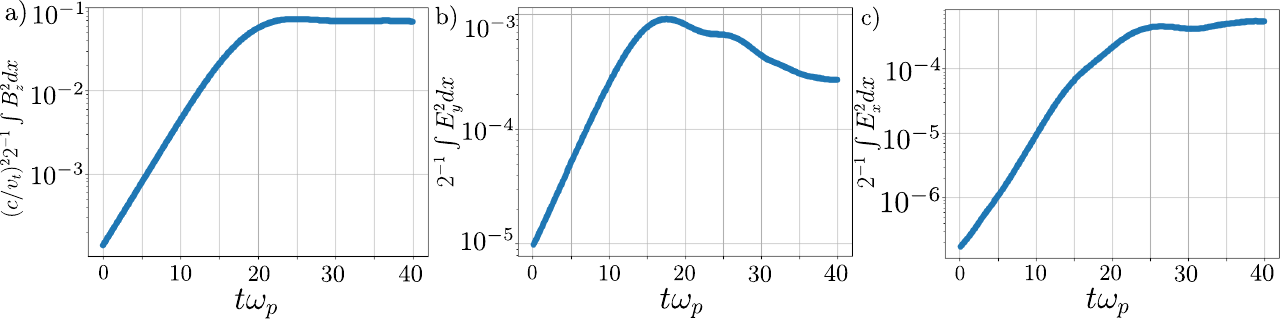}
        \caption{Evolution of domain-integrated energy for (a) the out-of-plane magnetic field $B_z$,
            (b) the electric field transverse to the maximum growth-rate axis, and (c) the electric field
        parallel to that axis (formerly the longitudinal field of the one-dimensional simulation).
        A difference from the one-dimensional simulation is a steadily decreasing transverse electric energy
        rather than a coherent oscillation after saturation time.
            The $\hat{x}$-directed electric energy increases at a rate faster than the growth of the linear modes due to
        the same space charge effects as in the single-mode simulation
            discussed in Section~\ref{subsec:single-mode-saturation-one-d}.
        }\label{fig:weibel_energy_2d}
    \end{figure}

    The problem described in Section~\ref{sec:ic_twod_weibel} was simulated using the methods
    of Section~\ref{sec:num_twod_weibel}.
    Figure~\ref{fig:weibel_energy_2d} shows the energy traces of the electrodynamic field during the linear phase
    and beyond instability saturation.
    Nonlinear saturation occurs around $t\omega_p=17$, as gauged by the $\hat{y}$-directed transverse electric field.
    The $\hat{x}$-directed electric field energy follows a similar trend as the one-dimensional problem,
    its growth composed of two effects:
    to first-order from the initialized oblique modes, and to second-order in the 
    magnetic field as discussed in Section~\ref{subsec:single-mode-saturation-one-d}.
    At nonlinear saturation, magnetic-trapping electron holes form between the
    counter-streaming mean flows.

\begin{figure}
        \centering
        \includegraphics[width=\columnwidth]{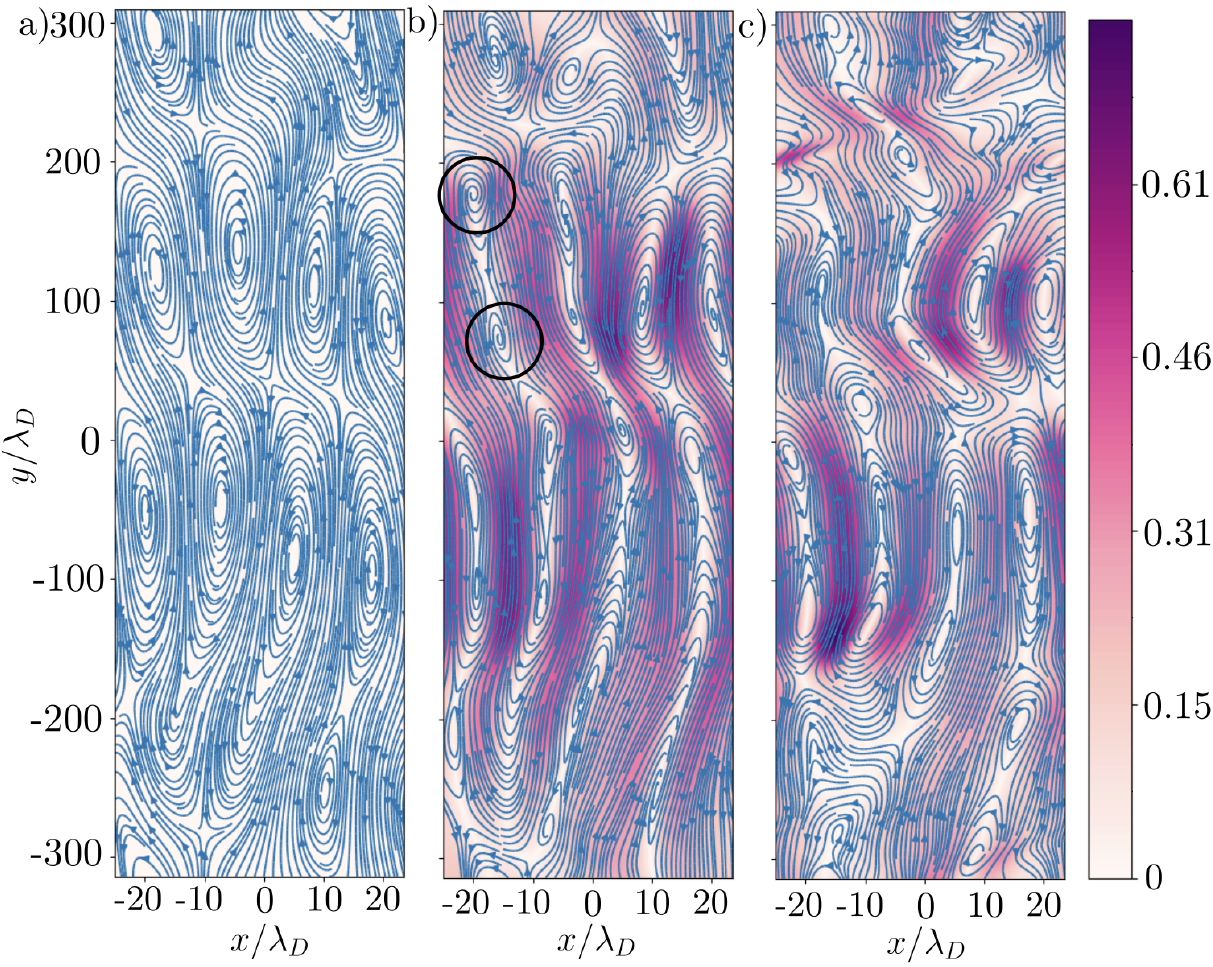}
        \caption{Evolution of current at times (a) $t\omega_p=0$, (b) $t\omega_p = 25$, and (c) $t\omega_p = 39.5$.
        Plotted are streamlines of $\bm{j}$ and its magnitude $|\bm{j}|$ as filled contours.
        Current tends to form closed paths at long wavelength in $y$.
        The indicated spiral vortices in the streamlines demonstrate rapid local space charge production ($\nabla\cdot\bm{j}\neq 0$)
            as expected from the filamentation and trapping dynamics of the one-dimensional problem.
            The large circulating configuration space electron current vortices manifest on small scales
            as counter-propagating electron streams which sustain a train of magnetic-trapping electron phase space vortices.
            In this unmagnetized model problem, these structures isotropize from four-dimensional phase space dynamics.}\label{fig:weibel_current_2d}
    \end{figure}

\begin{figure}
        \centering
        \includegraphics[width=\columnwidth]{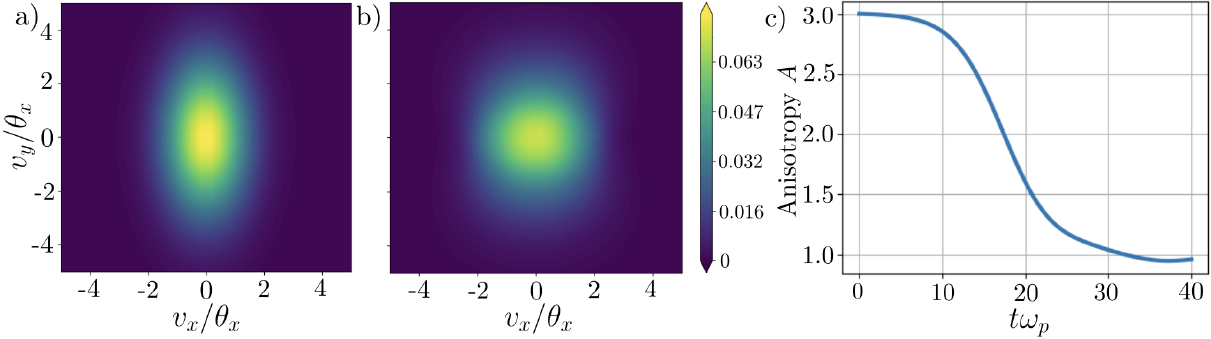}
        \caption{The relaxing spatially-averaged distribution function $\langle f\rangle_{(x,y)}(v_x, v_y)$ shown
        at two times:
            (a) the initial condition $t\omega_p=0$, (b) the stop-time $t\omega_p=40$,
            and in addition (c) the time evolution of the anisotropy parameter
        $A = \langle v_y^2\rangle / \langle v_x^2\rangle - 1$.
        The turbulence of magnetic trapping phase space vortices in the turbulent currents of the saturated instability isotropizes
            the distribution function.
        Note that $A=0$ would indicate isotropy, so that the saturated state consists of
        persistent anisotropy coexisting with the wavefield~(\cite{davidson1972nonlinear}).
    }\label{fig:weibel_avg_f_2d}
    \end{figure}

    Figure~\ref{fig:weibel_current_2d} illustrates the dynamics of the multidimensional instability through streamlines
    of the current density $\bm{j}(x,y)$ and filled contours of its magnitude for three times in the simulation output.
    Trigonometric interpolation is used to visualize the current density by zero-padding the spectrum and inverse Fourier
    transformation, as the spectrum is properly resolved up to the chosen spectral cut-off.
    In Fig.~\ref{fig:weibel_current_2d}b one can observe spiral current streamlines indicating the local production of
    space charge as the electron density filaments into two-dimensional analogues of the magnetic-trapping electron holes
    studied in the single-mode problem, also observed in~\cite{califano1997spatial}.
    By the simulation's end, nonlinear mixing has caused some of the filaments to rotate, as in the multidimensional
    electrostatic problem considered in Section~\ref{subsec:two_d_electrostatic_simulation},
    indicating a similar isotropization on averaged scales.
    Inspection of Fig.~\ref{fig:weibel_avg_f_2d}, which plots the spatially-averaged distribution function
    $\langle f\rangle_{(x,y)}(v_x, v_y)$ at $t\omega_p=0$ and $t\omega_p=40$ in Figs.~\ref{fig:weibel_avg_f_2d}a,b and
    in Fig.~\ref{fig:weibel_avg_f_2d}c the trace of the anisotropy parameter $A$, shows the anisotropic Maxwellian
    to relax towards isotropy through Weibel-induced turbulence.
    The initial anisotropy $A=3$ is observed to decrease monotonically until at saturation $A\approx 1$,
    meaning that the saturated state is weakly anisotropic.
    This persistent anisotropy coexists with the fluctuating filamentation currents,
    consistent with the observations of~\cite{davidson1972nonlinear}.
    Indeed, persistent anisotropy accompanying 
    sheared flows in the saturated state is expected for collisionless dynamics~(\cite{del2018shear}).

    \section{General discussion and summary}\label{sec:simulations_and_structures}
    This work advocates for the application of kinetic eigenfunctions
    to initialize Vlasov-Poisson, Vlasov-Maxwell, and quasilinear kinetic simulations.
    It reviews linearized kinetic theory
    and presents example simulations of the most commonly treated model problems.
    The historical discussion of Section 2 reviews the kinetic eigenvalue and initial-value problems,
    and highlights that the instabilities identified by Landau's initial-value analysis are indeed true
    eigenfunctions which may be utilized as simulation perturbations.
    Perturbation of a kinetic problem using its eigenfunctions provides several
    benefits, such as a controlled partition of perturbation energy, initialization of perturbations at
    close-to-nonlinear amplitudes, and measurement of linear instability growth rates up to machine precision
    unpolluted by linear Landau damping activity.
    
    Notable findings for researchers utilizing kinetic theory and simulation include:
    \begin{itemize}
    \item A historical overview of eigenfunctions for the Vlasov-Poisson system~(Section~\ref{subsec:historical_summary});
    \item Worked examples of eigenfunction initialization for the Vlasov-Poisson and Vlasov-Maxwell systems to
      illustrate the method's advantages
      (Sections~\ref{subsec:the-one-dimensional-two-stream-instability}
      (single-mode two-stream),~\ref{subsec:two_d_electrostatic_simulation}
      (two-dimensional multi-mode two-stream),~\ref{subsec:two_case_studies}
      (single-mode loss-cone),~\ref{subsec:single-mode-saturation-one-d}
      (single-mode Weibel),
      and~\ref{subsec:two_dimensional_weibel_simulation} (two-dimensional multi-mode Weibel));
    \item Eigenfunction initialization of quasilinear kinetic simulations
      (Section~\ref{subsec:quasilinear-theory} as applied to quasilinear bump-on-tail dynamics)
      for which the phase space fluctuation is always an eigenfunction of the instantaneous state;
    \item Power series representations of the dielectric function for ring distributions in strongly magnetized plasmas
      at arbitrary propagation angles (Section~\ref{subsec:dielectric_for_losscone} and Appendix~\ref{sec:integral_appendix});
    \item Closed form hypergeometric and trigonometric integral representations of the dielectric for ring distributions
      in strongly magnetized plasmas (Section~\ref{subsec:perpendicular_propagation} and Appendix~\ref{subsec:hypergeomtric_closed_form});
    \item Helical geometry of phase space fluctuations in magnetized plasmas (Section~\ref{subsec:visualization_and_eigenmodes});
    \item Description of the generation of electron density holes by magnetic trapping in the saturating Weibel instability
      (Section~\ref{subsec:nonlinear_single_weibel} and Appendix~\ref{sec:magnetic_well_app}) occuring without invoking
      electrostatic streaming effects as hypothesized in~\cite{cagas2017nonlinear}.
    \end{itemize}
    
    This work is by no means an exhaustive overview of the possible applications of kinetic eigenfunctions.
    Indeed, many other applications are noted in Section~\ref{sec:intro}.
    Nevertheless, a few notable problems are demonstrated here to significantly benefit from eigenfunction initialization.
    Namely, the electrostatic problem in a static magnetic field is treated in detail for ring distributions, and new
    analytic results for the ring dielectric function are presented.
    Nonlinear saturation of the multi-dimensional Weibel instability of an anisotropic Maxwellian,
    originally treated by~\cite{davidson1972nonlinear} with the particle-in-cell method,
    is revisited and new light shed with a phase portrait analysis and nonlinear continuum-kinetic simulations.
    However, a simple and important model problem is left for future work, namely anisotropy-induced
    field-parallel whistler emission.

    A few further notes are in order regarding the importance of phase space eigenfunctions in numerical plasma theory.
    In this work the emphasis is on observing the evolution of strongly unstable linear eigenfunctions into nonlinear structures.
    Another important class of problems are the weakly unstable distributions that are
    commonly treated by quasilinear theory.
    In numerical solutions of quasilinear theory there is no transition to nonlinear structures and the phase space
    dynamics of the obtained spectra are necessarily linear eigenfunctions.
    As another way of phrasing this, one can say that in weak turbulence the linear eigenfunctions contain most of the
    fluctuation energy.
    The application of kinetic phase space eigenfunctions to numerically study quasilinear theory potentially offers a
    multitude of interesting and valuable avenues of further exploration.

\subsection*{Acknowledgements}
The authors would like to thank J.~B.~Coughlin, I.~A.~M.~Datta, A.~Ho, E.~T.~Meier, A.~D.~Stepanov, Y.~Takagaki, W.~H.~Thomas, and
G.~V.~Vogman for helpful discussions.
The authors would also like to thank the anonymous peer-reviewers for their extensive and essential assistance with revision of the manuscript.
The information, data, or work presented herein was funded in part by the Air Force Office of Scientific Research under
Award No. FA9550--15--1--0271.
This material is based upon work supported by the National Science Foundation under Grant No. PHY-2108419.

\bibliography{kinetic_modes}

\begin{thebibliography}{}

\bibitem[Allmann-Rahn et~al., 2022]{ruhr_university_code}
Allmann-Rahn, F., Lautenbach, S., and Grauer, R. (2022).
\newblock {An Energy Conserving Vlasov Solver That Tolerates Coarse Velocity
  Space Resolutions: Simulation of MMS Reconnection Events}.
\newblock {\em Journal of Geophysical Research: Space Physics},
  127(2):e2021JA029976.
\newblock e2021JA029976 2021JA029976.

\bibitem[Balescu, 1997]{balescu1997statistical}
Balescu, R. (1997).
\newblock {\em Statistical dynamics: matter out of equilibrium}.
\newblock World Scientific.

\bibitem[Barnes and Chacón, 2021]{barnes_pic}
Barnes, D.~C. and Chacón, L. (2021).
\newblock Finite spatial-grid effects in energy-conserving particle-in-cell
  algorithms.
\newblock {\em Computer Physics Communications}, 258:107560.

\bibitem[Bernstein, 1958]{bernstein1958waves}
Bernstein, I.~B. (1958).
\newblock Waves in a plasma in a magnetic field.
\newblock {\em Physical Review}, 109(1):10.

\bibitem[Bertrand and Feix, 1968]{bertrand_waterbag}
Bertrand, P. and Feix, M.~R. (1968).
\newblock {Nonlinear electron plasma oscillation: the “water bag model”}.
\newblock {\em Physics Letters A}, 28(1):68--69.

\bibitem[Birdsall and Langdon, 1991]{birdsall2004plasma}
Birdsall, C.~K. and Langdon, A.~B. (1991).
\newblock {\em {Plasma Physics via Computer Simulation}}.
\newblock Series in Plasma Physics and Fluid Dynamics. Taylor \& Francis.

\bibitem[Bohm and Gross, 1949]{bohmgross}
Bohm, D. and Gross, E.~P. (1949).
\newblock {Theory of plasma oscillations. A. Origin of medium-like behavior}.
\newblock {\em Phys. Rev.}, 75.

\bibitem[Boyd, 2001]{boyd2001chebyshev}
Boyd, J.~P. (2001).
\newblock {\em Chebyshev and Fourier spectral methods}.
\newblock Courier Corporation.

\bibitem[Bratanov et~al., 2013]{bratanov}
Bratanov, V., Jenko, F., Hatch, D., and Brunner, S. (2013).
\newblock {Aspects of linear Landau damping in discretized systems}.
\newblock {\em Physics of Plasmas}, 20(2):022108.

\bibitem[Cagas et~al., 2017]{cagas2017nonlinear}
Cagas, P., Hakim, A., Scales, W., and Srinivasan, B. (2017).
\newblock {Nonlinear saturation of the Weibel instability}.
\newblock {\em Physics of Plasmas}, 24(11):112116.

\bibitem[Califano et~al., 1997]{califano1997spatial}
Califano, F., Pegoraro, F., and Bulanov, S.~V. (1997).
\newblock {Spatial structure and time evolution of the Weibel instability in
  collisionless inhomogeneous plasmas}.
\newblock {\em Physical review E}, 56(1):963.

\bibitem[Case, 1959]{case1959plasma}
Case, K.~M. (1959).
\newblock Plasma oscillations.
\newblock {\em Annals of physics}, 7(3):349--364.

\bibitem[Che, 2016]{che2016electron}
Che, H. (2016).
\newblock Electron two-stream instability and its application in solar and
  heliophysics.
\newblock {\em Modern Physics Letters A}, 31(19):1630018.

\bibitem[Cheng and Knorr, 1976]{cheng_knorr}
Cheng, C.~Z. and Knorr, G. (1976).
\newblock {The integration of the Vlasov equation in configuration space}.
\newblock {\em Journal of Computational Physics}, 22(3):330--351.

\bibitem[Choi et~al., 2021]{choi2021high}
Choi, B., Christlieb, A., and Wang, Y. (2021).
\newblock {High-dimensional sparse Fourier algorithms}.
\newblock {\em Numerical Algorithms}, 87:161--186.

\bibitem[Conner and Wilson, 1994]{conner1994survey}
Conner, J.~W. and Wilson, H.~R. (1994).
\newblock Survey of theories of anomalous transport.
\newblock {\em Plasma physics and controlled fusion}, 36(5):719.

\bibitem[Crews, 2022]{crews2022numerical}
Crews, D.~W. (2022).
\newblock {\em Numerical simulation of collisionless kinetic plasma
  turbulence}.
\newblock PhD thesis, University of Washington.

\bibitem[Crews et~al., 2024]{crews2023kadomtsev}
Crews, D.~W., Datta, I. A.~M., Meier, E.~T., and Shumlak, U. (2024).
\newblock {The Kadomtsev Pinch Revisited for Sheared-Flow-Stabilized Z-Pinch
  Modeling}.
\newblock {\em IEEE Transactions on Plasma Science}, pages 1--13.

\bibitem[Crews and Shumlak, 2022]{crews_kinetic}
Crews, D.~W. and Shumlak, U. (2022).
\newblock {On the validity of quasilinear theory applied to the electron
  bump-on-tail instability}.
\newblock {\em Physics of Plasmas}, 29(4):043902.

\bibitem[Datta et~al., 2021]{datta2021electromagnetic}
Datta, I. A.~M., Crews, D.~W., and Shumlak, U. (2021).
\newblock {Electromagnetic extension of the Dory--Guest--Harris instability as
  a benchmark for Vlasov--Maxwell continuum kinetic simulations of magnetized
  plasmas}.
\newblock {\em Physics of Plasmas}, 28(7):072112.

\bibitem[Datta and Shumlak, 2023]{datta_kinetic}
Datta, I. A.~M. and Shumlak, U. (2023).
\newblock Computationally efficient high-fidelity plasma simulations by
  coupling multi-species kinetic and multi-fluid models on decomposed domains.
\newblock {\em Journal of Computational Physics}, 483:112073.

\bibitem[Davidson and Chen, 1998]{davidson1998}
Davidson, R.~C. and Chen, C. (1998).
\newblock {Kinetic description of high intensity beam propagation through a
  periodic focusing field based on the nonlinear Vlasov-Maxwell equations}.
\newblock {\em Part. Accel.}, 59:175--250.

\bibitem[Davidson et~al., 1972]{davidson1972nonlinear}
Davidson, R.~C., Hammer, D.~A., Haber, I., and Wagner, C.~E. (1972).
\newblock Nonlinear development of electromagnetic instabilities in anisotropic
  plasmas.
\newblock {\em The Physics of Fluids}, 15(2):317--333.

\bibitem[Del~Sarto and Pegoraro, 2018]{del2018shear}
Del~Sarto, D. and Pegoraro, F. (2018).
\newblock Shear-induced pressure anisotropization and correlation with fluid
  vorticity in a low collisionality plasma.
\newblock {\em Monthly notices of the royal astronomical society},
  475(1):181--192.

\bibitem[Dodin, 2022]{dodin2022quasilinear}
Dodin, I.~Y. (2022).
\newblock Quasilinear theory for inhomogeneous plasma.
\newblock {\em Journal of Plasma Physics}, 88(4):905880407.

\bibitem[Dodin and Fisch, 2006]{dodin2006}
Dodin, I.~Y. and Fisch, N.~J. (2006).
\newblock {Correction to the Alfvén-Lawson criterion for relativistic electron
  beams}.
\newblock {\em Physics of Plasmas}, 13(10):103104.

\bibitem[Dory et~al., 1965]{dory1965unstable}
Dory, R.~A., Guest, G.~E., and Harris, E.~G. (1965).
\newblock {Unstable electrostatic plasma waves propagating perpendicular to a
  magnetic field}.
\newblock {\em Physical Review Letters}, 14(5):131.

\bibitem[Drummond and Rosenbluth, 1962]{drummond_micro_diffusion}
Drummond, W.~E. and Rosenbluth, M.~N. (1962).
\newblock {Anomalous Diffusion Arising from Microinstabilities in a Plasma}.
\newblock {\em The Physics of Fluids}, 5(12):1507--1513.

\bibitem[Dyabilin and Razumova, 2015]{dyabilin2015thermodynamic}
Dyabilin, K.~S. and Razumova, K.~A. (2015).
\newblock Thermodynamic approach to the interpretation of self-consistent
  pressure profiles in a tokamak.
\newblock {\em Plasma Physics Reports}, 41:685--695.

\bibitem[Einkemmer, 2019]{einkemmer2019performance}
Einkemmer, L. (2019).
\newblock A performance comparison of semi-{L}agrangian discontinuous
  {G}alerkin and spline based {V}lasov solvers in four dimensions.
\newblock {\em Journal of Computational Physics}, 376:937--951.

\bibitem[Escande, 2016]{escande_contrib}
Escande, D.~F. (2016).
\newblock Contributions of plasma physics to chaos and nonlinear dynamics.
\newblock {\em Plasma Physics and Controlled Fusion}, 58(11):113001.

\bibitem[Ewart et~al., 2022]{ewart_2022}
Ewart, R.~J., Brown, A., Adkins, T., and Schekochihin, A.~A. (2022).
\newblock {Collisionless relaxation of a Lynden-Bell plasma}.
\newblock {\em Journal of Plasma Physics}, 88(5):925880501.

\bibitem[Fox et~al., 2013]{fox2013filamentation}
Fox, W., Fiksel, G., Bhattacharjee, A., Chang, P.-Y., Germaschewski, K., Hu,
  S.~X., and Nilson, P.~M. (2013).
\newblock Filamentation instability of counterstreaming laser-driven plasmas.
\newblock {\em Physical review letters}, 111(22):225002.

\bibitem[Fried and Conte, 1961]{fried_plasma}
Fried, B.~D. and Conte, S.~D. (1961).
\newblock {\em The plasma dispersion function: the {H}ilbert transform of the
  {G}aussian}.
\newblock Academic press.

\bibitem[Glasser and Qin, 2020]{glasser_qin_2020}
Glasser, A.~S. and Qin, H. (2020).
\newblock The geometric theory of charge conservation in particle-in-cell
  simulations.
\newblock {\em Journal of Plasma Physics}, 86(3):835860303.

\bibitem[Gradshteyn and Ryzhik, 2015]{gradshteyn}
Gradshteyn, I.~S. and Ryzhik, I.~M. (2015).
\newblock {\em Table of {I}ntegrals, {S}eries, and {P}roducts}.
\newblock Academic {P}ress, {S}eventh edition.

\bibitem[Griffiths, 2012]{griffiths_momentum}
Griffiths, D.~J. (2012).
\newblock {Resource Letter EM-1: Electromagnetic Momentum}.
\newblock {\em American Journal of Physics}, 80(1):7--18.

\bibitem[Gurnett and Bhattacharjee, 2017]{gurnett2017introduction}
Gurnett, D.~A. and Bhattacharjee, A. (2017).
\newblock {\em {Introduction to plasma physics: {W}ith space, laboratory and
  astrophysical applications}}.
\newblock Cambridge University Press.

\bibitem[Hakim and Juno, 2020]{gkeyll2}
Hakim, A. and Juno, J. (2020).
\newblock {Alias-Free, Matrix-Free, and Quadrature-Free Discontinuous Galerkin
  Algorithms for (Plasma) Kinetic Equations}.
\newblock In {\em SC20: International Conference for High Performance
  Computing, Networking, Storage and Analysis}, pages 1--15.

\bibitem[He et~al., 2016]{ham_pic}
He, Y., Sun, Y., Qin, H., and Liu, J. (2016).
\newblock {Hamiltonian particle-in-cell methods for Vlasov-Maxwell equations}.
\newblock {\em Physics of Plasmas}, 23(9):092108.

\bibitem[Heath et~al., 2012]{heath_dg}
Heath, R.~E., Gamba, I.~M., Morrison, P.~J., and Michler, C. (2012).
\newblock {A discontinuous Galerkin method for the Vlasov–Poisson system}.
\newblock {\em Journal of Computational Physics}, 231(4):1140--1174.

\bibitem[Heninger and Morrison, 2018]{heninger_morrison_2018}
Heninger, J.~M. and Morrison, P.~J. (2018).
\newblock {An integral transform technique for kinetic systems with
  collisions}.
\newblock {\em Physics of Plasmas}, 25(8):082118.

\bibitem[Hesthaven and Warburton, 2007]{hesthaven2007nodal}
Hesthaven, J.~S. and Warburton, T. (2007).
\newblock {\em {Nodal discontinuous Galerkin methods: algorithms, analysis, and
  applications}}.
\newblock Springer Science \& Business Media.

\bibitem[Hill et~al., 2005]{hill2005beam}
Hill, J.~M., Key, M.~H., Hatchett, S.~P., and Freeman, R.~R. (2005).
\newblock {Beam-Weibel filamentation instability in near-term and fast-ignition
  experiments}.
\newblock {\em Physics of plasmas}, 12(8):082304.

\bibitem[Ho et~al., 2018]{ho2018physics}
Ho, A., Datta, I. A.~M., and Shumlak, U. (2018).
\newblock Physics-based-adaptive plasma model for high-fidelity numerical
  simulations.
\newblock {\em Frontiers in Physics}, 6:105.

\bibitem[Howes et~al., 2006]{gyrokinetics1}
Howes, G.~G., Cowley, S.~C., Dorland, W., Hammett, G.~W., Quataert, E., and
  Schekochihin, A.~A. (2006).
\newblock {Astrophysical Gyrokinetics: Basic Equations and Linear Theory}.
\newblock {\em The Astrophysical Journal}, 651(1):590.

\bibitem[Howes et~al., 2011]{gyrokinetics2}
Howes, G.~G., TenBarge, J.~M., Dorland, W., Quataert, E., Schekochihin, A.~A.,
  Numata, R., and Tatsuno, T. (2011).
\newblock {Gyrokinetic Simulations of Solar Wind Turbulence from Ion to
  Electron Scales}.
\newblock {\em Phys. Rev. Lett.}, 107:035004.

\bibitem[Huntington et~al., 2015]{huntington2015observation}
Huntington, C.~M., Fiuza, F., Ross, J.~S., Zylstra, A.~B., Drake, R.~P.,
  Froula, D.~H., Gregori, G., Kugland, N.~L., Kuranz, C.~C., Levy, M.~C.,
  et~al. (2015).
\newblock {Observation of magnetic field generation via the Weibel instability
  in interpenetrating plasma flows}.
\newblock {\em Nature Physics}, 11(2):173--176.

\bibitem[Hutchinson, 2020]{hutchinson2020particle}
Hutchinson, I.~H. (2020).
\newblock Particle trapping in axisymmetric electron holes.
\newblock {\em Journal of Geophysical Research: Space Physics},
  125(8):e2020JA028093.

\bibitem[Hutchinson, 2021]{hutchinson2021synthetic}
Hutchinson, I.~H. (2021).
\newblock {Synthetic multidimensional plasma electron hole equilibria}.
\newblock {\em Physics of Plasmas}, 28(6):062306.

\bibitem[Ichimaru, 1992]{ichimaru1992statistical}
Ichimaru, S. (1992).
\newblock {\em Statistical {P}lasma {P}hysics, {V}olume {I}: {B}asic
  {P}rinciples}.
\newblock CRC Press.

\bibitem[Juno et~al., 2018]{gkeyll1}
Juno, J., Hakim, A., TenBarge, J., Shi, E., and Dorland, W. (2018).
\newblock {Discontinuous Galerkin algorithms for fully kinetic plasmas}.
\newblock {\em Journal of Computational Physics}, 353:110--147.

\bibitem[Kempf et~al., 2013]{vlasiator_kempf1}
Kempf, Y., Pokhotelov, D., von Alfthan, S., Vaivads, A., Palmroth, M., and
  Koskinen, H. E.~J. (2013).
\newblock {Wave dispersion in the hybrid-Vlasov model: Verification of
  Vlasiator}.
\newblock {\em Physics of Plasmas}, 20(11):112114.

\bibitem[Kraus et~al., 2017]{kraus_2017}
Kraus, M., Kormann, K., Morrison, P., and Sonnendrücker, E. (2017).
\newblock Gempic: geometric electromagnetic particle-in-cell methods.
\newblock {\em Journal of Plasma Physics}, 83(4):905830401.

\bibitem[Landau, 1946]{landau_damping}
Landau, L.~D. (1946).
\newblock On the vibrations of the electronic plasma.
\newblock {\em J. Phys. USSR}, 10(26).

\bibitem[Landau and Lifshitz, 1946]{landau1946electrodynamics}
Landau, L.~D. and Lifshitz, E.~M. (1946).
\newblock {\em Electrodynamics of continuous media}.
\newblock Pergamon press Oxford.

\bibitem[Lenard and Bernstein, 1958]{lenard}
Lenard, A. and Bernstein, I.~B. (1958).
\newblock {Plasma oscillations with diffusion in velocity space}.
\newblock {\em Physical Review}, 112(5):1456.

\bibitem[Lerche, 1974]{lerche1974note}
Lerche, I. (1974).
\newblock {A note on summing series of Bessel functions occurring in certain
  plasma astrophysical situations}.
\newblock {\em Astrophysical Journal, Vol. 190, pp. 165-166 (1974)},
  190:165--166.

\bibitem[Leubner, 2004]{leubner2004fundamental}
Leubner, M.~P. (2004).
\newblock {Fundamental issues on kappa-distributions in space plasmas and
  interplanetary proton distributions}.
\newblock {\em Physics of Plasmas}, 11(4):1308--1316.

\bibitem[Leubner and Schupfer, 2001]{leubner2001general}
Leubner, M.~P. and Schupfer, N. (2001).
\newblock A general kinetic mirror instability criterion for space
  applications.
\newblock {\em Journal of Geophysical Research: Space Physics},
  106(A7):12993--12998.

\bibitem[Livadiotis and McComas, 2023]{livadiotis2023entropy}
Livadiotis, G. and McComas, D.~J. (2023).
\newblock Entropy defect in thermodynamics.
\newblock {\em Scientific Reports}, 13(1):9033.

\bibitem[Mace, 2004]{mace2004generalized}
Mace, R.~L. (2004).
\newblock {Generalized electron Bernstein modes in a plasma with a kappa
  velocity distribution}.
\newblock {\em Physics of Plasmas}, 11(2):507--522.

\bibitem[Mace and Hellberg, 2009]{mace2009new}
Mace, R.~L. and Hellberg, M.~A. (2009).
\newblock {A new formulation and simplified derivation of the dispersion
  function for a plasma with a kappa velocity distribution}.
\newblock {\em Physics of Plasmas}, 16(7):072113.

\bibitem[Mahajan and Hazeltine, 2000]{mahajan2000sheared}
Mahajan, S.~M. and Hazeltine, R.~D. (2000).
\newblock {Sheared-flow generalization of the Harris sheet}.
\newblock {\em Physics of Plasmas}, 7(4):1287--1293.

\bibitem[Morrison, 2000]{morrison2000hamiltonian}
Morrison, P.~J. (2000).
\newblock Hamiltonian description of vlasov dynamics: action-angle variables
  for the continuous spectrum.
\newblock {\em Transport theory and statistical physics}, 29(3-5):397--414.

\bibitem[Morrison, 2017]{morrison_metriplectic}
Morrison, P.~J. (2017).
\newblock {Structure and structure-preserving algorithms for plasma physics}.
\newblock {\em Physics of Plasmas}, 24(5):055502.

\bibitem[Morrison and Pfirsch, 1992]{morrison1992dielectric}
Morrison, P.~J. and Pfirsch, D. (1992).
\newblock {Dielectric energy versus plasma energy, and Hamiltonian action-angle
  variables for the Vlasov equation}.
\newblock {\em Physics of Fluids B: Plasma Physics}, 4(10):3038--3057.

\bibitem[Morse and Nielson, 1971]{morse1971numerical}
Morse, R.~L. and Nielson, C.~W. (1971).
\newblock {Numerical simulation of the Weibel instability in one and two
  dimensions}.
\newblock {\em The Physics of Fluids}, 14(4):830--840.

\bibitem[Mouhot and Villani, 2011]{villani_2011}
Mouhot, C. and Villani, C. (2011).
\newblock {On {L}andau damping}.
\newblock {\em Acta Mathematica}, 207(1):29 -- 201.

\bibitem[Newberger, 1982]{newberger1982new}
Newberger, B.~S. (1982).
\newblock {New sum rule for products of Bessel functions with application to
  plasma physics}.
\newblock {\em Journal of Mathematical Physics}, 23(7):1278--1281.

\bibitem[Ng et~al., 1999]{numerical_ng}
Ng, C.~S., Bhattacharjee, A., and Skiff, F. (1999).
\newblock {Kinetic eigenmodes and discrete spectrum of plasma oscillations in a
  weakly collisional plasma}.
\newblock {\em Physical review letters}, 83(10):1974.

\bibitem[Ng et~al., 2004]{ng_eigenmodes_collisional_2004}
Ng, C.~S., Bhattacharjee, A., and Skiff, F. (2004).
\newblock Complete spectrum of kinetic eigenmodes for plasma oscillations in a
  weakly collisional plasma.
\newblock {\em Phys. Rev. Lett.}, 92:065002.

\bibitem[Ng et~al., 2019]{gkeyll_clarify_spatial_mode}
Ng, J., Hakim, A., Juno, J., and Bhattacharjee, A. (2019).
\newblock {Drift Instabilities in Thin Current Sheets Using a Two-Fluid Model
  With Pressure Tensor Effects}.
\newblock {\em Journal of Geophysical Research: Space Physics},
  124(5):3331--3346.

\bibitem[Nicholson, 1983]{nicholson1983introduction}
Nicholson, D.~R. (1983).
\newblock {\em Introduction to plasma theory}.
\newblock Wiley New York.

\bibitem[Palmroth et~al., 2018]{vlasiator_kempf2}
Palmroth, M., Ganse, U., Pfau-Kempf, Y., Battarbee, M., Turc, L., Brito, T.,
  Grandin, M., Hoilijoki, S., Sandroos, A., and von Alfthan, S. (2018).
\newblock Vlasov methods in space physics and astrophysics.
\newblock {\em Living Reviews in Computational Astrophysics}, 4(1):1.

\bibitem[Paul and Sharma, 2024]{paul_and_sharma}
Paul, A. and Sharma, D. (2024).
\newblock {Kinetic instability of whistlers in electron beam-plasma systems}.
\newblock {\em Physics of Plasmas}, 31(3):032117.

\bibitem[Penrose, 1960]{penrose}
Penrose, O. (1960).
\newblock {Electrostatic Instabilities of a Uniform Non‐Maxwellian Plasma}.
\newblock {\em The Physics of Fluids}, 3(2):258--265.

\bibitem[Perse et~al., 2021]{benedikt_geometric}
Perse, B., Kormann, K., and Sonnendr\"{u}cker, E. (2021).
\newblock {Geometric Particle-in-Cell Simulations of the Vlasov--Maxwell System
  in Curvilinear Coordinates}.
\newblock {\em SIAM Journal on Scientific Computing}, 43(1):B194--B218.

\bibitem[Pezzi et~al., 2019]{vida}
Pezzi, O., Cozzani, G., Califano, F., Valentini, F., Guarrasi, M., Camporeale,
  E., Brunetti, G., Retinò, A., and Veltri, P. (2019).
\newblock {ViDA: a Vlasov–DArwin solver for plasma physics at electron
  scales}.
\newblock {\em Journal of Plasma Physics}, 85(5):905850506.

\bibitem[Pierrard and Lazar, 2010]{pierrard2010kappa}
Pierrard, V. and Lazar, M. (2010).
\newblock Kappa distributions: {T}heory and applications in space plasmas.
\newblock {\em Solar physics}, 267:153--174.

\bibitem[Pokhotelov et~al., 2004]{pokhotelov2004mirror}
Pokhotelov, O.~A., Sagdeev, R.~Z., Balikhin, M.~A., and Treumann, R.~A. (2004).
\newblock {Mirror instability at finite ion-Larmor radius wavelengths}.
\newblock {\em Journal of Geophysical Research: Space Physics}, 109(A9).

\bibitem[Pokhotelov et~al., 2002]{pokhotelov2002linear}
Pokhotelov, O.~A., Treumann, R.~A., Sagdeev, R.~Z., Balikhin, M.~A.,
  Onishchenko, O.~G., Pavlenko, V.~P., and Sandberg, I. (2002).
\newblock Linear theory of the mirror instability in non-{M}axwellian space
  plasmas.
\newblock {\em Journal of Geophysical Research: Space Physics},
  107(A10):SMP--18.

\bibitem[Rosenbluth and Post, 1965]{rosenbluth_losscone}
Rosenbluth, M.~N. and Post, R.~F. (1965).
\newblock {High‐Frequency Electrostatic Plasma Instability Inherent to
  ``Loss‐Cone'' Particle Distributions}.
\newblock {\em The Physics of Fluids}, 8(3):547--550.

\bibitem[Roytershteyn and Delzanno, 2018]{los_alamos_spectral2}
Roytershteyn, V. and Delzanno, G.~L. (2018).
\newblock {Spectral Approach to Plasma Kinetic Simulations Based on Hermite
  Decomposition in the Velocity Space}.
\newblock {\em Frontiers in Astronomy and Space Sciences}, 5.

\bibitem[Rudakov and Tsytovich, 1978]{rudakov1978strong}
Rudakov, L.~I. and Tsytovich, V.~N. (1978).
\newblock {Strong Langmuir turbulence}.
\newblock {\em Physics Reports}, 40(1):1--73.

\bibitem[Schamel, 2023]{schamel2023pattern}
Schamel, H. (2023).
\newblock {Pattern formation in Vlasov--Poisson plasmas beyond Landau caused by
  the continuous spectra of electron and ion hole equilibria}.
\newblock {\em Reviews of Modern Plasma Physics}, 7(1):11.

\bibitem[Sharma and Bhatnagar, 1976]{sharma1976wave}
Sharma, S.~R. and Bhatnagar, T.~N. (1976).
\newblock Wave propagation along an arbitrary direction in a bi-maxwellian
  plasma.
\newblock {\em Plasma Physics}, 18(2):95.

\bibitem[Short and Simon, 2002]{analytical_short}
Short, R.~W. and Simon, A. (2002).
\newblock {Damping of perturbations in weakly collisional plasmas}.
\newblock {\em Physics of Plasmas}, 9(8):3245--3253.

\bibitem[Shukla et~al., 2018]{shukla2018conditions}
Shukla, N., Vieira, J., Muggli, P., Sarri, G., Fonseca, R., and Silva, L.
  (2018).
\newblock Conditions for the onset of the current filamentation instability in
  the laboratory.
\newblock {\em Journal of Plasma Physics}, 84(3).

\bibitem[Shumlak et~al., 2011]{SHUMLAK20111767}
Shumlak, U., Lilly, R., Reddell, N., Sousa, E., and Srinivasan, B. (2011).
\newblock Advanced physics calculations using a multi-fluid plasma model.
\newblock {\em Computer Physics Communications}, 182(9):1767--1770.
\newblock Computer Physics Communications Special Edition for Conference on
  Computational Physics Trondheim, Norway, June 23-26, 2010.

\bibitem[Skoutnev et~al., 2019]{Skoutnev_2019}
Skoutnev, V., Hakim, A., Juno, J., and TenBarge, J.~M. (2019).
\newblock {Temperature-dependent Saturation of Weibel-type Instabilities in
  Counter-streaming Plasmas}.
\newblock {\em The Astrophysical Journal Letters}, 872(2):L28.

\bibitem[Taggart et~al., 1972]{taggart1972second}
Taggart, K.~A., Godrey, B.~B., Rhoades~Jr, C.~E., and Ives, H.~C. (1972).
\newblock {Second-Order Effects in the Weibel Instability}.
\newblock {\em Physical Review Letters}, 29(26):1729.

\bibitem[Tataronis and Crawford, 1970a]{tataronis1970_cyclotron1}
Tataronis, J.~A. and Crawford, F.~W. (1970a).
\newblock Cyclotron harmonic wave propagation and instabilities: {I}.
  {P}erpendicular propagation.
\newblock {\em Journal of Plasma Physics}, 4(2):231--248.

\bibitem[Tataronis and Crawford, 1970b]{tataronis1970_cyclotron2}
Tataronis, J.~A. and Crawford, F.~W. (1970b).
\newblock {Cyclotron harmonic wave propagation and instabilities: II. Oblique
  propagation}.
\newblock {\em Journal of Plasma Physics}, 4(2):249--264.

\bibitem[Tzoufras et~al., 2006]{tzoufras2006space}
Tzoufras, M., Ren, C., Tsung, F.~S., Tonge, J.~W., Mori, W.~B., Fiore, M.,
  Fonseca, R., and Silva, L.~O. (2006).
\newblock {Space-charge effects in the current-filamentation or Weibel
  instability}.
\newblock {\em Physical review letters}, 96(10):105002.

\bibitem[Valentini et~al., 2007]{hvm_valentini}
Valentini, F., Trávníček, P., Califano, F., Hellinger, P., and Mangeney, A.
  (2007).
\newblock {A hybrid-Vlasov model based on the current advance method for the
  simulation of collisionless magnetized plasma}.
\newblock {\em Journal of Computational Physics}, 225(1):753--770.

\bibitem[van Kampen, 1955]{van1955theory}
van Kampen, N.~G. (1955).
\newblock On the theory of stationary waves in plasmas.
\newblock {\em Physica}, 21(6-10):949--963.

\bibitem[Vedenov, 1963]{vedenov1963quasi}
Vedenov, A.~A. (1963).
\newblock Quasi-linear plasma theory (theory of a weakly turbulent plasma).
\newblock {\em Journal of Nuclear Energy. Part C, Plasma Physics, Accelerators,
  Thermonuclear Research}, 5(3):169.

\bibitem[Vencels et~al., 2016]{los_alamos_spectral1}
Vencels, J., Delzanno, G.~L., Manzini, G., Markidis, S., Peng, I.~B., and
  Roytershteyn, V. (2016).
\newblock {SpectralPlasmaSolver: a Spectral Code for Multiscale Simulations of
  Collisionless, Magnetized Plasmas}.
\newblock {\em Journal of Physics: Conference Series}, 719(1):012022.

\bibitem[Verniero et~al., 2018]{gyrokinetics4}
Verniero, J.~L., Howes, G.~G., and Klein, K.~G. (2018).
\newblock {Nonlinear energy transfer and current sheet development in localized
  Alfvén wavepacket collisions in the strong turbulence limit}.
\newblock {\em Journal of Plasma Physics}, 84(1):905840103.

\bibitem[Verscharen et~al., 2016]{dsp_kinetic_slow_modes1}
Verscharen, D., Chandran, B. D.~G., Klein, K.~G., and Quataert, E. (2016).
\newblock Collisionless isotropization of the solar-wind protons by compressive
  fluctuations and plasma instabilities.
\newblock {\em The Astrophysical Journal}, 831(2):128.

\bibitem[Verscharen et~al., 2017]{dsp_kinetic_slow_modes2}
Verscharen, D., Chen, C. H.~K., and Wicks, R.~T. (2017).
\newblock {On Kinetic Slow Modes, Fluid Slow Modes, and Pressure-balanced
  Structures in the Solar Wind}.
\newblock {\em The Astrophysical Journal}, 840(2):106.

\bibitem[Vogman et~al., 2018]{LLNL_code}
Vogman, G., Shumlak, U., and Colella, P. (2018).
\newblock {Conservative fourth-order finite-volume Vlasov–Poisson solver for
  axisymmetric plasmas in cylindrical phase space coordinates}.
\newblock {\em Journal of Computational Physics}, 373:877--899.

\bibitem[Vogman et~al., 2014]{vogman2014dory}
Vogman, G.~V., Colella, P., and Shumlak, U. (2014).
\newblock {Dory--Guest--Harris instability as a benchmark for continuum kinetic
  Vlasov--Poisson simulations of magnetized plasmas}.
\newblock {\em Journal of Computational Physics}, 277:101--120.

\bibitem[Vásconez et~al., 2014]{hvm_clarify_spatial_mode}
Vásconez, C.~L., Valentini, F., Camporeale, E., and Veltri, P. (2014).
\newblock {Vlasov simulations of kinetic Alfvén waves at proton kinetic
  scales}.
\newblock {\em Physics of Plasmas}, 21(11):112107.

\bibitem[Watanabe et~al., 2014]{gyrokinetics3}
Watanabe, T.-H., Idomura, Y., Maeyama, S., Nakata, M., Sugama, H., Nunami, M.,
  and Ishizawa, A. (2014).
\newblock Exploring phase space turbulence in magnetic fusion plasmas.
\newblock {\em Journal of Physics: Conference Series}, 510(1):012045.

\bibitem[Weibel, 1959]{weibel_orig}
Weibel, E.~S. (1959).
\newblock {Spontaneously Growing Transverse Waves in a Plasma Due to an
  Anisotropic Velocity Distribution}.
\newblock {\em Phys. Rev. Lett.}, 2:83--84.

\bibitem[Wu et~al., 2019]{dsp_kinetic_alfven_turbulence}
Wu, H., Verscharen, D., Wicks, R.~T., Chen, C. H.~K., He, J., and Nicolaou, G.
  (2019).
\newblock {The Fluid-like and Kinetic Behavior of Kinetic Alfvén Turbulence in
  Space Plasma}.
\newblock {\em The Astrophysical Journal}, 870(2):106.

\bibitem[Yoon, 2007]{yoon2007spontaneous}
Yoon, P.~H. (2007).
\newblock Spontaneous thermal magnetic field fluctuation.
\newblock {\em Physics of plasmas}, 14(6).

\bibitem[Yoon and Lui, 2006]{yoon_anom}
Yoon, P.~H. and Lui, A. T.~Y. (2006).
\newblock Quasi-linear theory of anomalous resistivity.
\newblock {\em Journal of Geophysical Research: Space Physics}, 111(A2).

\bibitem[Zhao et~al., 2022]{dsp_alfven_energy_transfer}
Zhao, J., Lee, L., Xie, H., Yao, Y., Wu, D., Voitenko, Y., and Viviane, P.
  (2022).
\newblock {Quantifying Wave–Particle Interactions in Collisionless Plasmas:
  Theory and Its Application to the Alfvén-mode Wave}.
\newblock {\em The Astrophysical Journal}, 930(1):95.

\bibitem[Zhdankin, 2023]{zhdankin2023dimensional}
Zhdankin, V. (2023).
\newblock Dimensional measures of generalized entropy.
\newblock {\em Journal of Physics A: Mathematical and Theoretical},
  56(38):385002.

\end{thebibliography}
\bibliographystyle{apalike}


\appendix

    \section{General description of our numerical methods}\label{app:summary_of_numerical}
    Our numerical method is a combination of the pseudospectral method for
    spatial fluxes~(\cite{boyd2001chebyshev}) with high-order discontinuous
    Galerkin method for velocity space~(\cite{crews_kinetic}).
    The Vlasov-Poisson system is discretized in configuration space by Galerkin
    projection onto a truncated multi-dimensional Fourier basis, for example $(x,y)\to (k_n, k_m)$.
    That is, if the original kinetic equation is
    \begin{equation}\label{eq:basic_vlasov}
    \partial_t f + \bm{v}\cdot\nabla_x f + \bm{F}\cdot\nabla_v f = 0
    \end{equation}
    with $\bm{F}$ the vector of momentum fluxes, then the Galerkin-projected kinetic equation is
    \begin{equation}\label{eq:fourier_2d}
        \frac{\partial f_{nm}}{\partial t} + ik_n u f_{nm} + ik_m v f_{nm} + \nabla_v\cdot(\bm{\mathcal{F}}_{nm}) = 0
    \end{equation}
    with $\bm{\mathcal{F}}_{nm}$ the vector of spectral momentum fluxes.
    Eq.~\ref{eq:fourier_2d} is discretized by truncating the velocity domain and applying
    the discontinuous Galerkin method $(u,v)\to (u_{jk}, v_{pq})$, with $(j,p)$ the
    element indices and $(k,q)$ the sub-element indices (that is, the collocation nodes), and a set of basis
    functions $\psi_k$ are chosen.
    We use the Legendre-Gauss-Lobatto basis~(\cite{hesthaven2007nodal}).
    The semi-discrete equation obtained is
    \begin{equation}\label{eq:discretized_kinetic}
        \frac{df_{nmjkpq}}{dt} = \mathcal{N}_{nmjkpq}[f] + \mathcal{L}_{nmjkpq}^{\ell r}f_{nmj\ell pr}
    \end{equation}
    where $\mathcal{N}$ is a nonlinear operator representing the discretized velocity
    flux divergences, while $\mathcal{L}$ is the advective part of the semi-discrete operator and is linear in $f$.
    Advection is not linear in $\bm{v}$, but is integrated over the basis functions $\psi_k$ to form a
    linear operator
    \begin{equation}\label{eq:spectral_advection_operator}
        \mathbb{T}_{njk}^{\ell} \equiv -ik_n (\bar{v}_j I_k^\ell +
    J_m \langle \psi_k | \psi^s\rangle^{-1}\langle \xi \psi_s | \psi^\ell\rangle),
    \end{equation}
    with $\langle\cdot|\cdot\rangle$ an inner product over the reference element with coordinate
    $\xi\in[-1, 1]$, and $\bar{v}_j$ the velocity in the midpoint of element $j$
    with transformation $v_j(\xi) = \bar{v}_j + (\Delta v)_j\xi/2$ with $(\Delta v)_j$ the element width.
    Then advection is represented by a linear operator
    \begin{equation}\label{eq:linear_advection_operator}
        \mathcal{L}_{nmjkpq}^{\ell r}f_{nmj\ell pr} \equiv \mathbb{T}_{njk}^{\ell}f_{nmj\ell pq} + \mathbb{T}_{mpq}^r f_{nmjkpr}.
    \end{equation}
    Poisson's equation is solved algebraically in Fourier space, and
    the velocity fluxes $\mathcal{N}$ are computed by the pseudospectral method.
    That is, the spectral field and distribution are both zero-padded using Orszag's two-thirds rule and transformed to nodal
    values on the spatial collocation points where the fluxes are calculated.
    The velocity element boundary fluxes are then calculated in $(\bm{x}, \bm{v})$
    space by upwind method and transformed back to spectral space for time integration.
    However, element-internal fluxes are calculated in $(\bm{k}, \bm{v})$ space by inverse transformation as the spectral
    symmetry from $\text{Im}(f(\bm{x}, \bm{v}))=0$ reduces the operations by a factor of two.
    Finally, the semi-discrete equation is advanced through $\Delta t=h$ using second-order Strang splitting
    of linear $\mathcal{L}$ and nonlinear $\mathcal{N}$ fluxes,
    \begin{equation}\label{eq:solve_strang_split}
        f^{n+1} = \mathcal{S}[\mathcal{L}]^{h/2}\mathcal{S}[\mathcal{N}]^h\mathcal{S}[\mathcal{L}]^{h/2}f^n
    \end{equation}
    where $\mathcal{S}[\cdot]$ represents evaluation of the separated operator by any method of approximation.
    In this simulation the explicit third-order Adams-Bashforth method is used for the nonlinear flux $\mathcal{S}[\mathcal{N}]^h$
    while advection is advanced implicitly by Crank-Nicholson stepping,
    \begin{equation}\label{eq:split_advection}
        \Big(\mathcal{S}[\mathcal{L}]^h\Big)^{\ell r}_{nmjkpq} =
        \Big(I^{\ell}_{njk} - \frac{h}{2}\mathbb{T}^{\ell}_{njk}\Big)^{-1}
        \Big(I^{\ell}_{njk} + \frac{h}{2}\mathbb{T}^{\ell}_{njk}\Big)
        \Big(I^r_{mpq} - \frac{h}{2}\mathbb{T}^r_{mpq}\Big)^{-1}
        \Big(I^r_{mpq} + \frac{h}{2}\mathbb{T}^r_{mpq}\Big)
    \end{equation}
    with $I_{mpq}^r$ the identity matrix $I_q^r$ for each $(m,n)$.
    Fractional stepping for advection is advantageous as it reduces the $\mathcal{O}(N^2)$ operations per cell into
    two $\mathcal{O}(N)$ operations.
    The calculation of the nonlinear momentum flux operator $\mathcal{N}$ is treated in~\cite{crews2022numerical}, and
    we briefly recapitulate this here.
    In this work, the quadratically nonlinear fluxes are \textit{not} integrated consistently as described in~\cite{crews2022numerical}
    Section 2.5 and~\cite{gkeyll2}, and instead an alias error is incurred.
    Consistent integration of the quadratically nonlinear fluxes in discretization of the Vlasov equation greatly improves
    the conservation of Casimirs (nonlinear functionals) such as phase space entropy density.
    However, in this work an alias error is accepted
    because the smoothing spatial hyperviscosity used here breaks the Casimirs near the aliased grid scales.
    Thus, inaccuracy in conservation of nonlinear functionals is accepted in exchange for less computational effort.

 \section{Polar Fourier integrals of ring distributions}\label{sec:integral_appendix}
 This appendix integrates the ring distribution of Eq.~\ref{eq:ring_distribution} over perpendicular velocities,
 \begin{equation}\label{eq:considered_integral}
     \mathbb{F}_{n,m,\gamma}(k) \equiv \int_0^{\infty}f_{\gamma}(v)J_n(kv)J_m(kv)2\pi v dv
 \end{equation}
 for integer $n, m$.
 By this formula one can also determine the integral of products, $J_n(v)J_n'(v)$ etc.,
 by recursion of the derivatives $J_n'(z)$.
 The result is that Eq.~\ref{eq:considered_integral} is a type-$_3 F_3$ hypergeometric function,
 \begin{equation}\label{eq:key_integral_proof}
     \begin{split}
         \mathbb{F}_{n,m,\gamma}(k)=&\frac{\Gamma(\gamma + \frac{n+m}{2} + 1)(k\alpha)^{n+m}}{\Gamma(n+1)\Gamma(m+1)\Gamma(\gamma+1)}\times\\
         &\pFq{3}{3}{\frac{n+m}{2}+\frac{1}{2}, \frac{n+m}{2} + 1, \gamma + \frac{n+m}{2} + 1}{n+1,\quad m+1,\quad n+m+1}(-(2k\alpha)^2)
     \end{split}
 \end{equation}
 which is equivalently written as a power series with Gamma function coefficients,
 \begin{equation}\label{eq:alternative_power_series}
     \mathbb{F}_{n, m, \gamma}(k) = \frac{1}{\Gamma(\gamma+1)}
     \sum_{\ell=0}^{\infty}\frac{\Gamma(n+m+2\gamma+2\ell+1)}{\Gamma(n+\ell+1)\Gamma(m+\ell+1)\Gamma(n+m+\ell+1)}
 \frac{(-1)^{\ell}}{\ell!}(k\alpha)^{2\ell + n + m}.
 \end{equation}
 Setting $n=m$ as in electrostatic theory reduces to a more manageable $_2 F_2$ function,
 \begin{equation}\label{eq:2f2_function}
     \mathbb{F}_{n,\gamma}(k)=
     \frac{\Gamma(\gamma + n + 1)(k\alpha)^{2n}}{\Gamma^2(n+1)\Gamma(\gamma+1)}
         \pFq{2}{2}{n+\frac{1}{2}, \gamma + n + 1}{n+1, 2n+1}(-(2k\alpha)^2)
 \end{equation}
 with the power series for practical computation given in Eq.~\ref{eq:hypergeometric}.

 \subsection{The product of integer-order Bessel functions}\label{sec:bessel}
 The product of integer-order Bessel functions is a generalized
 $_2 F_3$-type hypergeometric,
 \begin{equation}\label{eq:bessel_product}
     J_n(z)J_m(z) = \frac{1}{n!m!}\Big(\frac{z}{2}\Big)^{n+m}\pFq{2}{3}{\frac{n+m}{2} + \frac{1}{2},
         \frac{n+m}{2}+1}{n+1, m+1, n+m+1}(-z^2)
 \end{equation}
 which for $n=m$ reduces to a type $_1 F_2$ function,
 \begin{equation}\label{eq:bessel_product2}
     J_n^2(z) = \frac{1}{(n!)^2}\Big(\frac{z}{2}\Big)^{2n}\pFq{1}{2}{n + \frac{1}{2}}{n+1, 2n+1}(-z^2).
 \end{equation}

 \subsubsection{Demonstration of Equation~\ref{eq:bessel_product}}
 Multiplying term-by-term the power series of $J_n$, $J_m$ and diagonalizing by $\ell = j+k$,
 \begin{equation}\label{eq:proof_1}
     \begin{split}
         J_n(z)J_m(z) &= \Big(\frac{z}{2}\Big)^{n+m}\sum_{j,k=0}^{\infty}\frac{1}{\Gamma(n+k+1)\Gamma(m+j+1)}\frac{(-z^2/4)^{j+k}}{j!k!}\\
                      &= \Big(\frac{z}{2}\Big)^{n+m}\sum_{\ell=0}^{\infty}
         \Big[\sum_{k=0}^{\infty}\frac{1}{\Gamma(k+n+1)\Gamma(\ell-k+m+1)\Gamma(\ell-k+1)\Gamma(k+1)}
             \Big]\Big(\frac{-z^2}{4}\Big)^\ell.
     \end{split}
 \end{equation}
 Using properties of Pochhammer symbols and Gauss's hypergeometric theorem,
 \begin{equation}\label{eq:gauss_summation}
     \begin{split}
         \sum_{k=0}^{\infty}&\frac{1}{\Gamma(k+n+1)\Gamma(\ell-k+m+1)\Gamma(\ell-k+1)\Gamma(k+1)}\\
             =&\frac{1}{\Gamma(n+1)\Gamma(\ell+1)\Gamma(\ell+m+1)}\pFq{2}{1}{-\ell, -\ell-m}{n+1}(1)\\
             =&\frac{\Gamma(n+m+2\ell+1)}{\Gamma(\ell+1)\Gamma(m+\ell+1)\Gamma(n+\ell+1)\Gamma(n+m+\ell+1)}
     \end{split}
 \end{equation}
 one obtains a single summation for $J_n(z)J_m(z)$,
 \begin{equation}\label{eq:summation2}
     J_n(z)J_m(z) = \Big(\frac{z}{2}\Big)^{n+m}\sum_{\ell=0}^{\infty}
     \frac{\Gamma(n+m+2\ell+1)}{\Gamma(m+\ell+1)\Gamma(n+\ell+1)\Gamma(n+m+\ell+1)}\frac{(-z^2/4)^\ell}{\ell!}.
 \end{equation}
 The numerator factorial is then written in Pochhammer symbols by the formulas
 \begin{align}
     (x+\ell)_{\ell} &= \frac{(x)_{2\ell}}{(x)_\ell},\\
 (x)_{2\ell} &= 2^{2\ell}\Big(\frac{x}{2}\Big)_{\ell}\Big(\frac{1+x}{2}\Big)_{\ell},
 \end{align}
 yielding a series identical to Eq.~\ref{eq:bessel_product} and proving the identity,
 \begin{equation}\label{eq:proved}
     J_n(z)J_m(z) = \frac{1}{n!m!}\Big(\frac{z}{2}\Big)^{n+m}\sum_{\ell=0}^{\infty}
     \frac{\big(\frac{n+m}{2}+\frac{1}{2}\big)_\ell\big(\frac{n+m}{2}+1\big)_\ell}{(m+1)_\ell(n+1)_\ell(n+m+1)_\ell}
 \frac{(-z^2)^\ell}{\ell!}.
 \end{equation}

 \subsection{Integration of Eq.~\ref{eq:bessel_product} with a loss-cone distribution}\label{subsec:bessel_integral}
 Having developed the product of two Bessel functions in terms of a single entire function the
 integration over perpendicular velocities is now shown to be Eq.~\ref{eq:key_integral_proof}.

 \subsubsection{Demonstration of Eq.~\ref{eq:key_integral_proof}}
 The ring distribution $f_\gamma(v)$ is combined with Eq.~\ref{eq:proved}
 and integrated term-by-term,
 \begin{equation}\label{eq:proof_part_1}
     \begin{split}
     \mathbb{F}_{n,m,\gamma}(k) = \frac{k^{n+m}}{n!m!\gamma!(\alpha^2)^{\gamma+1}}\Big(\frac{k}{2}\Big)^{n+m}\sum_{\ell=0}^{\infty}&
     \frac{\big(\frac{n+m}{2}+\frac{1}{2}\big)_\ell\big(\frac{n+m}{2}+1\big)_\ell}{(m+1)_\ell(n+1)_\ell(n+m+1)_\ell}
     \frac{(-k^2)^{\ell}}{\ell!}\times\\
     &\int_0^{\infty}v^{2(\ell + \gamma + \frac{n+m}{2})}e^{-v^2/\alpha^2}2vdv.
     \end{split}
 \end{equation}
 The interior integral is Euler's form of the Gamma function upon substitution $u=\frac{v^2}{\alpha^2}$,
 \begin{equation}\label{eq:proof_part_2}
 \begin{split}
     \mathbb{F}_{n,m,\gamma} = &\frac{\Gamma(\gamma + \frac{n+m}{2} + 1)}{\Gamma(n+1)\Gamma(m+1)\Gamma(\gamma+1)}
     (k\alpha)^{n+m}\times\\
     &\sum_{\ell=0}^{\infty}
     \frac{\big(\frac{n+m}{2}+\frac{1}{2}\big)_\ell\big(\frac{n+m}{2}+1\big)_\ell
     (\gamma + \frac{n+m}{2}+1)_\ell}{(m+1)_\ell(n+1)_\ell(n+m+1)_\ell}
     \frac{(-4\alpha^2 k^2)^{\ell}}{\ell!}.
 \end{split}
 \end{equation}
 The series is a hypergeometric function of type-$_3F_3$, demonstrating Eq.~\ref{eq:key_integral_proof}.
 While the form of the integral as a type-$_3F_3$ function reveals connections to the theory of special functions,
 in a practical computation it is more practical to use the Gamma function form of the series as in
 Eq.~\ref{eq:alternative_power_series}.

\section{The Harris dispersion function for ring distributions}\label{sec:perp_app}
This appendix demonstrates Eqs.~\ref{eq:closed_form} and~\ref{eq:trig_form} for electrostatic
modes in a plasma with zero-order cyclotron orbits distributed as a loss cone, beginning from Eq.~\ref{eq:harris_dispersion}
known as the Harris dispersion function.
We begin by constructing the function in closed form as a hypergeometric function, and then consider
a convenient trigonometric integral form by calculation of the loss cone's Hankel transform.
These forms of the dielectric function $\varepsilon(\omega, k_\perp)$, which sum the contribution of the cyclotron resonances
to all orders, enable precise numerics to determine eigenvalues in continuum kinetic simulations.

\subsection{The perpendicular wave dielectric function in closed form}\label{subsec:hypergeomtric_closed_form}
We demonstrate Eq.~\ref{eq:closed_form} for the closed form of the dielectric function for perpendicular
cyclotron waves in a loss-cone distributed plasma using the Lerche-Newberger summation
    theorem~(\cite{lerche1974note, newberger1982new}),
 which sums the Bessel series in Eqs.~\ref{eq:upsilon2} and~\ref{eq:lambda2} to all
orders in the cyclotron harmonics,
\begin{align}
    \Upsilon_2 &= \sum_{n=-\infty}^\infty\frac{n}{\omega'-n}J_n^2(\beta)
    = \frac{\pi \omega'}{\sin(\pi\omega')}J_{\omega'}(\beta)J_{-\omega'}(\beta) - 1\label{eq:newberger1}\\
    \Lambda_2 &= \sum_{n=-\infty}^{\infty}\frac{1}{\omega' - n}J_n^2(\beta)
    = \frac{\pi}{\sin(\pi\omega')}J_{\omega'}(\beta)J_{-\omega'}(\beta)\label{eq:newberger2}
\end{align}
with each sum limiting to a product of Bessel functions $J_z(\beta)J_{-z}(\beta)$ of complex order.
Here the auxiliary quantities are
$\omega' = (\omega - k_\parallel v_\parallel) / \omega_c$ and $\beta = k_\perp v_\perp / \omega_c$.
The polar velocity-space integrals over
$v_\perp$, when considering the closed forms of Eqs.~\ref{eq:newberger1} and~\ref{eq:newberger2},
result in a generalized hypergeometric function~(see the following for a demonstration),
\begin{equation}\label{eq:closed_form_hyper}
    \frac{1}{2^\gamma\alpha^{2\gamma+2}\Gamma(\gamma+1)}
    \int_0^{\infty}v^{2j+1}e^{-v^2/2\alpha^2}\frac{\pi\omega}{\sin(\pi\omega)}J_{\omega}(qv)J_{-\omega}(qv)dv
    = \pFq{2}{2}{\frac{1}{2},\quad \gamma + 1}{1+\omega, 1-\omega}(-2(\alpha q)^2).
\end{equation}
First note that with $k_\parallel=0$ the Harris dispersion function is
\begin{equation}\label{eq:perp_harris_to_integrate}
    \varepsilon(\omega, k_\perp) = 1 + \Big(\frac{\omega_p}{\omega_c}\Big)^2\frac{1}{(k\lambda_D)^2}
    \int_0^{\infty}\frac{\Upsilon_2}{v_\perp}\frac{\partial f_0}{\partial v_\perp} 2\pi v_\perp dv_\perp.
\end{equation}
Using the identity Eq.~\ref{eq:closed_form_hyper}, and the recurrence relation for $f_\gamma(v_\perp)$
of Eq.~\ref{eq:recurrence_relation},
one obtains
\begin{equation}\label{eq:h_perp}
    \int_0^{\infty}\frac{\Upsilon_2}{v_\perp}\frac{\partial f_0}{\partial v_\perp} 2\pi v_\perp dv_\perp
    = \pFq{2}{2}{\frac{1}{2},\quad\gamma+1}{1+\omega', 1-\omega'}(-2(k_\perp r_L)^2) -
    \pFq{2}{2}{\frac{1}{2},\quad\quad\gamma}{1+\omega', 1-\omega'}(-2(k_\perp r_L)^2)
\end{equation}
which was to be shown.

\subsubsection{Demonstration of Eq.~\ref{eq:closed_form_hyper}}
Observe that term-by-term multiplication of the power series for both $J_{\omega}$ and $J_{-\omega}$, and diagonalization
of the double sum with $\ell = m+k$, leads to the expression
\begin{align}
    \frac{\pi\omega}{\sin(\pi\omega)}J_{\omega}(z)J_{-\omega}(z) &=
    \sum_{m=0}^{\infty}\sum_{k=0}^{\infty}\frac{\Gamma(1+\omega)\Gamma(1-\omega)}{\Gamma(m+\omega+1)\Gamma(k-\omega+1)}\frac{(-1)^{m+k}}{m!k!}
    \Big(\frac{z}{2}\Big)^{2(m+k)}\\
    &= \sum_{\ell=0}^{\infty}
    \Big[\sum_{m=0}^{\infty}\frac{\Gamma(1+\omega)\Gamma(1-\omega)}{\Gamma(m+\omega+1)\Gamma(\ell-m-\omega+1)}
    \frac{1}{m!(\ell-m)!}\Big](-1)^{\ell}\Big(\frac{z}{2}\Big)^{2\ell}\label{eq:to_manipulate}
\end{align}
having used Euler's reflection formula $\pi z \csc(\pi z) = \Gamma(1+z)\Gamma(1-z)$ with $\Gamma(z)$ the Gamma function.
Recall an identity for the rising factorial $(z)_n$ (or Pochhammer symbol),
\begin{equation}\label{eq:pochhammer_gamma_relation}
    (z)_n = \frac{\Gamma(z+n)}{\Gamma(z)} = (-1)^n\frac{\Gamma(z+1)}{\Gamma(z-n+1)}
\end{equation}
as well as Gauss's hypergeometric summation theorem,
\begin{equation}\label{eq:gauss_hypergeometric}
    \pFq{2}{1}{a, b}{c}(1) = \frac{\Gamma(c)\Gamma(c-a-b)}{\Gamma(c-a)\Gamma(c-b)},\quad \text{Re}(c) > \text{Re}(a+b).
\end{equation}
Application of Eqs.~\ref{eq:pochhammer_gamma_relation} and~\ref{eq:gauss_hypergeometric} to Eq.~\ref{eq:to_manipulate}
    shows that the inner summation becomes
\begin{equation}\label{eq:inner_summation}
    \sum_{m=0}^{\infty}\frac{\Gamma(1+\omega)\Gamma(1-\omega)}{\Gamma(m+\omega+1)\Gamma(\ell-m-\omega+1)}
    \frac{1}{m!(\ell-m)!} = 2^{2\ell}\frac{(\frac{1}{2})_\ell}{(1+\omega)_\ell (1-\omega)_\ell}\frac{1}{\ell!}.
\end{equation}
Therefore, the function expressed by the Lerche-Newberger theorem is a hypergeometric,
\begin{equation}\label{eq:specified_integral}
    \frac{\pi\omega}{\sin(\pi\omega)}J_{\omega}(z)J_{-\omega}(z) =
    \sum_{\ell=0}^{\infty}\frac{(-1)^{\ell}(\frac{1}{2})_\ell}{(1+\omega)_\ell(1-\omega)_\ell}\frac{z^{2\ell}}{\ell!}
    = \pFq{1}{2}{\frac{1}{2}}{1+\omega, 1-\omega}(-z^2).
\end{equation}
Now to compute the integral, the power series in Eq.~\ref{eq:specified_integral} is integrated term-by-term,
\begin{align}
    \frac{1}{2^\gamma\alpha^{2\gamma+2}\Gamma(\gamma+1)}
    \int_0^{\infty}&v^{2j+1}e^{-v^2/2\alpha^2}\frac{\pi\omega}{\sin(\pi\omega)}J_{\omega}(qv)J_{-\omega}(qv)dv
    =\\
    &=\frac{1}{2^\gamma\alpha^{2\gamma+2}\Gamma(\gamma+1)}
    \int_0^{\infty} v^{2\gamma + 1}e^{-v^2/2\alpha^2} \pFq{1}{2}{\frac{1}{2}}{1+\omega, 1-\omega}(-(qv)^2) dv\\
    &=\sum_{\ell=0}^{\infty}\frac{(\frac{1}{2})_\ell (\gamma+1)_\ell}{(1+\omega)_\ell (1-\omega)_\ell}
    \frac{(-2(\alpha q)^2)^\ell}{\ell!}
\end{align}
as the coefficients reduce to Euler's integral $\Gamma(1+z) = \int_0^\infty x^z e^{-x}dx$,
    establishing Eq.~\ref{eq:closed_form_hyper}.

\subsection{The trigonometric form of the dielectric function}
The closed form of complex order connects the characteristic frequencies to the theory of special functions.
However, presently there are limited practical options to calculate with complex index.
For this reason $\varepsilon(\omega, k)$ 
represented as a trigonometric integral~(\cite{tataronis1970_cyclotron1,vogman2014dory,datta2021electromagnetic}),
\begin{equation}\label{eq:vogman_integral_form}
    \varepsilon(\omega, k) = 1 + \frac{\omega_p^2}{\omega_c^2}\int_0^{\pi}
    \frac{\sin(\theta)\sin(\theta\omega)}{\sin(\pi\omega)}\Big[\int_0^\infty f_\perp(v)J_0(\lambda(\theta)v)2\pi vdv\Big]d\theta
\end{equation}
with $\lambda = 2k\cos(\theta/2)/\omega_c$.
The inner integration is
a zero-order Hankel transform
$\mathcal{H}_0[f(r);k] = \int_0^\infty f(r)J_0(kr)rdr$.
The Hankel transform has a Fourier-multiplier property, 
\begin{equation}\label{eq:fourier_multiplier_property}
    \mathcal{H}_0[r^{2n}f(r);k] = (-\nabla_k^2)^n\mathcal{H}_0[f(r);k].
\end{equation}
Further, the radial Laplacians of the Gaussian are precisely the Laguerre functions,
\begin{equation}\label{eq:laguerre_formula}
    (\nabla_k^2)^n[\exp(-\alpha^2 k^2 / 2)] = (-2\alpha^2)^n L_n(\alpha^2 k^2/2)\exp(-\alpha^2k^2/2).
\end{equation}
Therefore the Hankel transform of $f_\gamma(v)$ is the family of polar Hermite functions
\begin{equation}\label{eq:polar_hermite}
    \mathcal{H}_0[f_\gamma(v); q] = L_{\gamma}\Big(\frac{\alpha^2 q^2}{2}\Big)\exp\Big(-\frac{\alpha^2 q^2}{2}\Big).
\end{equation}
Substitution of Eq.~\ref{eq:polar_hermite} into Eq.~\ref{eq:vogman_integral_form} gives Eq.~\ref{eq:trig_form},
as was to be shown.

\section{Magnetic potential well of the Weibel instability}\label{sec:magnetic_well_app}
The concept of a potential energy well is familiar to every physicist,
and essential, among other things, to understanding nonlinear phase space dynamics.
Unfortunately, the companion concept of potential momentum has been neglected historically due to confusing issues
which arose in the development of electromagnetic and relativistic theory,
and this has hindered an analogous understanding of magnetic trapping as a potential momentum well.
The conceptual consistency of potential momentum is reviewed positively in~\cite{griffiths_momentum} Section III.
Fortunately, simple magnetic trapping in a potential momentum well
is reducible to an effective potential energy in a lower dimensional phase space.
This effective potential is useful to model the collisionless trajectories of any magnetic trap,
and is commonly utilized to analyze particle beams in magnetic fields~(\cite{davidson1998, dodin2006}).
Here the effective potential method is applied to describe the saturation of Weibel instability.
\begin{figure}
    \centering
    \includegraphics[width=\columnwidth]{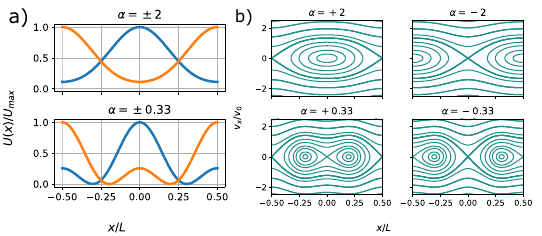}
    \caption{(a) Effective potentials of Eq.~\ref{eq:effective_potential2}
      and (b) phase portraits of trajectories in reduced phase space $(x,v_x)$ associated
      with motion in the unreduced phase space $(x,v_x,v_y)$
      from magnetic trapping in a transverse magnetic vector potential $A_y(x)=A_0\cos(kx)$.
      The parameter $\alpha\equiv mv_{y0}/(eA_0)$ 
      measures the ratio of kinetic-to-potential momentum
      and the $v_x$-axis is scaled to the transverse momentum $P_y$
      of the particle by $v_0\equiv (1 + |\alpha|)(eA_0/m)/\sqrt{2}$. 
      The opposite side of the phase space across the plane $v_y=0$ is observed through the inversion $v_{y0}\to -v_{y0}$
      taking $\alpha\to -\alpha$.
      The trapping potential bifurcates through $|\alpha|=1$ from a single-well around $\alpha\to\infty$ into a double-well around $\alpha=0$, as
      trajectories transition from bounce orbits to cyclotron orbits.
      This transition accomplishes the density filamentation associated with nonlinear magnetic trapping.
      That is, for $|\alpha| > 1$ inversion exchanges the elliptic and hyperbolic fixed points,
      but for $|\alpha| < 1$ the elliptic fixed points and their reflection nearly coincide.
      Thus, for $eA_0 \ll mv_{th}$ the decrease of particles by trapping is exactly balanced by an increase of passing particles
      of opposite transverse momentum.
      For this reason the low-amplitude bounce trapping has no associated density perturbation and instead produces a velocity perturbation.
      However, for $eA_0 \approx mv_{th}$ the near coincidence of the elliptic fixed points with their reflections
      produces a coherent density perturbation and filamentation.
      The electron density evolution shown in Fig.~\ref{fig:1d_weibel_field_density} can be understood through this bifurcation
      in the phase space topology.}\label{fig:app_d}
\end{figure}

Ideal magnetic trapping occurs when only magnetic potential $\bm{A}$ is present in the lab frame.
For example, consider the one-dimensional chain of electron holes formed by magnetic trapping at nonlinear saturation of the Weibel instability
of Section~\ref{subsec:single-mode-saturation-one-d}.
The motion is periodic in the $x$-direction with a transverse vector potential $\bm{A}=A_y(x)\hat{y}$.
The two constants of motion for a particle of mass $m$ and charge $q$ are the energy $H$ and the momentum $P_y$
(both of which are fixed by the particle's initial conditions),
\begin{align}
  H &= \frac{1}{2}mv_x^2 + \frac{1}{2}mv_y^2,\label{eq:constant_energy}\\
  P_y &= mv_y + qA_y,\label{eq:constant_momentum}
\end{align}
where for simplicity we neglect the second-order electric potential energy associated with the charge density
of magnetic trapping in a fixed neutralizing background.
The energy consists of only kinetic energy and is constant because the Lorentz force $q\bm{v}\times\bm{B}$ does no work on the particle.
On the other hand, constancy of momentum $P_y$ implies an exchange between kinetic momentum and potential momentum with a change in position.
Eliminating $v_y$ between Eqs.~\ref{eq:constant_energy} and~\ref{eq:constant_momentum} results in
\begin{equation}
  H = \frac{1}{2}mv_x^2 + U(x),
\end{equation}
representing one-dimensional motion in an effective potential
\begin{equation}\label{eq:effective_potential}
  U(x) \equiv \frac{1}{2m}(P_y - qA_y)^2.
\end{equation}
In stark contrast to trapping in a potential energy well (for example, $\varphi(x)=\varphi_0\sin(kx)$),
the magnetic well depends on the initial position $x_0$ \textit{and} velocity $v_{y0}$ of the charge
because the integrable dynamics are characterized by two constants of motion.
Indeed, motion is perturbed but not trapped in the direction associated with $\bm{A}$.

To model Weibel instability saturation, take $q=-e$ and $A_y = A_0\cos(kx)$ of amplitude $A_0$ and wavenumber $k$.
As a quadratic, the potential $U(x)$ has harmonics at $k$ and $2k$, the self-consistent currents of which
generate the harmonic cascade of the saturating instability.
The potential takes its extremum when $A_y=A_0$ for any initial position $x_0$.
Thus, in terms of the parameter
$\alpha \equiv mv_{y0}/(eA_0)$ measuring the ratio of kinetic-to-potential momentum,
the normalized form of the potential is
\begin{equation}\label{eq:effective_potential2}
  \frac{U(x)}{U_\text{max}} = \frac{(\alpha + \cos(kx))^2}{(|\alpha| + 1)^2}
\end{equation}
with $U_\text{max} = \frac{m}{2}(eA_0/m)^2(|\alpha| + 1)^2$. 
Around $\alpha=0$ the second harmonic dominates as $U=U_\text{max}\cos^2(kx)$,
while with $\alpha\gg 1$ the first harmonic $k$ dominates.
Figure~\ref{fig:app_d} illustrates this effect by plotting phase portraits using the normalized Hamiltonian
\begin{equation}\label{eq:normalized_hamiltonian}
  \frac{H}{U_\text{max}} = \frac{1}{2}\Big(\frac{v_x}{v_0}\Big)^2 + \frac{(\alpha + \cos(kx))^2}{(|\alpha|+1)^2}
\end{equation}
where $v_0\equiv \frac{P_{y,\text{max}}}{\sqrt{2}m}$ measures the particle's momentum.
Equation~\ref{eq:normalized_hamiltonian} bifurcates from a double-well around $\alpha=0$ to a single-well through $\alpha=1$,
physically indicating trapped trajectories transitioning from closed cyclotron orbits into magnetic bounce orbits.
Passing orbits have $H > U_\text{max}$, 
which about $v_y=0$ occurs for $H > \frac{m}{2}(eA_0/m)^2$.
In plain words, cyclotron orbits fill the inner well, magnetic bounce orbits fill the outer well,
and purely passing trajectories exceed the well barrier altogether.
In a fully three-dimensional model, the third component of kinetic energy associated with the $z$-direction shifts the
energy level in the well by an amount $\frac{1}{2}mv_z^2$.
Equation~\ref{eq:effective_potential} is the main result of this appendix as a conceptual tool to understand the
nonlinear phase space dynamics in the saturating Weibel instability.

\end{document}